\documentclass[12pt]{article}
\usepackage{amssymb}
\usepackage{amsmath}
\usepackage[dvips]{graphicx}
\usepackage{cite}
\usepackage{epsfig}
\usepackage{psfrag}
\renewcommand{\Large}{\large}

\begin{document}

\baselineskip=18pt
\numberwithin{equation}{section}
\allowdisplaybreaks  

\pagestyle{myheadings}

\thispagestyle{empty}

\newcommand{\bea}{\begin{eqnarray}}
\newcommand{\eea}{\end{eqnarray}}
\newcommand{\be}{\begin{equation}}
\newcommand{\ee}{\end{equation}}
\newcommand{\eq}[1]{(\ref{#1})}

\newcommand{\del}{\partial}
\newcommand{\delbar}{\overline{\partial}}
\newcommand{\zbar}{\overline{z}}
\newcommand{\wbar}{\overline{w}}
\newcommand{\vbar}{\overline{\varphi}}

\newcommand{\hf}{\frac{1}{2}}
\newcommand{\qrt}{\frac{1}{4}}\bibliographystyle{ieeetr}

\newcommand{\Z}{{\mathbb Z}}
\newcommand{\R}{{\mathbb R}}
\newcommand{\M}{{\cal M}}
\newcommand{\Q}{{\mathbb Q}}
\newcommand{\tX}{\widetilde{X}}
\newcommand{\mO}{\Omega}
\newcommand{\mJ}{{\mathbb J}}
\newcommand{\ep}{{\epsilon}}
\def\taubar{\overline{\tau}}

\def\eg{{\it e.g.}}

\pagestyle{plain}
\baselineskip=19pt

\begin{titlepage}

\begin{center}

\hfill                  arXiv:xxxx.xxxxx


\hfill CERN-PH-TH/2010-252

\vskip 1.5 cm {\Large \bf DEGENERATE STARS AND GRAVITATIONAL \\ [1mm]
  COLLAPSE IN ADS/CFT}
\vskip 1 cm {Xerxes Arsiwalla\footnote{x.d.arsiwalla@gmail.com}$^{a}$, Jan
  de Boer\footnote{j.deboer@uva.nl}$^{ab}$,
 Kyriakos
  Papadodimas\footnote{kyriakos.papadodimas@cern.ch}$^{c}$ and Erik
  Verlinde\footnote{ e.p.verlinde@uva.nl}$^{ab}$ }\\ {\vskip 0.75cm $^{a}$ Institute
  for Theoretical Physics  University of Amsterdam\\ 
$^{b}$ Gravitation and Astro-Particle Physics Amsterdam\\Science Park
  904, Postbus 94485, 1090 GL \\ Amsterdam, The Netherlands\\}

 {\vskip .75cm $^{c}$ Theory Group, Physics Department, CERN\\
CH-1211 Geneva 23, Switzerland\\}
\vskip .8cm

\end{center}

\vskip  .25 cm

\begin{abstract}
\baselineskip=16pt

We construct composite CFT operators from a large number of fermionic
primary fields corresponding to states that are holographically dual
to a zero temperature Fermi gas in AdS space. We identify a large $N$
regime in which the fermions behave as free particles. In the
hydrodynamic limit the Fermi gas forms a degenerate star with a radius
determined by the Fermi level, and a mass and angular momentum that
exactly matches the boundary calculations. Next we consider an
interacting regime, and calculate the effect of the gravitational
back-reaction on the radius and the mass of the star using the
Tolman-Oppenheimer-Volkoff equations. Ignoring other interactions, we
determine the "Chandrasekhar limit" beyond which the degenerate star
(presumably) undergoes gravitational collapse towards a black
hole. This is interpreted on the boundary as a high density phase
transition from a cold baryonic phase to a hot deconfined phase.

\end{abstract}

\end{titlepage}

\tableofcontents
\vskip 40pt
\section{INTRODUCTION}

String theory contains all kinds of elementary particles and forces,
and thus in principle can be used to describe all forms of matter. In
the past two decades much of the progress in string theory has
centered around black holes, which may be viewed as the most extreme
form of matter. In fact, the matter that makes up a black hole is in
such an extreme form that it has entirely disappeared from the picture
and is replaced by pure geometry.  Holography may be seen as an
attempt to rescue some of this matter by assigning material properties
to the black hole horizon.  To clarify the relationship between the
holographic matter and the matter that made the black hole one needs
to investigate the process of black hole formation, in other words,
gravitational collapse.

A related question is whether string theory, in the context of the
AdS/CFT correspondence \cite{Maldacena:1997re}, can be used to
describe other extreme forms of matter, such as the degenerate matter
that makes up a neutron star. Indeed, neutron stars, or more generally
degenerate stars like "quark stars" or "strange stars", can be seen as
black hole precursors in the sense that they may undergo gravitational
collapse to form a black hole. An essential feature of neutron stars
is that they are made out of fermions: it is the exclusion principle
that keeps the star from undergoing gravitational collapse. Our aim is
to obtain a description of a "neutron star" (or more precisely a
"degenerate star") in anti de Sitter space in terms of the dual
conformal field theory on the boundary. In particular, we are
interested to investigate and interpret the criterion that determines
whether the star undergoes gravitational collapse.

One of the first and most famous equations that deals with the physics
of a neutron star is the Tolman-Oppenheimer-Volkoff (TOV) equation. In
four dimensional flat space time it takes the form 
\be
\label{tov}
\frac{dp(r)}{dr} = - 2\,G\,\frac{  p(r)+\rho(r)}{r^2} \cdot
     { M(r) + 4\pi r^2 p(r) \over 1 - 2GM(r)/r} 
\ee 
where the radial dependent mass $M(r)$ is obtained from the energy
density $\rho(r)$ through \be {dM(r)\over dr}=4\pi r^2 \rho(r).  \ee
The TOV equation follows from combining the energy-momentum
conservation with the Einstein equations and holds for spherically
symmetric neutron stars.  It describes the radial dependence of the
pressure $p(r)$ and energy density $\rho(r)$ for a given equation of
state.  Solutions only exist when the pressure is sufficient to
sustain the gravitational force to prevent the neutron star from
collapsing. Typically one finds that there is an upper limit for the
mass beyond which the neutron star will be too heavy and starts to
collapse to form a black hole (or in some cases another type of
degenerate star). In their original paper \cite{Oppenheimer:1939ne}
Oppenheimer and Volkoff used the equation of state of a free
relativistic Fermi gas to find a limiting mass of 0.7 times the solar
mass. However, more realistic equations of state lead to a higher
value of the limiting mass around 2 or 3 solar masses.

In this paper we construct states in the boundary CFT that are the
holographic dual of a degenerate star in anti de Sitter space.  Our
discussion will be quite general without reference to a specific
AdS/CFT setup and the qualitative aspects of our results are
independent of the number of space time dimensions. Using the
state-operator correspondence we consider degenerate states
corresponding to composite multi-trace operators constructed from a
large number of copies of a given fermionic single trace operator.
Using 't Hooft factorization it is argued that in a strict large $N$
limit these operators behave as (generalized) free fields. The bulk
description of these states is given by a zero temperature degenerate
gas of relativistic fermions. In the limit of a large number of
particles one can give a hydrodynamic description purely in terms of
their energy density and pressure.  We first consider the case without
bulk interactions so that the equation of state is simply that of a
free degenerate Fermi gas. We show that in the hydrodynamic
description the degenerate star is confined to a spherical region in
AdS space, due to the gravitational potential well. The radius in AdS
units turns out to be given by the ratio of the Fermi momentum in the
center of the star and the mass of the constituent fermions.  As a
verification of our hydrodynamic description we show that it precisely
reproduces the mass predicted from a boundary analysis for static
degenerate stars as well as for the case with rotation.

Next we consider a double scaling regime chosen so that the
gravitational self interactions of the fermions will become important,
while we can still ignore Planck scale physics.  Since our primary
interest is to understand first qualitatively the criterion for
gravitational collapse we focus our attention to including the self
gravity of the degenerate star, while ignoring the role of all other
interactions. Thus, following Oppenheimer and Volkoff, we use the
equation of state of a free degenerate Fermi gas in the hydrodynamic
description of the star, and take the gravitational self interaction
into account through a generalization of the TOV equation to five
dimensional anti de Sitter space. The resulting equations are studied
numerically. We find that as a function of the central density, the
mass of the star at first increases, but reaches a limiting mass at
some critical value after which the mass slightly reduces again. We
interpret the limiting mass as the point after which the star will
become unstable and will start to undergo gravitational collapse. The
value of the limiting mass translates in the boundary theory to a
limiting conformal dimension of the composite operator made from the
fermionic primary fields.

A basic motivation for our work was towards analyzing the process of
gravitational collapse and black hole formation in AdS/CFT. This is an
interesting problem, which has been addressed in several works in the
past\cite{Ross:1992ba, Peleg:1994wx, Danielsson:1998wt,
  Danielsson:1999zt, Giddings:1999zu, Danielsson:1999fa,
  Balasubramanian:2000rt, Giddings:2001ii, Alberghi:2003pr,
  Hubeny:2006yu, Lin:2008rw, Chesler:2008hg, Bhattacharyya:2009uu,
  Chesler:2009cy}. The basic intuition is that radial collapse
corresponds to motion in the ``scale direction'' of the CFT, while the
subsequent black hole formation corresponds to deconfinement and
thermalization on the boundary theory.  A more quantitative
description of this picture would clearly be desirable, but is
difficult since the boundary theory is strongly interacting, the
process time dependent and the radial holographic dimension is not
easily represented in the CFT. In this direction, it may be helpful to
identify simple initial conditions to study (at least the onset of)
the dynamical process of collapse.  A star, on the verge of
gravitational collapse, is a good candidate since it is a static
configuration and presumably simpler to understand from the boundary
point of view.

Another motivation was to make contact with the recent developments on
the holographic analysis of strongly interacting fermions. In several
works \cite{Lee:2008xf, Liu:2009dm, Cubrovic:2009ye, Hartnoll:2009ns}
correlators of fermionic fields have been evaluated in the presence of
a black hole in AdS. Such computations reveal the presence of a Fermi
surface in the bulk, of non-Fermi liquid type. For a better
understanding of these results it might be worthwhile to explore the
boundary meaning of a simpler system, that of a Fermi-gas in AdS
without the presence of a black hole\footnote{The interesting paper
\cite{Hartnoll:2010gu} appeared while this draft was being prepared, in which
fermion correlators were evaluated on a background with fermions, without 
a black hole horizon.}. One should keep in mind that
while the fermions are free in the bulk, they are not ordinary free
fermions on the boundary (i.e. they do not obey a Dirac equation on
the boundary) as the boundary theory is strongly coupled. Instead they
are ``generalized free fields'' i.e. their correlators factorize even
though they do not obey linear wave equations. 

Finally let us mention another motivation for constructing a
holographic degenerate star, one of a more conceptual nature.  Our
degenerate star in AdS resembles, for many purposes, a conventional
macroscopic object. Since it is embedded in AdS it has a holographic
description on the boundary theory. It would be interesting to
understand in more detail how the hologram works in this case. In
particular it might be easier to ``decode'' the hologram in the case
of a star than for a black hole, due to the absence of a horizon. Our
star is a static, spherically symmetric configuration in the bulk. The
only nontrivial information is encoded in the radial profile of the
density. It is not obvious how this radial profile is encoded in the
boundary theory, since the radial dimension is not manifest on the
boundary but rather entangled with the ``scale'' of excitations in the
CFT. We believe that a star in AdS is a good toy model to further
explore the holographic mapping.

We would like to emphasize the following point about the
aforementioned goal. One might think that decoding the hologram on
the boundary theory is difficult because of strong coupling effects in
the gauge theory and thus hopeless unless one manages to compute at
strong coupling. While this is true to some extent, we believe that
there are interesting questions to be understood, which are not
dependent on strong coupling dynamics. In a large $N$ gauge theory the
strong coupling effects can be ``hidden'' in the conformal dimensions
and 3-point functions of single-trace gauge invariant operators. These
quantities are difficult to compute. Nevertheless, let us assume that
in some magical way we knew their exact values at strong
coupling. Would that be sufficient to understand how the hologram
works?  In principle it should\footnote{At least for length scales
  which are not too small compared to the size of AdS and in the
  absence of black holes. In other words when the ${\cal O}(N^2)$
  degrees of freedom are not excited nontrivially above their ground
  state at strong coupling.}, because from this information we can
reproduce all correlators of the CFT. However in practice, even armed
with this information, we do not know yet how to reconstruct the bulk.
Understanding how the radial profile of a macroscopic object, such as
our degenerate star, is encoded on the boundary theory may be a good
starting point for addressing this question. Ultimately it would be
fascinating to understand the CFT meaning of the Einstein/TOV
equations\footnote{In CFTs with classical gravity
  dual.}. Unfortunately we do not have much to report on this yet but
we hope to revisit this and related questions in future work.

The main idea presented in this work was briefly described in
\cite{deBoer:2009wk}. In this paper we expand on various aspects that
were not covered in detail in \cite{deBoer:2009wk}, we consider the
case of charged fermions and that of a thermal star, and finally we
discuss some issues about the validity of our approximations and the
embedding in specific AdS/CFT dualities. The plan of the paper is as
follows: in section \ref{sec:freelimit} we consider a ball of Fermi
gas in AdS in the limit of infinite $N$ and construct its dual
holographic representation. In section \ref{sec:interactions} we
discuss our double scaling limit, estimate the importance of various
interactions and present the TOV equations for the fermionic fluid. In
section \ref{sec:numerical} we present our numerical results and
discuss the Chandrasekhar limit beyond which the star undergoes
gravitational collapse. In section \ref{sec:boundary} we try to
understand the boundary interpretation of the gravitational
interactions between the fermions. In section \ref{sec:validity} we
discuss the extent to which our various approximations are justified
and the possibility of embedding our construction in a specific
AdS/CFT correspondence. In section \ref{sec:charge} we analyze the
case where the fermions are charged under a $U(1)$ gauge field in the
bulk. In section \ref{sec:superheated} we present a similar system
corresponding to a star made out of thermal gas and discuss its
relevance as the dual of a ``superheated phase'' in certain gauge
theories.  Finally in section \ref{sec:discussions} we close with some
discussions.

\section{FREE FERMI GAS IN ADS}
\label{sec:freelimit}

In this section we will study a degenerate gas of free fermions in
AdS$_{d+1}$. We will assume that the bulk theory is holographically
dual to a conformal field theory which has the analogue of a large $N$
expansion. To be more general we will express the expansion parameter
in terms of the central charge $c$ of the CFT.  In this section we
work in the limit $c\rightarrow \infty$, keeping other quantities
fixed. In this limit all interactions between single-trace operators
are suppressed by powers of $1/\sqrt{c}$ and the fermions become
essentially free.  The only interaction which remains is the
gravitational attraction of the AdS background which acts as a
confining box.  We work in a regime of the gauge theory where there is
a classical gravity dual, such as the $\lambda\gg 1$ limit in the
${\cal N}=4$ SYM\footnote{However, as we will discuss in section
  \ref{sec:validity}, the ${\cal N}=4$ SYM may not be the best setup
  to realize the degenerate star.}.

\subsection{Boundary description}
\label{subsec:qsbound}

We will construct composite multi-trace operators that are made out of
a large number of single trace fermionic operators.  The composite
operators will be "degenerate" in the sense that they represent the
operators with the lowest possible conformal dimension that can be
made out of a given number of single trace fermionic operators of a
certain kind.  Through the state-operator correspondence these
degenerate composites represent the states that are holographically
dual to the degenerate fermionic gas in the bulk theory.

\subsubsection{Degenerate fermionic operators}

Gravity in AdS$_{d+1}$ is holographically dual to a conformal field
theory on $S^{d-1}\times {\rm time}$.  The Hilbert spaces of the two
theories are isomorphic, so any quantum state in the bulk is dual to a
state in the CFT. The states of the CFT can be conveniently labeled by
local operators on the plane via the state-operator correspondence
$$
\lim_{z\to 0} \Phi(z)\left | {\rm vac}\right \rangle= \left |\Phi\right\rangle
$$ The energy of the state $|\Phi \rangle$ is equal to $\Delta /\,\ell$
where $\ell$ is the radius of $S^{d-1}$ and $\Delta$ the dimension of
the operator $\Phi$. The other quantum numbers of the state (like the
angular momentum and R-charge) are also determined by those of $\Phi$.

In AdS/CFT one makes a distinction between single trace and multi
trace operators. The first kind may be thought of as single particle
states, while the latter are multi-particle states. However, this
distinction is only well defined at large $N$, or more generally for
certain classes of CFTs with a large central charge $c$. The reason
being that $1/N$ effects will lead to a mixing between single trace
and multi-trace operators.

For certain strongly coupled supersymmetric CFTs the bulk theory is
given by an AdS supergravity theory, which for large central charge
can be treated (semi-) classically.  The bulk theory contains bosons
and fermions, and accordingly the boundary single trace operators may
be distinguished into bosonic and fermionic operators. Whether a field
is fermionic or bosonic is determined according to the spin-statistics
theorem through its representations of the $SO(d,2)$ conformal
symmetry group. In particular, primary operators that transform in the
spinor representation correspond to fermions, and have correlation
functions that are anti-symmetric under the exchange of two identical
operators.

The fermionic nature of primary operators can in specific cases also
be understood from their construction in terms of the basic fields of
the underlying field theory. To give a concrete example, let us
consider the case of ${\cal N}=4$ SYM. The spectrum of operators
contains chiral primary fields that are holographically dual to the
Kaluza-Klein modes of the type IIB supergravity theory on AdS$_5\times
$S$^5$. These chiral primaries form representations of the $SO(6)$
R-symmetry group, and are arranged in short supermultiplets. The
lowest component is given by the bosonic single trace operators
$$ 
\Phi^{I_1I_2\ldots I_k} =\bigl [ {\rm tr}\left(
  \phi^{I_1}\phi^{I_2}\phi^{I_3}\ldots\phi^{I_k
  }\right)\bigr]_{{}_{\rm symmetric}\atop{\rm traceless}}
$$ 
By applying a supersymmetry transformation one learns that the same
supermultiplet contains fermionic single trace operators of the form
\be \Psi_{A\alpha}^{I_1I_2\ldots I_k} =[Q_{A\alpha},\Phi^{I_1I_2\ldots
    I_k}]=\bigl [\Gamma^{I_1}_{AB} {\rm tr}\left(
  \lambda_a^B\phi^{I_2}\phi^{I_3}\ldots\phi^{I_k
  }\right)\bigr]_{{}_{\rm symmetric}\atop{\rm traceless}} 
\ee 
Here $\Gamma^I_{AB}$ is an $SO(6)$ gamma matrix, with spinor indices
$A, B$ and vector index $I$.  We note that this operator is not
primary with respect to the superconformal group, but it is a
conformal primary operator with respect to its bosonic subgroup. The
BPS bound implies that the field $\Psi$ has conformal dimension
$k+{1\over 2}$ for all values of the 't Hooft coupling $\lambda$.  The
fermionic field $\Psi$ also carries an $SO(4)$ spinor index $\alpha$,
and a multi-index made from a symmetric traceless combination of
vector indices.  In the following we clearly wish to avoid using all
of these indices. Indeed, except for the fact that these fields are
fermionic, we have no need for any of these quantum numbers. This is
because we will work either in a regime in which we ignore all
interactions, or in an approximation in which we do the same except
for gravity\footnote{And even if we would try to keep track of
  R-charges, we can still construct neutral composite states by
  including all fermionic operators in a complete multiplet. The only
  complication this will lead to is a degeneracy factor equal to the
  dimension $g$ of the $SO(6)$ representation corresponding to the
  traceless symmetric $s$-tensors.}.  All this will not be very
relevant for what follows, but merely was meant to give a concrete
example of a fermionic primary field.

Let us now start with a fermionic single trace operator $\Psi$ of
dimension $\Delta_0$. For simplicity we ignore the spinor indices of
$\Psi$ since this will not affect any of our following results
qualitatively.  We emphasize that $\Psi$ is not one of the fundamental
field of the Lagrangian but rather a gauge invariant operator which
has finite conformal dimension at strong coupling.  We work in
normalization where the 2-point function is
$$
\left\langle \Psi(x) \overline{\Psi}(y)\right\rangle = {1\over |x-y|^{2\Delta_0}}
$$ 
The 't Hooft large $c$ factorization implies that for single trace
operators the connected correlators are suppressed as
$$
\bigl\langle \Psi(x_1)\Psi(x_2)... \Psi(x_n) \bigr\rangle_{con} ={\cal O}\left(c^{2-n\over 2}\right)
$$ which in turn implies that the field $\Psi(x)$ behaves effectively
like a ``free'' field at large $c$.  We can construct multi trace
operators by taking products of many of these fields without having to
worry about operator mixing with operators with a different number of
traces. In a sense, at large $c$ the fermionic operators generate a
Fock space, just like free fields\footnote{Though it is not a standard
  free field in $d$ dimensions, but rather a ``generalized free
  field''\cite{Jost}: it does not obey any linear wave equation since
  its conformal dimension is non-canonical but nevertheless its
  correlators factorize.}.  This fact will be true as long as the
number of operators, or more precisely the total conformal dimension,
is small compared to the central charge $c$ of the CFT.  This point
will become important when we discuss the possible role of
interactions.
 
We now turn to the multitrace operators that can be constructed from
the operator $\Psi$.  From the fact that correlation functions are
antisymmetric under the exchange of two fermionic primaries
$$
\bigl\langle\ldots \Psi(x)\Psi(y) \ldots \bigr\rangle =-\bigl\langle\ldots \Psi(y)\Psi(x) \ldots \bigr\rangle
$$ 
one derives that the operator product of two of these primaries is
also antisymmetric
$$
\Psi(x)\Psi(y)=-\Psi(y)\Psi(x).
$$
Normally one defines the normal ordered product $:\Psi^2(x):$ by
taking the limit $x\to y$ after subtracting possible divergent
terms. For fermionic fields the result is zero, due to the
antisymmetry. This is of course just a reflection of the Pauli
exclusion principle. Therefore, to construct a nontrivial operator out
of two fermionic primaries we have to include a derivative. This gives
operators of the kind $:\Psi\partial_i\Psi:$, where $\partial_i$
denotes the derivative w.r.t.\ $x_i$. This operator is the operator of
lowest conformal dimension that can be found in the regular part of
the operator product expansion of $\Psi(x) \Psi(y)$ \footnote{This
  statement is precise in the $c\rightarrow \infty$ limit.}. For a
$d$-dimensional CFT there are $d$ types of these operators.

We want to continue with the construction of multi-trace operators of
the lowest possible conformal dimension made out of three, four, all
the way to a very large number of copies of the basic fermionic
operator $\Psi$. Each time we have to consider the fact that the next
operator needs to be anti-symmetrized with the previous ones. So we
have to act again with derivatives every time making sure that the
combination of derivatives has not been used before and also that we
are using the smallest number of derivatives that are necessary. If we
have constructed an operator of this type with $N_F$ insertions of the
basic operator $\Psi$, we can inductively define the degenerate
operator with $N_F+1$ insertions as the operator with lowest conformal
dimension appearing in the regular part of the OPE between the
composite with $N_F$ insertions of $\Psi$'s and one additional
$\Psi$. This definition is unambiguous at infinite $c$. Continuing
like this one notices that one is filling up "shells", very much like
an atom, where each shell is labeled with the number $n$ of
derivatives that are being used.

In order to end up with a rotationally symmetric composite operator it
is natural to consider the "noble elements" with a completely filled
last shell. We will denote the number of derivatives in the last shell
by $n_F$, since it plays the role of a Fermi level. In this way we
arrive at the following form of the degenerate composite operators 
\be
\label{Phi}
{\bf \Phi}= \Psi \prod_i \partial_i \Psi\prod_{\lbrace i,j
  \rbrace}\partial_i\partial_j \Psi\prod_{\lbrace i,j,k
  \rbrace}\partial_i\partial_j \partial_k \Psi \ldots
\ldots\!\!\!\!\prod_{\lbrace i_1, i_2,\ldots
  i_{n}\rbrace}\!\!\!\partial_{i_1}\partial_{i_2}
\ldots\partial_{i_{n_F}}\!\!\Psi 
\ee 
where one takes in each shell the product over all possible $n$-tuples
made from $d$ derivatives.  We note that the composite operator only
depends on the choice of the number $n_F$. It does not carry any other
indices, and hence the corresponding state is unique. In other words,
we do not have to deal with counting states: the total entropy is
equal to zero. It should also be obvious why these composite operators
are called "degenerate".

In the bulk interpretation each insertion of the operator $\Psi$
represents the addition of a particle. Some simple combinatorics show
that a completely filled shell made from fields with $n$ derivatives
contains
$$
\left(\! \begin{array}{c}\! n \! +\! d\!-\! 1\! \\ \! d\!-\! 1\! \end{array}\! \right)
$$ 
different fields, and hence corresponds to that many particles in
the bulk. The total number of fields (or particles) is found by
summing over all shells 
\be
\label{NF}
N_F = \sum_{n=0}^{n_F} \left(\! \begin{array}{c}\! n \! +\! d\!-\! 1\!
  \\ \! d\!-\! 1\! \end{array}\! \right) = \left(\! \begin{array}{c}\!
  n_F \! +\! d\! \\ \! d\! \end{array}\! \right) 
\ee 
Next let us now determine the conformal dimension of a degenerate
composite operator $\bf \Phi$ with $n_F$ filled shells.  The conformal
dimension of the operators in the $n$-th shell is $n+\Delta_0$, since
each derivative adds one to the conformal dimension.  Hence the total
conformal dimension is 
\be
\label{DeltaF}
\Delta = \sum_{n=0}^{n_F}(n+\Delta_0) \left(\! \begin{array}{c}\! n \!
  +\! d\!-\! 1\! \\ \! d\!-\! 1\! \end{array}\! \right) =d
\left(\! \begin{array}{c}\! n_F \! +\! d\! \\ \! d\!+\!
  1\! \end{array}\! \right) +\Delta_0\left(\! \begin{array}{c}\! n_F
  \!  +\! d\! \\ \! d\! \end{array}\! \right) 
\ee 
The second term is proportional the the total number of fermions and
represents the contribution of their ``rest mass'' to the energy,
while the first term is the ``kinetic energy'' coming from the
derivatives.

In this discussion we have ignored so far the spinor index of the
field. This can be treated in a similar way as the electron spin in
atomic physics. It simply tells us that each state can be occupied by
a number of fermions, namely as many as the dimension of the spinor
representation. So if we would have included it, we would have to
write an additional product over $\alpha$ which would have multiplied
the counting formulas for $N_F$ and $\Delta$ by the dimension of the
spinor representation. Similarly, if we would have considered a number
of species $g$ of fermions, these formulas have an overall factor $g$.

Apart from these harmless simplifications we would like to emphasize
that our discussion so far has been exact for all values of the
coupling of the gauge theory, as long as $c\rightarrow \infty$. We
have assumed that the basic building block $\Psi$ of our operators has
conformal dimension $\Delta_0$. At the moment we do not care whether
this $\Delta_0$ is the same as the conformal dimension of $\Psi$ at
weak coupling.  For us $\Delta_0$ can be considered as an input
parameter which is determined by the (possibly strongly coupled)
dynamics. But even this issue could be settled by taking $\Psi$ to be
a chiral primary in a supersymmetric theory. On the other hand the
operator ${\bf \Phi}$ defined in \eqref{Phi} is definitely not
supersymmetric, even if $\Psi$ is, since the derivatives increase the
conformal dimension without adding R-charge. The reason that the
expression \eqref{DeltaF} is still reliable is based on the 't Hooft
factorization at large $c$: the conformal dimensions of multi-trace
operators are equal to the sum of the individual conformal dimensions
up to ${1\over c}$ corrections. This statement is based on the planar
expansion and does not depend on the value of the 't Hooft
coupling. Hence in the limit $c\rightarrow \infty$ our formula
\eqref{DeltaF} is exact. The 't Hooft factorization also guarantees
that at large $c$ we do not have mixing of the operator ${\bf \Phi}$
with other operators with the same quantum numbers. These statements
are reliable if we take $c\rightarrow \infty$ keeping other quantities
(like $\Delta_0$ and $n_f$) fixed.

Finally, let us finish this section with a comment on the $SO(d)$
decomposition of the states in the $n$-th shell. These states are
given by symmetric $n$-tensors, but since they are not traceless they
do not form an irreducible representation. The dimension of the
irreducible representation of the symmetric traceless $n$-tensors is
given by the number of symmetric $n$-tensors minus the number of
symmetric $(n\!-\!2)\,$-tensors. In this way we find that the states
in the $n$-th shell decompose into a sum of representations
corresponding to traceless symmetric $(n\!-\! 2k)\,$-tensors, with $k$
an integer that runs from\footnote{Here with [$\ .. \ $] we denote the
  integral part of a possibly fractional number.} $0$ to $[n/2]$
. Just as a check one verifies that 
\be
\label{decomp}
\left(\! \begin{array}{c}\! n \! +\! d\!-\! 1\! \\ \! d\!-\!
  1\! \end{array}\! \right) = {\sum_{k=0} ^{[n/2]}
}\left\lbrack\left( \begin{array}{c}\! n\!-\!2k\! +\!d\!-\! 1\\ \!
  d\!-\!1 \!\end{array}\! \!\right) - \left( \begin{array}{c}\!
    n\!-\!2k \! +\!d\!-\!3\\ \!d\!-\! 1\!\end{array}\!
    \!\right)\right\rbrack 
\ee

\subsubsection{``Free field'' mode expansion }

The state-operator correspondence is a convenient way to describe
states in a CFT. The relationship with the corresponding bulk state
becomes more manifest, however, in a slightly different representation
of the states.  Instead of creating the state at $\tau=-\infty$ one
can represent the same state by making use of a mode expansion of the
primary field on the cylinder $S^{d-1} \times R$.  So we go back to
Minkowski signature.  At large $c$ the primary operator can be
expanded in modes \cite{Banks:1998dd} on $S^{d-1}$ as \be
\label{modeexpansion}
\Psi(t,\Omega) = \sum_{n,l,m} {\bf \alpha}_{n,l,m}\,
Y_l(\Omega_{d-1}) e^{i E_n t} + {\bf \beta}^\dagger_{n,l,m}\,
Y^*_l(\Omega_{d-1}) e^{-i E_n t} 
\ee 
where the energies satisfy the condition $E_{nl} = (\Delta_0 + n)/
\ell$, $n=0,1,2...$, determined by the representations of the
conformal group. The expansion \eqref{modeexpansion} has to be
understood as a definition of the operators
$\alpha_{n,l,m},\beta_{n,l,m}$. Using the 2-point function of the
field $\Psi$ and the fact that its correlators factorize at large $c$
we find that the operators $\alpha_{n,l,m},\beta_{n,l,m}$ satisfy the
algebra of fermionic ladder operators 
\be
\label{comrel}
\{\alpha_{n,l,m},\alpha^\dagger_{n',l',m'}\} \sim
\delta_{nn'}\delta_{ll'}\delta_{m,m'}\qquad,\qquad
\{\beta_{n,l,m},\beta^\dagger_{n',l',m'}\} \sim
\delta_{nn'}\delta_{ll'}\delta_{m,m'} 
\ee 
and all other anticommutators are zero. The vacuum state satisfies the
condition $\alpha_{n,l,m}|vac \rangle =
\beta_{n,l,m}|vac\rangle=0$. So we can use these modes to build a
multi particle state from the vacuum.  In this way one finds the state
$|{\bf \Phi}\rangle$ corresponding to the operator (\ref{Phi}) has an
alternative representation as
\be
\label{Phib}
|{\bf \Phi}\rangle= \prod_{ n,l, m} \alpha^\dagger_{n,l,m}|vac\rangle
\ee 
However it should be stressed that the expansion \eqref{modeexpansion}
and the commutation relations \eqref{comrel} are only true in the
$c\rightarrow \infty$ limit, keeping other quantities fixed. If the
operator $\Psi$ is inserted between states whose conformal dimension
is of the order of the central charge, then factorization may not hold
any more and the commutation relations \eqref{comrel} will be
modified. Hence the repeated use of the modes of the operator $\Psi$
is only allowed in the limit of large central charge where
factorization is reliable.

\subsubsection{Density of states}

Let us now introduce the total mass $M$, the fermion mass $m_f$, and
the Fermi energy $\epsilon_F$ via 
\be
\label{identi}
m_f={\Delta_0\over \ell},\qquad M={\Delta\over \ell}, \qquad
\epsilon_F={ n_F+\Delta_0\over \ell}.  \ee where $\ell$ is the radius
of the boundary $S^{d-1}$.  We express the number of particles $N_F$
given in (\ref{NF}) and the mass $M$ in terms of the Fermi energy
$\epsilon_F$ and the fermion mass $m_f$ in the limit that
$n_F=(\epsilon_F-m_f)\ell$ and $\Delta_0=m_f\ell$ become large.  The
number of fermionic particles may be expressed as \be
\label{NEF}
N_F = {\ell^d \over d!} (\epsilon_F-m_f)^d  
\ee
where we ignored all terms that are subleading in $n_F$.  For the
total mass we use the expression (\ref{DeltaF}) for the total
conformal dimension $\Delta$.  Keeping the leading contributions of
both terms we find
\be
\label{Mtot}
M= {d\, {\ell^d}\over (d\!+\! 1)!}(\epsilon_F-m_f)^{d\!+\!1} +
{m_f\,\ell^d \over d!}(\epsilon_F-m_f)^d 
\ee 
It is important that we keep both $\epsilon_F$ as well as $m_f$ in
these expressions, since we want to keep the ratio of $\epsilon_F$ and
$m_f$ finite.  The expressions (\ref{NEF}) and {\ref{Mtot}) only hold
  in the specified regime, since we have not taken into account any
  interactions, gravitational or otherwise.

We will be interested in the limit of large $n_F$ (and large
$\Delta_0$), so it is more convenient to rewrite the sums in the
expressions (\ref{NF}) and (\ref{DeltaF}) as integrals. This gives 
\be
\label{relation}
N(\epsilon_F) = \int^{\epsilon_F}_{m_f} \!\!\!\! d\epsilon
\,g(\epsilon)\, ,\qquad M(\epsilon_F) =\int^{\epsilon_F}_{m_f}
\!\!\!\! d\epsilon \,\, \epsilon\, g(\epsilon)\, , 
\ee 
where the density of states $g(\epsilon)$ is given by
\be
\label{DOS}
g(\epsilon) = \frac{\ell^d}{(d\!-\!1)!}(\epsilon-m_f)^{d-1}
\ee
This expression is valid in the strict large $c$ limit. 

Notice that this density of states differs qualitatively from that of
a free fermion in $d$ spacetime dimensions. For the latter the density
of states grows like $\epsilon^{d-2}$. In other words, while the
fermionic operator $\Psi$ is defined in a $d$ dimensional CFT we have
found that the density of states created by it grows like that of a
$d+1$ dimensional gas. Of course this is related to the fact that the
operator $\Psi$ is not really a fundamental free field obeying a
linear wave equation, but rather a ``generalized free field'' which
naturally lives in one additional dimension. In a sense this dimension
is the scale in the conformal field theory.

\subsection{Quantum states in the bulk}
\label{subsec:qsbulk}

The fermionic single trace operator $\Psi $ corresponds to a fermionic
field $\psi$ in AdS spacetime.  On-shell this field satisfies a field
equation which in principle depends on the spin of the field. For
spin one-half it would be the Dirac equation, or for higher spin one
would get the Rarita-Schwinger equations and its generalizations. But
since eventually we are interested in taking the limit of a large
number of particles and making contact with a hydrodynamic
description, the distinction between the different type of fermionic
particles and field equations becomes irrelevant.  Therefore, to
simplify our discussion we will use the partial waves of the
Klein-Gordon equation, which are easier to deal with and closer to the
standard knowledge\footnote{The fact that the Klein-Gordon equation
  usually describes bosonic particles does not pose a problem, because
  we can implement Fermi statistics by hand and take the degeneracy of
  the spin degrees of freedom into account afterwards. Again, this is
  very analogous to the standard treatment of the electron spin in
  atomic physics and is a reliable approximation in the limit of large
  number of particles.}.

The global AdS$_{d+1}$ metric is given by 
\be\label{adsmetric} ds^2 =
-\left(1+{r^2\over \ell^2}\right)dt^2+ \left(1+{r^2\over
  \ell^2}\right)^{-1}dr^2+r^2d\Omega_{d-1}^2.  
\ee 
where $\ell$ is the AdS radius.

We will assume that the field $\psi$ satisfies the
$d\!+\!1$-dimensional Klein Gordon equation in AdS space, which in
full detail looks like \be \left\lbrack-\left(1+{r^2\over
  \ell^2}\right)^{\!\! -1}\!\partial_t^2+ r^{1-d}
\partial_r\left(\Bigl(1+{r^2\over
  \ell^2}\Bigr)r^{d-1}\partial_r\right)+r^{-2} \Delta_{\Omega}+
m_f^2\right\rbrack\psi(t,r,\Omega)=0, \ee where $\Delta_\Omega$
denotes the laplacian on $S^{d\!-\!1}$. The mass $m_f$ of the field
$\psi$ and the conformal dimension of the operator $\Delta_0$ are
related by the formula $\Delta_0 = {d\over 2} + \sqrt{ {d^2 \over 4} +
  (m_f\ell)^2}$. For large values of $m_f$ we can approximate
$\Delta_0 \approx m_f \ell$, which of course was the reason for the
identification in \eqref{identi}. Since we are interested in states
which can be considered as finite energy excitations above the
standard vacuum of the theory, we only focus on the normalizable modes
in AdS.  We then find that the partial wave solutions of the Klein
Gordon equation take the general form \be
\label{modes}
\psi_{n,l,{\bf m}}(t,r, \Omega )= e^{-i(n+\Delta_0)
  t/\ell}\,f_{n-2l,l}(r)\, Y_{l,{\bf m}}(\Omega).  \ee Here
$f_{n,l}(r)$ is some hypergeometric function \cite{Aharony:1999ti},
whose precise form will not be important for us.   The key point is
that both the bulk and the boundary fields have the same quantum
numbers. This makes the correspondence straightforward.  To be
precise, the relationship with the operator $\Psi$ in the CFT may be
expressed by \be \Psi(t,\Omega)| vac\rangle= \lim_{r\to \infty}
e^{\Delta_0 r /\ell} \, \psi (t,r, \Omega )|vac\rangle \ee This
implies that the creation modes that we defined in the CFT are indeed
directly related to those constructed out of the bulk fields. The
Hilbert spaces are therefore completely identified, at least in this
free field limit. This also implies that the composite operator that
we constructed indeed may be identified with a multiparticle state,
whose complete wave function may be represented in the bulk as a Slater
determinant in terms of the modes given in (\ref{modes}).
\be
\label{slaterbulk}
|{\bf \Phi}\rangle= \prod_{ n,l, m} \alpha^\dagger_{n,l,m}|vac\rangle
\ee

\subsection{Hydrodynamic description}
\label{subsec:hydrofree}

In this section we will give a hydrodynamic description of the
multi-particle state as a free degenerate Fermi gas in AdS space.  The
Fermi gas is found to be confined to a spherical region with a radius
determined by the ratio of the Fermi momentum and the fermion mass.
This fact is due to the gravitational potential well of the AdS space
and not due to self-gravity.

\subsubsection{The equation of state}

At first one may think that the equation of state for a degenerate
Fermi gas in AdS space is different from that in flat spacetime. For
instance, the spacing of the lowest energy levels are determined by
the AdS radius, and differ from that of flat space. But when the
number of particles within one AdS volume is very large, one can
safely ignore the effect of the curvature radius, and use a local
description of the gas as if it lived in flat spacetime. 
As we will see this approximation reproduces the exact results of
the previous subsection in the limit of large number of particles.

To make this point more clear, consider a degenerate free Fermi gas in
a box of size $L$ in $d$ spatial dimensions. The Fermi momentum $k_F$
is related to the chemical potential $\mu$ via \be
\label{ekf}
\mu = \sqrt{k_F^2+m_f^2}.
\ee
At zero temperature all states satisfying the bound
\be
\label{kF}
{2\pi |\vec{n}|\over L}\leq k_F \ee are being
occupied. Semi-classically we can count states by calculating the
volume of phase space. Hence, the particle number density ${\bf n}$ is
equal to the volume of the d-dimensional Fermi sphere: \be
\label{number}
{\bf n} = \frac{V_{d-1}k_F^{d}}{d (2\pi)^d}.  \ee Here $V_{d-1}$ is
the area of a unit $(d\!-\!1)$-sphere.

These are the usual relations that hold in flat space. But as long as
the number of particles per AdS volume is very large, the Fermi
momentum will be much larger than the inverse AdS radius
$\ell$. Therefore, in order to determine the equation of state one can
take the size $L$ of the box very small compared to $\ell$. This means
that the flat space relations given above also apply locally in the
curved AdS background.  The only difference with a translationally
invariant flat space situation is that in this case one should expect
that the Fermi momentum $k_F$ and chemical potential $\mu$ depend on
the position in the AdS space.

The energy density and pressure are given by the standard expressions
\bea
\label{textbook}
\rho &=& \frac {V_{d-1}}{(2\pi)^d}\int_0^{k_F}\!\!\!dk\,k^{d-1}
\sqrt{k^2+m_f^2} ,\nonumber \\ p & =& \frac {V_{d-1}}{d
  (2\pi)^d}\int_0^{k_F}\!\!\!dk\,\frac{k^{d+1} }{ \sqrt{k^2+m_f^2}} ,
\eea and obey the standard thermodynamic relations
\be
\label{identity}
 d\rho\ = \mu\, d {\bf n}, \qquad \rho+p = \mu\, {\bf n}.  \ee One
 small comment about the energy density $\rho$ will be useful. Namely,
 it can be written as \be
\label{rhon}
\rho=\int_{m_f}^{\epsilon_F} \epsilon\, {d{\bf n}\over d\epsilon}\,
d\epsilon\,\,
\ee Intuitively, the validity of this identity should be
obvious: one simply adds all energies $\epsilon$ with a weight given
by the local density of particles with that energy. Similarly the pressure
can be written as
$$
p = \int_{m_f}^{\epsilon_F} {\bf n}(\epsilon) d\epsilon
$$

\bigskip

\subsubsection{Energy momentum conservation.}

In the hydrodynamic limit we expect to be able to describe the free
Fermi gas purely in terms of the local energy density $\rho$, the
pressure $p$, and a local velocity field $u^\mu$ normalized so that
$u^\mu u_\mu=-1$. The energy momentum tensor is given in terms of
these quantities as \be T_{\mu\nu} = (\rho+p)u_{\mu} u_{\nu} + p
g_{\mu\nu} \ee We will first consider the non-rotating degenerate
state and postpone the discussion of the rotating state to section
\ref{subsec:rotatingcase}.  Without rotation the energy density and
pressure will only depend on the radius $r$. In this section we are
still ignoring the effect of self-gravity of the gas. However, we
still have to take into account the effect of the curved AdS
background. This is done by imposing conservation of the energy
momentum tensor. For a static configuration it is the equivalent of
imposing hydrostatic equilibrium.  Anticipating possible
generalizations let us write the metric in the general form \be
\label{Ansatz}
ds^2 = - A(r)^2 dt^2 + B(r)^2 dr^2 + r^2 d\Omega^2_{d-1}.  \ee The
Fermi gas is static, and hence we take $u_t=A(r)$ and all other
components of $u$ equal to zero.  The only non-trivial equation is
$\nabla^{\mu} T_{\mu r}=0$, which becomes \be \label{dpdr} {dp\over
  dr} +\frac{A'}{A} (\rho+p) =0.  \ee This equation is surprisingly
easy to solve. By making use of the identities (\ref{identity}) one
easily verifies that (\ref{dpdr}) is satisfied when the chemical
potential obeys \be
\label{redshift}
\mu(r)= \frac{\epsilon_F}{ A(r)}, \ee where at this stage $\epsilon_F$
is an arbitrary constant.  We conclude that the radial dependence of
the chemical potential is simply due to the gravitational
redshift. With some hindsight this result could have been anticipated:
it is an analogue of the familiar Tolman relation for the temperature,
which is known to hold for any static gravitational field.

So far, this discussion has been general in the sense that we have not
used the explicit form of the AdS metric.  But let us now go back to
the specific case of the AdS space time, and investigate its energy
and pressure profile in more detail. For the AdS metric
(\ref{adsmetric}) the relation (\ref{redshift}) becomes \be
\label{efrads}
\mu(r)= {\epsilon_F\over \sqrt{1+{r^2/ \ell^2}}}.  \ee Now that we
have determined the radial profile $\mu(r)$ of the chemical potential
we in principle also know the value of the energy density $\rho(r)$
and the pressure $p(r)$ through the relations (\ref{textbook}). It is
clear from this result that the chemical potential decreases as one
goes out to the boundary. It can not decrease arbitrarily, however,
since the energy per particle is always larger than the mass
$m_f$. Namely, when $\mu=m_f$ the Fermi momentum $k_F$ vanishes and so
do the energy density and pressure.  We conclude therefore that
$\rho(r)$ and $p(r)$ will go to zero at a finite value $R$ of the
radius, namely when \be
\label{edge}
\mu(R)=m_f.  \ee It will be interesting to re-express the value of the
radius $R$ in terms of the Fermi momentum $k_F(0)$.  Combining the
equations (\ref{efrads}) and (\ref{edge}) together with the standard
relation between $k_F$ and $\epsilon_F$ gives \be
\label{radius}
R= {k_F(0)\over m_f}\ell.  \ee Hence, the size of the star in AdS
units is a direct measure for how relativistic the fastest particles
are in its center.  When the size of the star is of the order of the
AdS radius $\ell$ these particles are mildly relativistic, while they
have ultra-relativistic velocities when the star is much larger than
$\ell$.  It is important to note that the Fermi gas has a sharp edge
only when the fermions are massive.  For massless fermions the
equation of state is simply that of a relativistic gas, and the star
will spread out with a dilute "tail" towards the boundary of
AdS\footnote{Of course when the density gets too low the hydrodynamic
  approximation breaks down.}.

\subsubsection{Total mass and particle number.}

The degenerate star-like object that we described in the previous
subsections is the holographic dual of the boundary state
$|\Phi\rangle$ constructed in section 2.1. As a non-trivial check on
our hydrodynamic description one can verify that the particle number
and mass agree with the boundary calculations.  We observe that the
constant $\epsilon_F$ equals the chemical potential defined with
respect to the time $t$. This means that it can be identified with the
Fermi energy $\epsilon_F$ of the CFT \be
\label{identification}
\epsilon_F^{{}^{{}_{CFT}}} = \epsilon^{{}^{{}_{AdS}}}_F \ee The total
particle number in the bulk is given by the integral \be
\label{integralN}
N^{{}^{{}_{AdS}}}_F= V_{d-1} \int_0^R\!\!  dr \, r^{d-1}{B(r)}\,{{\bf
    n}(r)}.  \ee The factor $B(r)$ comes from the volume form on the
spatial section.  Similarly, one obtains the mass $M^{{}^{{}_{AdS}}}$
by integrating the energy density. Because $M^{{}^{{}_{AdS}}}$ is
defined with respect to the time coordinate $t$ one has to include a
redshift factor $A(r)$. For the AdS metric \eqref{adsmetric} the redshift
factor $A(r)$ cancels the measure factor
$B(r)$, and hence \be
\label{integralM}
M^{{}^{{}_{AdS}}}= V_{d-1} \int_0^R\!\!  dr \,r^{d-1}\rho(r).\, \ee 
To calculate $N^{{}^{{}_{AdS}}}_F$ and
$M^{{}^{{}_{AdS}}}$ one has to insert the equations (\ref{number}) and
(\ref{textbook}) for the particle and energy density and include the
redshift effect (\ref{redshift}) into the chemical potential.  The
resulting integrals can be performed analytically and (perhaps not
surprisingly) precisely reproduce the results (\ref{relation}) and
(\ref{DOS}) in the limit of large number of particles.

\subsubsection{A comment on the validity of the hydrodynamic description}

In general, a basic assumption of the hydrodynamic approximation is
that the continuous system under investigation is in local
thermodynamic equilibrium. More precisely the hydrodynamic description
is an expansion in (space-time) derivatives of local thermodynamic
quantities. As such, hydrodynamics is valid when the typical variation
length is much larger than the microscopic scales of the fluid. In
particular it should be much larger than the mean free path. In our
case the ``fluid'' is actually a free gas, so the mean free path is
infinite or at most bounded by the AdS scale. This means that strictly
speaking we are not allowed to use hydrodynamics. In particular if we
had considered time dependent configurations, then there would clearly
be a problem since our fermions are non-interacting and there would be
no restoring agent to bring them back to equilibrium\footnote{Once 
we consider $1/N$ corrections the particles will
become interacting and they will of course reach equilibrium.}. However for our
static, spherically symmetric situation the results derived by the
hydrodynamic method are reliable. This is indicated by the agreement
of \eqref{NEF}, \eqref{Mtot} with \eqref{integralN}, \eqref{integralM}
and could perhaps be made more quantitative along the lines of
\cite{Lieb:1983ri,Lieb:1987rb,Ruffini:1969qy}.

\subsubsection{Radial Reconstruction}
\label{subsec:radialrecon}

In the limit of a large number of fermions the hydrodynamic
description in the bulk is more convenient than keeping track of the
exact quantum state \eqref{slaterbulk}. We saw in sections
\ref{subsec:qsbound} and \ref{subsec:qsbulk} that at the level of
multiparticle quantum states we have a one-to-one correspondence
between the boundary and the bulk. This means that we should be able
to construct the analogue of the ``hydrodynamic description'' in the
boundary CFT. In other words, we would like to understand what is the
CFT meaning of a hydrodynamic description of the multitrace operator
\eqref{Phi}.

The hydrodynamic description of the gas in the bulk is partly encoded
in the radial density profile $\rho(r)$ given by \eqref{textbook} and
\eqref{redshift}. How is $\rho(r)$ related to the operator
\eqref{Phi}? The radial direction $r$ does not have a direct meaning
in the CFT, though in a sense it corresponds to the ``scale'' of the
excitation. One way to probe the scale direction is to consider
2-point functions of operators in the CFT, evaluated on the state
\eqref{Phi}. The separation of the two insertions of operators on the
boundary controls how deeply in the bulk we are probing: when the two
operators are close to each other we are probing the region near the
boundary, while at larger separations of the boundary operators we
probe the interior and in particular the region where the fermionic
gas lives. Hence if we compute the 2-point function
$$
\langle {\bf \overline{\Phi}}|\,\, \overline{\Psi}(x)\,\Psi(y)\,\, |{\bf \Phi}\rangle
$$
in the limit where the number of particles in $|{\bf \Phi}\rangle$ is large, 
we should be able to relate the radial profile $\rho(r)$ to the $(x-y)$ 
dependence of this correlator.

Computing this correlator should be straightforward in the free
$c\rightarrow \infty$ limit and it would be interesting to perform
this computation, as this would be the first step before considering
the same problem including backreaction. We hope to revisit this
question in future work.

\subsubsection{Fermions at finite temperature}

It is not difficult to generalize the story to the case where the
fermions are at non-zero temperature. Let us assume that the fermions
are in an ensemble characterized by the average occupation number
$\xi(\epsilon)$ for the one-particle states which are multiplied
together to make the analogue of the multi-trace operator
\eqref{Phi}\footnote{For simplicity we assume that we are in an
  ensemble where the average occupation level $\xi(\epsilon)$ depends
  only on the energy of the state. This assumption guarantees that the
  configuration will be spherically symmetric in the bulk.}.  In the
zero temperature case the occupation numbers are given by the zero
temperature Fermi-Dirac distribution $\xi(\epsilon) =
\Theta(\epsilon_F-\epsilon)$, where $\Theta$ is the unit-step
function. This distribution corresponds to the radial profile
\eqref{efrads} of the local chemical potential in the bulk.  We denote
the corresponding density and pressure profiles by
$\rho_{\epsilon_F}(r),p_{\epsilon_F}(r)$.

What happens if we consider a more general distribution
$\xi(\epsilon)$ for the average occupation level of the single-trace
operators ?  Notice that formally we can write
any function $\xi(\epsilon)$ defined for $\epsilon \geq 0$ as a linear
combination of $\Theta$-functions in the following way
\begin{equation}
\xi(\epsilon) = -\int_0^\infty d \epsilon_F \,\xi'(\epsilon_F)\, \Theta(\epsilon_F - \epsilon)
\label{transs}
\end{equation}
which can be proven by a partial integration, assuming that $\xi$ vanishes at $\epsilon_F=\infty$.
 Since the fermions are non-interacting this implies that the bulk 
profile can be written as
\begin{equation}
\rho_\xi(r) = -\int_0^\infty d\epsilon_F\, \xi'(\epsilon_F) \,\rho_{\epsilon_F}(r)\qquad,\qquad p_\xi(r) = -\int_0^\infty d\epsilon_F\, \xi'(\epsilon_F) \,p_{\epsilon_F}(r)
\label{denssup}
\end{equation}
In particular for a thermal distribution in the grand-canonical ensemble at temperature $\beta$ and
chemical potential $\mu$ we have
\begin{equation}
\label{FDdist}
\xi(\epsilon) = {1\over e^{\beta(\epsilon - \mu)}+1}
\end{equation}
Using \eqref{efrads}, \eqref{denssup} and \eqref{FDdist} we can
compute the radial profile of the density (and pressure) corresponding
to a finite temperature fermion gas in AdS, before we consider any
backreaction. Let us notice that while the equations \eqref{denssup}
seem to work for any choice of $\xi$, the profile in the bulk will
have a meaningful description in terms of a density profile
$\rho_\xi(r)$ only if $\xi$ is sufficiently smooth.

It would be interesting to compute the entropy of a thermal ensemble from the
fluid configuration in the bulk. For large quantum numbers this entropy should
agree with the exact boundary computation of the entropy in terms of the quantum
states of the single-trace operators.

Finally, from the relation \eqref{denssup} it should be clear that the
mapping between perturbations of the bulk density profile and the
boundary is non-local, as expected from the general framework of
AdS/CFT.

\subsection{Rotating Star}
\label{subsec:rotatingcase}

As a straightforward generalization and as a further test of the
hydrodynamic approximation we will now try to incorporate rotation
into the system. For simplicity we focus on AdS$_5$ and we take the
metric to be \be ds^2 = -\left( 1 + \frac{r^2}{\ell^2}\right) dt^2
+\left( 1 + \frac{r^2}{\ell^2}\right)^{-1} dr^2 + r^2 (d\theta^2 +
\sin^2 \theta d\phi^2 + \cos^2 \theta d\psi^2).  \ee The range of the
variable $\theta$ is $[0,\pi/2]$. In order to put angular momentum into the
system we need to take a rotating fluid, and there is an obvious
question whether or not the rotation is uniform. In four dimensional
flat space it is claimed in \cite{rotstara},\cite{rotstarb} that
uniform rotation minimizes the mass of the star. We are going to try
to do the same thing for the five-dimensional AdS case.

The energy momentum tensor still equals
\be
T_{\mu\nu} = (\rho+p)u_{\mu} u_{\nu} + p g_{\mu\nu}
\ee
where now $p$ and $\rho$ will be functions of both $r$ and
$\theta$, and
\be
u^{\mu} = U^{-1} (1,0,0,\Omega_{\phi},\Omega_{\psi})
\ee
where
\be
U=\left( \left( 1 + \frac{r^2}{\ell^2}\right) - r^2 \sin^2 \theta
\Omega_{\phi}^2 - r^2 \cos^2 \theta \Omega_{\psi}^2 \right)^{1/2}.
\ee
Conservation of $T$ implies
\bea
0 & = &  U^{2} \partial_r p + \frac{r}{\ell^2}(p+\rho)(1-\ell^2
\Omega_{\psi}^2 \cos^2 \theta - \ell^2 \Omega_{\phi}^2 \sin^2
\theta) \nonumber \\ 0 & = & U^{2} \partial_{\theta} p + r^2
(p+\rho) (\Omega_{\psi}^2 - \Omega_{\phi}^2) \sin\theta\cos\theta
.
\eea
These equations are very simple, they are equivalent to
\bea 0 & = & \partial_r p + \frac{\partial_r U}{U}(p+\rho)
\nonumber \\
0 & = & \partial_{\theta} p + \frac{\partial_{\theta}
U}{U}(p+\rho).
\eea
We can now immediately solve for the star, the solution simply
reads
\be
\mu(r) = {\epsilon_F \over U(r)}
\ee
If $\Omega_{\phi},\Omega_{\psi}$ depend non-trivially on
$r,\theta$ this simple solution is no longer correct.

Given the exact solution, we wish to write expressions for the
mass and two angular momenta. These are all related to Killing
vectors $\xi^{\mu}$ and the conserved charges boil down to
\be
Q[\xi] =4\pi^2 \int  \xi^{\mu}\,T_{\mu
0} \left(1 + \frac{r^2}{\ell^2}\right)^{-1}\, r^3 \sin\theta\cos\theta\, dr\, d\theta .
\ee
In particular, we obtain for the mass
\be
M =4\pi^2 \int r^3 \sin\theta\cos\theta dr d\theta \left(-p +
\frac{1+r^2/\ell^2}{U^2}(p+\rho) \right)
\ee
and for the angular momenta
\be
J_{\phi} = -4\pi^2 \int r^3 \sin\theta\cos\theta dr d\theta
\frac{\Omega_{\phi} r^2 \sin^2 \theta}{U^2}(p+\rho) .
\ee
and
\be
J_{\psi} = -4\pi^2 \int r^3 \sin\theta\cos\theta dr d\theta
\frac{\Omega_{\psi} r^2 \cos^2 \theta}{U^2}(p+\rho) .
\ee
These are rather difficult expressions, which simplify quite a bit
if we put
\be
\Omega_{\phi}=\Omega_{\psi}={\omega\over \ell}.
\ee
One can then explicitly do the relevant integrals, and we obtain
\be
\label{Mhydr}
M=\frac{\ell^4}{120} \frac{(\epsilon_F-m_f)^4(4\epsilon_F+m_f-5m_f 
\omega^2)}{(1-\omega^2)^3}
\ee
and
\be
\label{Jhydr}
J = -\frac{\ell^6 (\epsilon_F-m_f)^5 \omega}{60(1- \omega^2)^3} .
\ee

Let us now describe the rotating degenerate state in the boundary
CFT. For this purpose we choose a $U(1)$ subgroup in one of the two
$SU(2)$ factors. To be specific, let us consider a rotation associated
with, say, a simultaneous right handed rotation in $\phi_1$ and
$\phi_2$. The corresponding quantum number is \be j=m_1+m_2.  \ee
Single particle states are characterized by their angular momentum $l$
and energy $E_0=\Delta_0+n$. When $n$ is fixed but $l$ is left free
$j$ takes the following values \be
\label{jrange}
\!\! -n\leq j\leq n,\  \mbox{$ $}
\ee
but remains even (odd) when $n$ is even (odd).

We describe the rotating degenerate state by introducing an angular
velocity $\omega$ which acts as a "chemical potential" for the quantum
number $j$.  This means that for a given Fermi energy $\epsilon_F$ all
states are being occupied which obey the inequality \be
\label{tilted}
{n+ \Delta_0 - \omega j} \leq \epsilon_F\,\ell.  \ee Thus the effect
of the rotation is to tilt the Fermi surface to favor positive values
of $j$.

To count the total number of particles contained in this rotating
degenerate many particle state we have to add all one particle states
labeled by $n$ and $j$ that satisfy the inequality (\ref{tilted}).  As
a preparation, let us count how many single particle states exist for
given values of $n$ and $j$. This is most easily done by making use of
a generating function for the symmetrized tensor products of the
vector representation of $SO(4)$. One can easily convince oneself that
this generating function is given by \be
\frac{1}{(1-qz)^2(1-qz^{-1})^2}, \ee 
Here $q$ keeps track of $n$, while $z$
keeps track of the $j$ quantum number. Picking the term proportional
to $q^nz^j$ gives the following degeneracy \be {\cal N}(n,j)= {1\over
  4} \left((n+2)^2-j^2\right) \ee Note that for even (or odd) $n$ the
highest number of states occurs at $j=0$ (or $j=\pm 1$).  For $j=\pm
n$ one finds precisely one state, as expected.

This result can now be used to count the number of particles in the
rotating degenerate state for a given chemical potential
$\epsilon_F$. We will do so in the limit of many particles, so that we
can replace the summation over the quantum numbers $n$ and $j$ by
integrals.  In fact, for this purpose it is convenient to replace
$n+2$ simply by $n$.  This leads to the following expression \be
N(\epsilon_F,\omega)={1\over 2} \int_{\strut {\cal
    D}_{\epsilon_F,\omega}}\!\! \!\!\!\!\!\! {\cal N}(n,j)\, dn\,dj .
\ee where ${\cal D}_{\epsilon_F,\omega}$ denotes the integration
domain bounded by (\ref{tilted}), (\ref{jrange}) and $n\geq 0$.  The
factor of $1/2$ is due to the fact that the values of $j$ are only
even or odd.  The integrals are straightforward but somewhat
tedious. After some work we find
 \be
\label{NFO}
N(\epsilon_F,\omega)={{\ell^4 }\over 24} {(\epsilon_F-m_0)^4\over
  (1-\omega^2)^2} \ee We see that for a given chemical potential the
number of particles increases due to the rotation. Note further that
in the limit $\omega\to \pm1$ the number of particles diverges. This
can be understood from the fact that in this case all states with
$j=\pm n$ are being counted for arbitrary values of $n$.

The mass and angular momentum can be obtained in a similar way as
integrals over $n$ and $j$ over the domain ${\cal
  D}_{\epsilon_F,\omega}$.  For the mass one has to sum up the energy
$\epsilon=(n+\Delta_0)/\ell$ for all particles contained in the
rotating degenerate state. This leads to the integral expression \be
M={1\over 2\ell} \int_{\strut{\cal
    D}_{\epsilon_F,\omega}}\!\!\!\!\!\!(n+\Delta_0)\,{\cal N}(n,j)\,
dn\,dj .  \ee Performing this integral is again straightforward. The
final result can be organized in the following form \be
\label{Mrot}
M= {\ell^4 \over 30} {(\epsilon_F-m_0)^5\over (1-\omega^2)^3} + {m_0
  \ell^4 \over 24} {(\epsilon_F-m_0)^4\over (1-\omega^2)^2}.  \ee This
result should be compared to the case without rotation given in
(\ref{Mtot}).  The last term represents the contribution of the mass
$m_0$ for all the particles, whose number is given in (\ref{NFO}),
while the first term gives the sum of all the $n$ quantum numbers. We
note that the modification in the mass due to the rotation is again
given in terms of inverse powers of $(1\!-\!\omega^2)$, and diverges
for $\omega\to \pm 1$.

Finally, let us come to the angular momentum $J$. The total angular
momentum carried by the particles in the rotating degenerate state is
represented by again a tedious but fortunately straightforward
integral \be J ={1\over 4\ell} \int_{\strut {\cal
    D}_{\epsilon_F,\omega}} \!\!\! j \ {\cal N}(n,j)\, dn\,dj .  \ee
It yields the following result \be
\label{J}
J= {\ell^4 \over 60} {(\epsilon_F-m)^5 \over(1-\omega^2)^3} \omega \ee
Note that $|J|< M$ in general, and only when $\omega\to \pm 1$ the
ratio $|J|/M\to \pm 1$.

The results \eqref{Mrot}, \eqref{J} for the mass and angular momentum
computed in the conformal field theory agree with the hydrodynamic
computations \eqref{Mhydr}, \eqref{Jhydr}.

\section{INCLUDING INTERACTIONS}
\label{sec:interactions}

\subsection{The scaling limit}
\label{subsec:scalinglimit}

The discussions of the previous section are accurate if we take the
$c\rightarrow \infty$ limit keeping $\Delta$, $N_F$ and $\Delta_0$
fixed.  In that limit the system is free as interactions can be
ignored due to 't Hooft factorization. However what we have
constructed so far is a trapped fermionic gas rather than a star. To
turn the gas into a star we have to include gravitational
backreaction. This can be achieved by scaling $\Delta_0$ and $N_F$
(and as a result $\Delta$) appropriately at the same time that we send
$c$ to infinity.

To determine what is the precise scaling limit we should take, we
consider the picture in the bulk.  The first condition we impose is
that the radius of the star is of the order of the AdS scale
$\ell$ \footnote{By this we mean that as $c\rightarrow \infty$ the
radius of the star scales like as $R \sim c^0
  \ell$.}. Assuming that equation \eqref{radius} is a good estimate
for the order of magnitude of the size of the star even after
backreaction\footnote{This assumption will be proven to be
  self-consistent after we actually compute the backreaction.}, we see
that this condition implies \be
\label{limita}
{k_F \over m_f} ={n_F \over \Delta_0} = {\rm fixed} \ee The second
condition is that we want to have appreciable gravitational
backreaction. This can be translated to the statement $G M \ell^{2-d}
\sim {\cal O}(1)$. Newton's constant is related to the central charge
by $G\sim c^{-1}\ell^{d-1}$. So we find that the conformal dimension
of the composite operator must satisfy \be
\label{limitb}
{\Delta \over c } = {\rm fixed} \ee where we used the identification
$\Delta=M \ell$. From equation \eqref{DeltaF} and
\eqref{limita} we find that for this to be true we have to take
$$
n_F,\Delta_0 \sim c^{1\over d+1}
$$ 
In this limit the number of particles \eqref{NF} goes like 
$$
N_F \sim c^{d\over d+1}
$$ In other words we want to study a specific class of composite
multitrace operators in the limit $ c\rightarrow \infty \,,\, \Delta =
\varepsilon c $ where $\varepsilon$ is kept fixed. When
$\varepsilon\ll 1$ the system is weakly interacting. We want to turn
on $\varepsilon$ up to values of order 1 where interactions become
important\footnote{Let us note that even though in our limit the
  particles become very heavy (since their rest mass scales like $
  \ell^{-1} c^{1\over d+1}$), they are parametrically lighter than the
  Planck mass which scales like $\ell^{-1} c^{1\over d-1}$.}.

Since in this limit the number of particles becomes large, it is
conceivable that the 't Hooft factorization might fail in a drastic
way. In terms of Feynman diagrams in a gauge theory this can happen
because we are scaling the number of external lines of a diagram with
$N$ so the usual $N$-counting rules may not apply. Let us describe the
problem in some more detail: consider the correlation functions of
multitrace operators whose size we keep fixed in the large $N$
limit. As far as interactions between distinct single trace
constituents are concerned, the dominant diagrams are the disconnected
ones, which correspond to the free limit discussed in section
\ref{sec:freelimit}. Connected planar diagrams involving interactions
between various single trace components of the multitrace operator are
suppressed by powers of ${1\over N}$ relative to the disconnected ones
and non-planar diagrams by even higher powers of ${1\over N}$.  Now
let us imagine scaling the size of the multitrace operators as we send
$N\rightarrow \infty$. In order to have nontrivial interactions at
large $N$ we want to scale the size of the multitrace operators in
such a way as to compensate for the ${1\over N}$ suppression of
connected planar diagrams (connecting the various single trace
components of the multi-trace operators). Our goal is to work in a
limit where these leading connected planar diagrams are of the same
order as the disconnected ones.  The danger now is that since in this
limit the planar connected diagrams became as important as the
disconnected ones, then the same could happen for non-planar
diagrams. This would imply that effects of quantum-gravity would
become important in the bulk. However we will argue below that this is
not the case and the planar diagrams remain the important ones.

It is perhaps more intuitive to address this question from the bulk
point of view. In theories with a gravitational dual the planar
interactions on the boundary are mapped to tree level supergravity
diagrams in the bulk. The $c\rightarrow \infty$ limit corresponds to
sending $G\rightarrow 0$ in AdS units. In this limit the tree level
interaction between any two particles (whose mass is kept fixed) will
vanish.  In our double scaling limit we are taking the number and
masses of particles to be large so that even though $G\rightarrow 0$
the total backreaction remains finite. Moreover the various factors
contribute in such a way that the backreaction receives contributions
from tree level diagrams only, as we explain in the next subsection.
This implies that the planar approximation remains reliable in this
limit.

\subsection{Interacting Quantum states in AdS}
\label{subsec:qsads}

Let us now consider the many-body state \eqref{slaterbulk} in the
scaling limit of the previous subsection, in which we have to take
into account the gravitational interactions between the fermions. It
is intuitively clear that in the limit of large number of fermions per
unit volume, the most efficient way to analyze the interactions is by
working in the hydrodynamic description of section
\ref{subsec:hydrofree}, but this time incorporating the gravitational
backreaction of the fermionic fluid. This approach will be pursued in
subsection \ref{ssec:TOV}. For now we will consider the interactions
from a more microscopic point of view, in order to understand the
validity of approximations involved in going from the exact to the
coarse grained hydrodynamic description.

A first simplification is that in our scaling limit (and for theories
with a semi-classical gravity dual) we can ignore processes which
change the number of fermions. In a relativistic system the number of
particles does not have to be conserved. If we put a certain number of
fermions in AdS and then turn on interactions it is not guaranteed
that their number will remain constant.  Unless the fermions are
protected by a conservation law, there will be processes which can
change their number, as well as processes which produce all kinds of
other particles such as gravitons etc. In principle all of these have
to be taken into account. However in our scaling limit the fermionic
gas becomes effectively long lived i.e. the rate of creation and
annihilation of particles becomes negligible. This is due to the fact
that at large $c$ interactions are suppressed. In principle this
suppression could be compensated by large combinatoric factors due to
the large number of particles, but as we will discuss in more detail
in section \ref{sec:validity}, these factors work out in such a way
that the processes which change the particle number become subleading
at large $c$, relative to processes responsible for the total
gravitational backreaction of the fermions.  Because of this, in the
rest of this section we will only focus on interactions which do not
change the number of particles, i.e. on processes with only $N_F$
incoming and $N_F$ outgoing fermions and no other external particles.

Second, we will assume that the only low lying fields in the bulk are
the graviton, described by the Einstein-Hilbert action, minimally
coupled to a fermionic field $\psi$ of mass $m_f$ \footnote{There may
  be interaction terms for the field $\psi$, but if their strength is
  controlled by $c$ in the standard way (i.e. a vertex with $n$
  fermions is suppressed by a factor of $c^{2-n\over 2}$) then the
  qualitative results of this section remain the same. For simplicity
  we will ignore such terms in what follows.}.  We will assume that
there are no other types of particles or forces in the bulk. Clearly
this is a very drastic approximation which however captures the
qualitative behavior of the phenomena we want to study without
unnecessary complications. It would be very interesting to study
further whether a specific AdS/CFT setup can be found where such a
simple bulk Lagrangian can be realized (at low energies), or at least
a setup in which it correctly captures the qualitative physics of
degenerate fermionic states. We present some first discussions along
these lines in section \ref{sec:validity}.

We now want to see that in our scaling limit the ${1\over c}$
suppression of interactions combines with the enhancement from the
large number of particles in such a way that only the tree level
interactions (or equivalently planar interactions) become
important. To illustrate this point we focus on one particular
physical observable as an example: we will try to estimate the total
energy of the fermions including their gravitational interactions.  So
let us consider a bulk state $|\Phi\rangle$ corresponding to $N_F$ fermions
in the free theory. It satisfies
$$
H_0 |\Phi\rangle = {\Delta_0\over \ell} |\Phi\rangle
$$ where $H_0$ is the Hamiltonian for a free fermionic field in
AdS$_{d+1}$. When we include interactions we have to use the full
Hamiltonian $H = H_0 + H_{int}$ where $H_{int}$ includes the coupling
between fermions and gravitons and the self-interactions of the
gravitons and $H_0$ has to be extended to include the quadratic part
of the Einstein-Hilbert action for propagating gravitons. The
corrected energy $\Delta/\ell$ can be computed by an expansion in
Feynman diagrams in the bulk.

For example, consider two fermions in AdS with energies $E_1,E_2$ and
wavefunctions $\psi_1(x),\psi_2(x)$.  To leading order the total
energy is the sum of the energies. To compute the first correction to
this energy we have to compute the tree level gravitational
interaction between them.  In the Born approximation this is given by
the diagram shown in figure 1, 
\begin{figure}[h]
\begin{center}
 \label{twoferm} 
 \epsfig{file=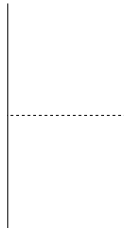,height=3cm,trim= 0 0 0 0}
\caption{Tree level gravitational interaction between two fermions.}
\end{center}
\end{figure}
where we have to use the graviton and fermion propagators in
AdS$_{d+1}$. In our scaling limit this diagram is proportional to
$$
 G {m_1 m_2 \over \ell^{d-2}} \sim c^{-{d-1 \over d+1}}
$$ so it goes to zero.  Higher order diagrams between two fermions are
 suppressed by further powers of ${1\over c}$, so they all become
 negligible in the large $c$ limit. Hence if we only had a finite
 number of particles the backreaction would be unimportant.

If we have $N_F$ fermions then to compute the correction to the energy
(or the conformal dimension of the dual operator) $\delta
\Delta=\Delta-\Delta_0$ we have to sum over many kinds of diagrams as
shown in figure 2 
\footnote{Only connected diagrams have to
  be considered for the computation of $\Delta$. The disconnected
  diagrams can be resummed into the exponential in $e^{-it (H_0
    +H_{int})} |\Phi\rangle = e^{-i (\Delta/\ell) t} |\Phi\rangle$.},
where as we explained before we only consider diagrams with the same
number of fermions in the past and future infinity.  Due to the
non-linearities of general relativity we not only have 2-particle
interactions but also n-particle ones for all n as depicted in the
diagram. We have to sum up all these diagrams to get the right
answer. The correction to the energy of the state can be schematically
written as
$$
 \delta\Delta =\sum_{k=2\,,\,L=0}^\infty \delta \Delta_k^L
$$ where $\delta \Delta_k^L$ denotes the contribution of diagrams
where $k$ fermions are involved in the interaction and $L$ is the
number of loops.
\begin{figure}[h]
\begin{center}
  \epsfig{file=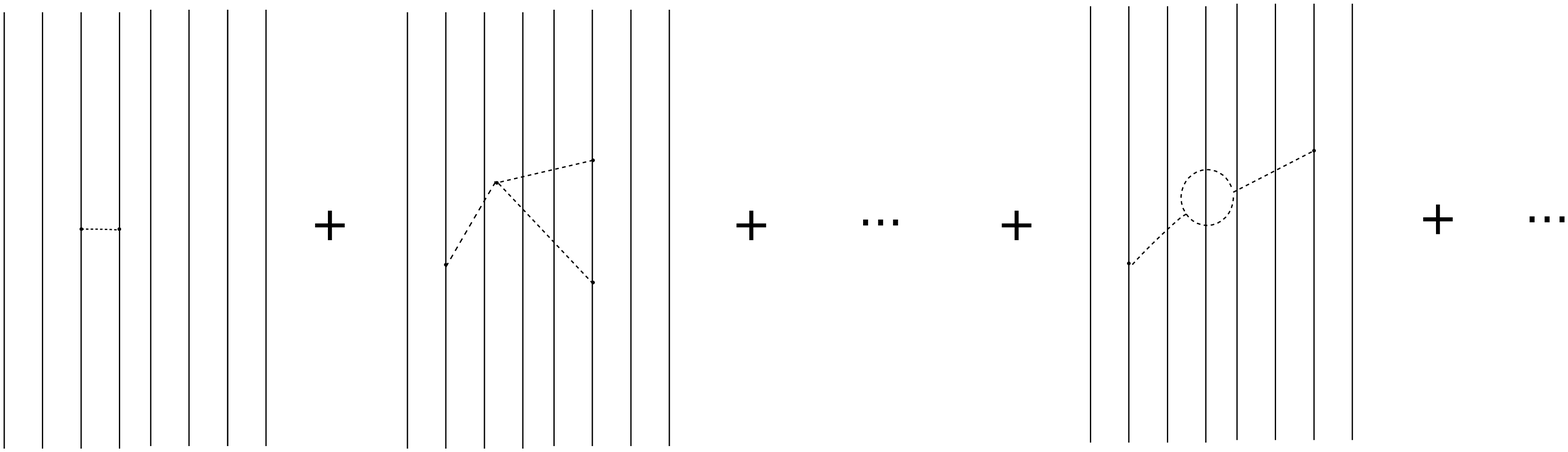,height=3cm,trim= 0 0 0 0}
\caption{2-body, 3-body etc. interactions at tree level and higher loop interactions.}
\end{center}
  \label{manyf}
\end{figure}

We will now see how diagrams of different types scale in our
limit. Going back to the single graviton exchange between a pair of
fermions, but now summing over all possible pairs, we find that the
correction to the energy is
$$
\delta \Delta^{0}_2  \sim \sum_{i\neq j} G {m_i m_j \over \ell^{d-2}}
\sim {c \over \ell}
$$ where we used that $N_F \sim c^{d \over d+1}$ and $m \sim c^{1\over
  d+1}$. In other words
$$
{\delta \Delta^0_{2} \over \Delta_0} \sim {\cal O}(1)
$$ so the correction is of the same order as the original energy of
the system $\Delta_0$ and remains important even in the large $c$
limit. The suppression that we found earlier has been compensated by
the large number of pairs that we have to sum over.  On the other
hand, loop diagrams between pairs of fermions are suppressed by
additional power of ${1\over c}$ and are unimportant in our scaling
limit.

For a general diagram with $L$ loops, $k$ fermions involved in the
interaction (i.e. $k$ ingoing and $k$ outgoing ones), $V_F$
fermion-fermion-graviton vertices, and $V_n$ vertices of $n$-graviton
type, we have the following topological constraints
$$ V_F+\sum n V_n = 2I_g\qquad ,\qquad 2V_F = 2I_F + 2k\qquad,\qquad L
= I_F + I_g - V_F - \sum V_n +1
$$ where $I_g,I_F$ denote the number of internal graviton and fermion
lines.  Now let us count powers of $c$ of these diagrams. Each
external line contributes a factor of $(N_F\, m_f)$ which in our scaling
limit is proportional to $c$. Each n-graviton vertex contributes a
power of $c^{1 - {n \over 2}}$ and each of the $V_F$ vertices a factor
of $c^{-{1\over 2}}$. So in total the powers for a diagram are
$$ \delta \Delta^{L}_k \sim c^{k+\sum \left(1 - {n\over
    2}\right)V_n-{V_F\over 2}}
$$ 
Using the three conditions that we found above this becomes
$$
{\delta \Delta^{L}_k\over \Delta_0} \sim c^{-L}
$$ So we find that that only $L=0$ (tree level) diagrams contribute
significantly at large $c$. The conclusion is the following: in our
double scaling limit we have contributions from diagrams of all orders
due to the non-linear nature of general relativity, but only tree
level ones. In this sense our system becomes (semi)-classical, though
nonlinear.

\subsection{From quantum states to fermionic fluid}
\label{subsec:qutofl}

Clearly summing up all such diagrams between fermions is not the most
efficient way to analyze the problem and an approximation method would
be desirable. In this section we review some basic ideas which are
often used in treatments of many-body systems and which may clarify
the conceptual steps in going from the quantum many-particle
description in the bulk to the hydrodynamic description and the TOV
equations of the next subsection. The reader who is not interested in
these issues can skip this subsection and go directly to section
\ref{ssec:TOV}.

In general, in a many-body quantum problem we first want to
determine the ground state wavefunction and energy, and then perhaps
to study small excitations around it. The simplest example to consider
is $N_F$ non-interacting fermions moving in an external potential.  In
this case we first compute the energy levels $\ep_i$ for a single
fermion in this potential and the corresponding wavefunctions
$\psi_i(x)$. The ground state of the many-fermion system is then
described simply by filling up these orbitals consistently with the
Pauli exclusion principle.  The wavefunction is a Slater determinant
of the one-particle wave functions
$$ 
\Psi_{N_F}(x_1,...,x_N) = {1\over \sqrt{{N_F}!}}\left| \begin{array}{cccc}
\psi_1(x_1)\ \ \psi_2(x_1)\ \  ...\ \  \psi_{N_F}(x_1)   \\
\psi_1(x_2)\ \ \psi_2(x_2)\ \  ...\ \  \psi_{N_F}(x_2)   \\
 \   \ \  ...\ \    \\
\ \psi_1(x_{N_F})\ \ \psi_2(x_{N_F})\ \  ...\ \  \psi_{N_F}(x_{N_F})   \\
\end{array} \right|
$$ 
and the total energy is given by the sum of the individual energies
$$
E_{tot} = \sum_{i=1}^{N_F} \ep_i
$$ 

When we turn on interactions between the fermions the situation is not
as simple, since it is no longer true that the many-body state can
be constructed from the knowledge of single particle energy
states. More precisely, single particle energy states, or ``orbitals''
are not even well defined in an interacting system of fermions. By
adding or removing a fermion we affect the wavefunctions of all other
fermions in a complicated way, which in turn influences the available
energy levels for the fermion under consideration.  Hence in principle
we have to solve the coupled problem of determining the wavefunction
for all fermions at once. This also means that the ground state cannot
necessarily be written as a Slater determinant\footnote{However it can
  be written as a superposition of Slater determinants.}.

A typical problem of this type is finding the ground state of a
polyelectronic atom. That problem is quite similar to finding the
ground state of our fermionic star, where the role of the attractive
Coulomb force from the nucleus is played by the AdS gravitational
potential and the electromagnetic interactions of the electrons
correspond to the gravitational force between the fermions (though the
latter is attractive). As we know, solving Schroedinger's equation for
a polyelectronic atom is not possible analytically and we have to
address the problem using various approximation methods. We have to do
the same for our system.

In our system the interaction between any two particular fermions is
negligible in the large $c$ limit due to the ${1\over c}$
suppression. However since the number of fermions is also very large
the effect of all of the other fermions on a single fermion can be
appreciable. This type of many-body problem can usually be treated by
mean field theory methods. We replace all interactions on a given
fermion by an effective external field, which has to be determined in
a self-consistent way, as we explain below.

\subsubsection{The Hartree-Fock approximation}

In many-body problems the Hartree-Fock approximation is often
employed. Its basic assumption is that the ground state of the
many-fermion system can be written as a single Slater determinant
$|\Psi\rangle$ of one-fermion orbitals in some appropriately chosen
external potential, which is determined in a self-consistent
manner. One then minimizes the norm of $\langle\Psi| H |\Psi\rangle$
with respect to the single-particle orbitals, where $H$ is the full
Hamiltonian.  This leads to the Hartree-Fock equations for the
single-particle orbitals $\psi_i$ and the corresponding energies
$\ep_i$.

In general the physical meaning of the single-particle energies
$\ep_i$ in the Hartree-Fock approximation is not transparent: for
example if we try to add one more fermion to a previously unoccupied
state $\ep_j$, then the energy of the new state will {\it not} be
exactly equal to $E_{tot} + \ep_j$. This is due to the fact that when
we add a fermion all other fermions will rearrange themselves and in
principle one has to solve the Hartree-Fock problem again for $N_F+1$
fermions, which will introduce corrections to the energy, additional
to the naive estimate $E_{tot}+\ep_j$. For the same reason exciting a
fermion from the state $\ep_i$ to the state $\ep_j$ will cost energy
equal to $\ep_j - \ep_i $ plus corrections due to rearrangement.

Nevertheless in the limit of large number of fermions the
aforementioned corrections due to rearrangement become negligible, at
least for certain many-body systems. This is sometimes called
``Koopmans theorem'', which basically states that in the limit of
large number of particles we can treat any given one of them as a
``probe'' and define its energy as if it had negligible backreaction
on the rest of the particles. Hence in this limit the excitation
energies of a single fermion are simply determined by the Hartree-Fock
spectrum $\ep_i$, by taking the difference of the energy of the final
minus the initial state. Similarly if we think of constructing the
ground state by gradually adding fermions and rearranging them to
their lowest available energy state, then $\ep_F$ would be the energy of
the last added fermion. Notice that these statements hold only if we
talk about changes to a few numbers of fermions relative to the total
number of them.  

Within the context of the Hartree-Fock approximation we can define a
notion of single particle density of states, even for a system of
interacting fermions
\be
\label{dofff}
g(\epsilon,\epsilon_F) \equiv {dN \over d\ep} (\ep,\ep_F) \ee which is
now a function of the ``Fermi energy'' $\ep_F$. This function encodes
the density of the single-particle wavefunctions of the Hartree-Fock
solution with corresponding energies $\ep_i$. We call Fermi energy
$\ep_F$ the highest energy of the occupied states and $N(\ep,\ep_F)$
is the number of occupied single-particle states with energy lower
than or equal to $\ep$. Clearly we have $N_F = N(\ep_F,\ep_F)$. In the
limit of large number of particles, where Koopmans theorem or the
probe approximation holds, the function $g(\ep,\ep_F)$ can be
determined operationally by doing ``spectroscopy'' on the state, as
follows: we consider the system in its ground state. We hit the state
by a photon coupled to the fermions and measure the absorption
spectrum. This gives information about the energy differences between
the single fermion states (assuming that Koopmans theorem holds) and
in this way we can reconstruct the function $g(\ep,\ep_F)$.

The total number of fermions can be written as
\be
\label{noff}
N_F(\ep_F) = \int^{\ep_F}_{\ep_{min}} d\ep \,g(\ep,\ep_F) \ee where
$\ep_{min}$ is the lowest of the single-particle
states\footnote{Notice that due to the interactions the energy
  $\ep_{min}$ can be low, for example lower than the
  rest mass of the fermions.}.  But now it is important that the total
energy of the system $E_{tot}\neq \int_{\ep_{min}}^{\ep_F} d\ep \,\ep
\,g(\ep,\ep_F)$.  This is due to the interactions between the fermions
and the rearrangement issues which cannot be ignored if one tries to
build up the ground state by adding fermions gradually, since this
process involves all of the fermions and clearly goes beyond the
``probe'' limit. However this can still be done for the last added
fermion, so the following is still true
\begin{equation}
{d E_{tot} \over d \ep_F} = \ep_F \,g(\ep_F,\ep_F)
\label{eosinter}
\end{equation}

It is clear that for our problem, that is, for determining the ground
state of a large number of gravitationally interacting fermions in
AdS, the Hartree-Fock method is applicable and moreover the
assumptions of Koopmans theorem are satisfied. This means that the
Hartree-Fock orbitals should be really thought as the possible quantum
states of a fermion moving in the gravitational background produced by
rest of the fermions and the density of states \eqref{dofff} describes
the distribution of fermions with various energies, generalizing the
non-interacting density of state \eqref{DOS}.

In practice we will not apply the Hartree-Fock method, since it is
still too complicated when the number of fermions is large. As we
explain below, a further approximation in addition to the Hartree-Fock
can be made, which leads naturally to the TOV equations.

\subsubsection{Fermion fluid and Thomas-Fermi approximation}

Even after imposing the Hartree-Fock approximation we can make a
further simplification as follows: instead of solving for the ground
state Slater-determinant wavefunction, we can only look for the
fermion number density ${\bf n}(x)$ in the ground state\footnote{In this section we
  think of fermions moving in flat $d$-dimensional space. If we are in
  curved spacetime then the appropriate measure factors have to be
  included.}
\be
\label{fermiondtf}
{\bf n}(x) = \int d^dx_2\, ...\, d^dx_N |\Psi(x,x_2,...,x_{N_F})|^2
\ee
which satisfies $\int d^dx\, {\bf n}(x) = N_F$. Of course ${\bf n}(x)$ carries
less information than the actual wavefunction but it is easier to
determine than $\Psi$. We express the kinetic and potential energy of
the fermions in terms of ${\bf n}(x)$.\footnote{For example for a system of
  many (non-relativistic) electrons moving in an external potential
  $V(x)$ the total energy in the Thomas-Fermi approximation would be
$$
E_{TF}(n) =K \int d^3x \,{\bf n}(x)^{5/3}+ \int d^3x \,V(x) {\bf n}(x) + {e^2\over 2} \int\int d^3x d^3y {{\bf n}(x) {\bf n}(y) \over |x-y|}
$$.} Minimizing the energy with respect to ${\bf n}(x)$ determines the
fermion density profile and the energy of the ground state. This is
the Thomas-Fermi approximation. In many systems the Thomas-Fermi
approximation becomes asymptotically exact in the limit of large
number of particles. As we will explain later, the hydrodynamic
approximation and the TOV equations in the bulk is the equivalent of
the Thomas-Fermi approximation (or rather its relativistic
generalization) for the quantum interacting fermions of the previous
sections, also see references \cite{Lieb:1983ri,Lieb:1987rb,Ruffini:1969qy}.

\subsection{TOV equations in the bulk}
\label{ssec:TOV}

After these general comments we return to what we would have naively
done to deal with the gravitational backreaction of a large number of
fermions: we revisit our hydrodynamic formalism of section
\ref{subsec:hydrofree}, but now taking into account the gravitational
backreaction of the fermionic fluid.  We will explain how the
hydrodynamic approximation is related to the Hartree-Fock and
Thomas-Fermi approximations in the next subsection

We have to solve Einstein's equations with cosmological constant
coupled to matter
$$
R_{\mu\nu}-{1\over 2}g_{\mu\nu}R + \Lambda g_{\mu\nu} = 8 \pi G T_{\mu\nu}
$$
where the source is a perfect fluid
$$
T_{\mu\nu} = (\rho + p) u_\mu u_\nu + p g_{\mu\nu}
$$ and $\rho,p$ are determined by the flat space equation of state
\eqref{textbook} for a degenerate fermionic fluid.

As before we look for static, spherically symmetric solutions, so we
write the metric in the general form \be
\label{Ansatzb}
ds^2 = - A(r)^2 dt^2 + B(r)^2 dr^2 + r^2 d\Omega^2_{d-1}.  \ee We
parametrize the functions $A(r)$ and $B(r)$ in terms of two new
functions $M(r)$ and $\chi(r)$ as \bea {A^2(r)} &=
&e^{2\chi(r)}\left(1-\frac{C_d
  M(r)\,}{\ \ r^{d-2}}+\frac{r^2}{\ell^2}\right),\nonumber \\ {B^2(r)}
&= & \left(1-\frac{C_d
  M(r)\,}{\ \ r^{d-2}}+\frac{r^2}{\ell^2}\right)^{\!\!
  -1}\!\!\!,\label{definitions} \eea where Newton's constant is
contained in the coefficient
$$
C_d =\frac{16\pi \,G\ }{(d\!-\!1)V_{d-1} }.
$$ The Fermi gas is static, and hence we take $u_t=A(r)$ and all other
components of $u_\mu$ equal to zero. In terms of $M(r)$ and $\chi(r)$
the Einstein equations read
$$
{d M\over dr}=  \, V_{d-1} \, r^{d-1}\,\rho
$$
\be
\label{toveq}
{d\chi \over d r} = V_{d-1} \,\frac{C_d}{2}{\Bigl(\rho+p\Bigr)} \, r{B^2}
\ee
$$
 {dp\over dr} +\frac{A'}{A} (\rho+p) =0
$$ These constitute the Tolman-Oppenheimer-Volkoff (TOV) equations.

We choose as boundary condition $M(0)\!=\!0$, so that $M(r)$
represents the contribution to the mass from the energy density inside
a ball of radius $r$, and the total mass is equal to $M(R)$.  By
Birkhoff's theorem the metric outside the star is given by
AdS-Schwarzschild.  Hence,
$$
\chi(r)=0, \qquad M(r)=M,\qquad  \mbox{ for $r\geq R$}.
$$ From (\ref{edge}) it follows that the radius $R$ is determined by
\be 1-\frac{C_d
  M\,}{\ \ R^{d-2}}+\frac{R^2}{\ell^2}=\left({\epsilon_F\over
  m}\right)^2.  \label{sradius}
\ee 
Finally the total fermion number can be computed by the integral
\be
N_F = V_{d-1}\int_0^R dr \,r^{d-1}\,B(r)\, {\bf n}(r)
\label{bulkfn}
\ee

Let us now see how we can compute the density of single-particle
states $g(\ep,\ep_F)$ from the hydrodynamic configuration in the
bulk. The energy $\ep$ of a fermion should be identified with the
redshifted energy of a fermionic particle in the bulk. At a given
point in the bulk we have fermions with various values of the local
energy ranging from the fermion mass $m$ up to the local Fermi energy
$\mu(r)$.  So we can write the following formula for the density of
states on the boundary
\be
N(\ep,\ep_F) = V_{d-1}\int_0^R dr\, B(r)\, r^{d-1}\, {\bf n}\left({\ep\over A(r)}\right)
$$
$$
g(\ep,\ep_F) = {d N \over d \ep}(\ep,\ep_F)
\label{bbbb}
\ee where ${\bf n}(\epsilon)$ was defined in \eqref{number}. Notice
that the statement that $E_{tot} \neq \int d\ep \, \ep\, g(\ep,\ep_F)$
is related to the statement in the bulk that the total mass is not
equal to the sum of the redshifted masses of the particles.

In fact, to show that the relation \eqref{eosinter} is still valid one
needs to use the Einstein equations!  One has \bea \mbox{${}$}\!\!\!
\epsilon_F\frac{dN}{d\epsilon_F}&\!=\!& V_{d\!-\!1} \int^R_0 \!\!\! dr
\, r^{d-1} \left( B \, \epsilon_F \frac{d {\bf n}}{d \epsilon_F}+
\frac{d B}{d \epsilon_F}\epsilon_F{\bf n}\right) \nonumber\\ &\!=\!&
V_{d\!-\!1} \int^R_0 \!\!\! dr \, r^{d-1} e^{\chi} \left(\, \frac{d
  \rho}{d \epsilon_F} + \frac{d B}{d \epsilon_F}
\frac{\rho+p}{B}\right) \nonumber \\ &\!=\!& \int^R_0 \!\! dr \,
e^{\chi } \left(\frac{d M'}{d \epsilon_F}+
\frac{dM}{d\epsilon_F}{\chi'}\right)=\frac{dM}{d\epsilon_F}.  \eea
Here we used the identity \eqref{identity}, the relation
\eqref{redshift}, the TOV equations \eqref{toveq}, the definitions of
$A$ and $B$ in terms of the functions $M$ and $\chi$ and the fact that
at the edge of the star ${\bf n}(R) = 0$. Notice that above the local
particle number density depends on the radius as ${\bf n}(r) = {\bf
  n}\left({\epsilon_F / A(r)}\right)$.

In the next section we will use the TOV equations to calculate the
corrections to the mass, the particle number and the density of states
defined through the relations (\ref{relation}).

\subsection{Relation between quantum description and TOV approach}

What is the precise relation between the analysis in terms of quantum
states and Feynman diagrams in section \ref{subsec:qsads} and the
hydrodynamic TOV equations \eqref{toveq}? The Einstein equations
coupled to a perfect fluid are in a certain sense the mean
field/Thomas-Fermi approximation for a system of fermions interacting
gravitationally. This correspondence is of course implicitly used in
the standard treatment of astrophysical neutron stars, in which the
star is treated as a classical self-gravitating ball of fluid governed
by a fermionic equation of state and not as a quantum bound state of
individual fermions. The logic in going from the quantum bound state
to the fluid description was outlined in subsection
\ref{subsec:qutofl}.

The equivalence between these two approaches can be made more
precise. In \cite{Lieb:1983ri,Lieb:1987rb} the following problem was
analyzed: the authors considered the Schroedinger equation for $N_F$
fermions mutually interacting with a Newtonian attractive
gravitational potential.  They passed to the Thomas-Fermi
description\footnote{With a relativistic kinetic term.} replacing the
ground state wavefunction by the fermion density $n(x)$. They then
showed that in the ground state the fermion density satisfies the
(non-relativistic limit of) the TOV equations.  In general relativity
one should also include higher order forces in the Schroedinger
equation. Presumably the Thomas-Fermi approximation would then
reproduce the full relativistic TOV equations.\footnote{However we
  should mention that one qualitative feature of the solution, the
  existence of a critical mass - the Chandrasekhar bound, is also
  visible in the Newtonian gravity approximation.}

In \cite{Ruffini:1969qy} a related analysis was followed: the starting
point is a static spherically symmetric metric parametrized by two
arbitrary functions $A(r),B(r)$ as in \eqref{Ansatzb}. One solves the
Dirac equation for fermions moving in this background and places $N_F$
of them in the lowest available states. Then one computes the stress
energy tensor produced by these fermions and uses it as a source for
Einstein equations for the metric \eqref{Ansatzb}. Then one shows that
in the limit of large $N_F$ solving this problem self-consistently is
equivalent to solving Einstein equations coupled to a perfect
fermionic fluid, that is the TOV equations \eqref{toveq}.

The reason that we mention all these issues is the following: as
explained above, in the bulk we have an intuitive understanding of how
to start with the microscopic many-fermion quantum description and
truncate it to the Thomas-Fermi type approximation of the TOV
equations coupled to a fermionic fluid. What is the equivalent
procedure on the boundary? In the boundary CFT the description in
terms of individual quantum states for the fermions is easier to
understand, at least in the $c\rightarrow \infty$ limit before taking
the backreaction into account. It would be very interesting to perform
the equivalent ``coarse graining'' on the boundary and to introduce the
boundary analogue of the density of fermions \eqref{fermiondtf} and
the TOV equations. The main obstacle in this direction is
understanding how to deal with the ``radial direction'' in boundary
language, see also the discussions in subsection
\ref{subsec:radialrecon}. We hope to revisit these issues in future
work.

\subsection{Quantum states vs classical geometries in AdS/CFT}

Before we close this section we would like to explain an important
conceptual point. In AdS/CFT we have an equivalence of two quantum
systems, which means that their Hilbert spaces are isomorphic. In
particular every quantum state on the boundary should be dual to a
quantum state in the bulk. On the gravity side and at low energies,
the Hilbert space can be approximated by the Fock space generated by
the semiclassical quantization of the supergravity fields. At infinite
$N$ these fields become free, so we simply have to canonically
quantize free fields around an AdS background. This semiclassical
quantization leads to a Fock space which is isomorphic to the
low-conformal dimension Hilbert space of the boundary theory, as we
discussed in detail in the previous subsections.

Our many-particle state is a {\it quantum} state: it is a
many-particle state constructed by acting on the vacuum with many
creation operators of the semi-classically quantized free fields in
AdS. Hence it should not be confused with a {\it classical}
configuration in the bulk, i.e. a classical supergravity solution. The
latter should be thought of as a coherent state where certain modes of
the bulk fields have been coherently excited. In this sense our star
differs from the ``boson stars'' (see \cite{Schunck:2003kk} and
references therein) and other solitonic configurations in the
bulk. Instead, a bosonic analogue of our fermionic star would be, for
example, a finite temperature bosonic gas like the radiation star
considered by Page and Phillips in \cite{Page:1985em} or the ``geons''
of Wheeler \cite{Wheeler:1955zz}.

Let us explain this in some more detail: the large $N$ limit is
usually thought of as a classical limit. Of course no matter how large
$N$ is, if the energy of the state under consideration is low enough
then one would still see the quantization and the discreteness of the
spectrum. Let us now consider classical supergravity solutions in the
bulk.  Such solutions have energy of order $N^2$ as can be seen by the
overall normalization of the bulk action.  However the opposite is not
true: a state whose energy is of order $N^2$ (at large $N$) does not
necessarily correspond to a classical supergravity solution. Our
quantum states are examples of this kind. The difference between the
two types of states (coherent vs multi-particle quantum states) is
that while the total energy is the same (order $N^2$), the
distribution of the energy to various modes is different: in the
coherent states we have few modes excited many times, while in our
quantum states we have many modes excited few times.\footnote{We are
  grateful to S. Minwalla for discussions on these points.}. The
latter cannot be described by a classical field. More generally, it is
important to remember that the fermionic perfect fluid is not a
``classical field'' in the bulk, even thought it is described by a
classical density profile.

Perhaps a useful analogy to keep in mind is the difference between a
classical electric field and black body radiation. The former has
non-vanishing energy density at the classical level. By tuning the
temperature of the black body radiation appropriately as $\hbar
\rightarrow 0$ we can end up with a finite classical energy density
for the radiation. However there is no sense in which the black body
radiation can be described by a classical electromagnetic field. The
reason is that the energy is distributed in too many modes unlike what
happens in a coherent state.

\section{NUMERICAL SOLUTIONS AND GRAVITATIONAL COLLAPSE}
\label{sec:numerical}

\subsection{Numerical Results}
\label{subsec:numericalresults}

We will now look for static, spherically symmetric solutions of the
TOV equations \eqref{toveq}. These equations cannot be solved
analytically and we have to resort to numerical integration. We
present the numerical analysis for AdS$_5$, i.e. $d=4$. The
qualitative features of our results are the same for other values of
$d$.  As we explained in section \ref{subsec:scalinglimit} we are
working in the limit $G_5\, \ell^{-3} \rightarrow 0\,,\, M \ell\,
\rightarrow \infty$, or in CFT language $c \rightarrow \infty, \Delta
\rightarrow \infty$. It is more convenient to present the graphs in
terms of rescaled dimensionless quantities which stay finite in this
limit. For the total mass and the total fermion number we define
$$
\hat{M}\equiv {\Delta\over c} = {8 \over \pi}{G_5 \over \ell^2}\, M\qquad, \qquad
\hat{N}_F \equiv {N_F  \over c^{4/5}}
$$
and for the fermion mass and local chemical potential in the bulk
$$
\hat{m}_f \equiv {\Delta_0 \over c^{1/5}} =
\left( {8 \over \pi }{G_5 \over \ell^3}\right)^{1/5}\,
( m_f\,\ell)\qquad,\qquad \hat{\mu} \equiv\left( {8 \over \pi }{G_5 \over 
\ell^3}\right)^{1/5}\, (\mu \,\ell)
$$ 
All hatted quantities are kept fixed as $c={\pi \ell^3 \over 8
  G_5}\rightarrow \infty$.

\subsubsection{A typical fermionic star}

We first describe a star of typical values for the total number of
fermions $\hat{N}_F$ and fermion mass $\hat{m}_f$. We plot the radial
profile of the local chemical potential in the bulk and the fermion
number density in figures \ref{chemicalpotential} and
\ref{fermionnumber} respectively. The dotted lines in the same graphs
show what would have been the profiles for the same number of fermions
in AdS without gravitational backreaction. The edge of the star is the
point where \eqref{edge} is satisfied and the density goes to zero. We
can see from the graphs that turning on self-gravity moves the
fermions towards the center of AdS, as expected.

\begin{figure}[h]
\begin{minipage}[t]{0.5\linewidth} 
\centering
\psfrag{a}[t]{ $r/\ell$}
\psfrag{b}[r]{$\hat{\mu}(r)-\hat{m}_f$}
\includegraphics[totalheight=0.18\textheight]{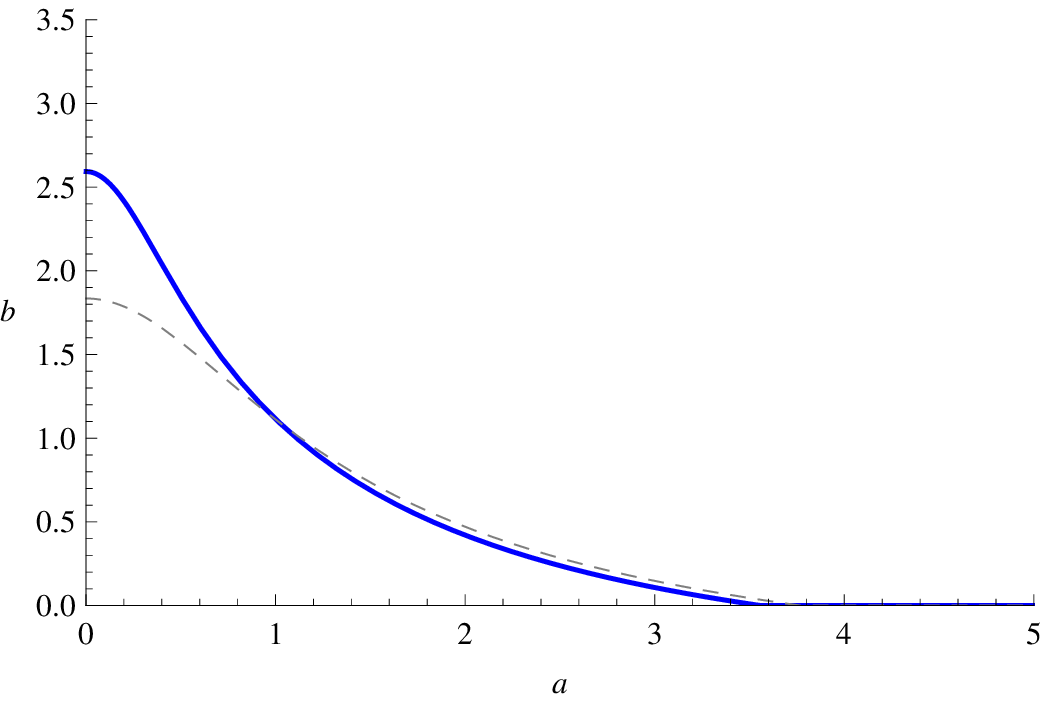}
\caption{Radial profile of the local chemical potential with (solid line) and without
gravitational backreaction (dashed line).}
\label{chemicalpotential}
\end{minipage}
\hspace{0.5cm} 
\begin{minipage}[t]{0.5\linewidth}
\centering
\psfrag{x3}[t]{$r / \ell$}
\psfrag{y3}[r]{$d \hat{N}_F / dr$}
\includegraphics[totalheight=0.18\textheight]{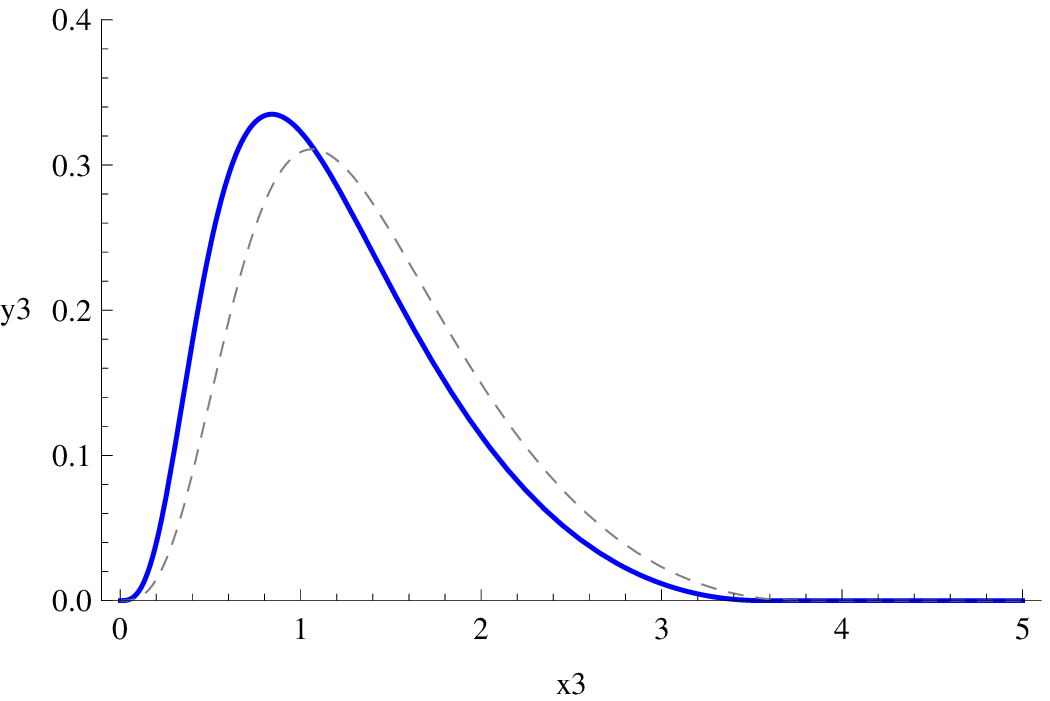}
\caption{ Fermion number density, with and without backreaction.}
\label{fermionnumber}
\end{minipage}
\begin{minipage}[t]{0.5\linewidth} 
\centering
\psfrag{x5}[t]{$r/ \ell$}
\psfrag{y5}[r]{$\hat{M}(r) $}
\includegraphics[totalheight=0.18\textheight]{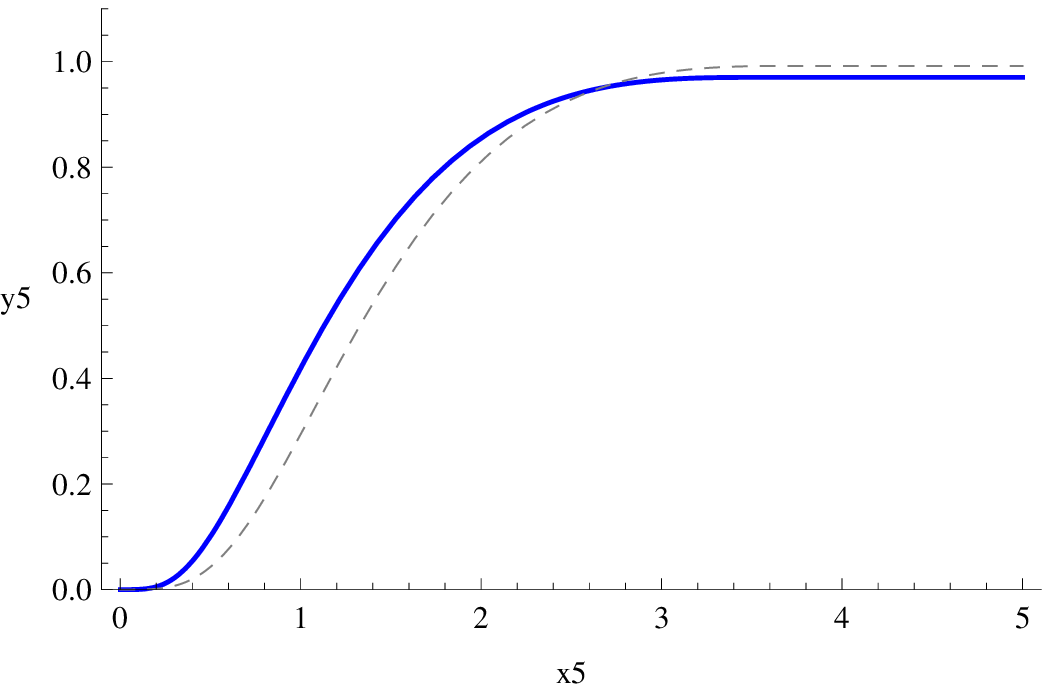}
\caption{Mass function with and without backreaction. The limiting value as $r/\ell\rightarrow\infty$ is
the total mass $M$ of the system.}
\label{ADMsingle}
\end{minipage}
\hspace{0.5cm}
\begin{minipage}[t]{0.5\linewidth}
\centering
\psfrag{x13}[t]{$r/ \ell$}
\psfrag{y13}[r]{$g_{00}(r) $}
\includegraphics[totalheight=0.18\textheight]{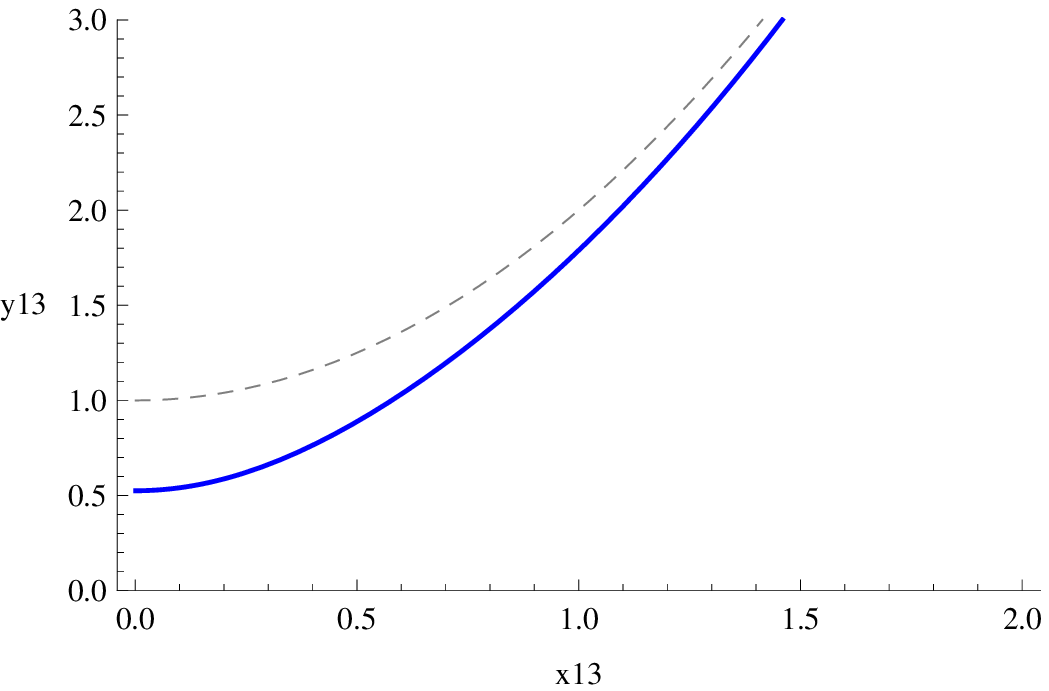}
\caption{Gravitational potential of the star (solid) vs that of empty AdS (dashed).}
\label{gravwell}
\end{minipage}
\end{figure}

In figure \ref{ADMsingle} we plot the function $\hat{M}(r)$ introduced
in \eqref{definitions}, which asymptotes to the total mass of the
solution as $r/\ell \rightarrow \infty$. We see that turning on
gravitational interactions lowers the total mass. The difference
between the two curves can be understood as the binding
energy. Remarkably, we see that the binding energy is a very small
percentage of the total mass. This fact is not a peculiarity of the
solution for the the particular choice of $\hat{N}_F, \hat{m}_F$ but
is generally true for all of their possible values. In figure
\ref{gravwell} we show the $g_{00}$ component of the metric in our
solution relative to that of empty AdS where we can see the
gravitational well produced by the star.

\subsubsection{Families of stars and a critical mass}
\label{ssecFCM}

We now want to study a family of stars of increasing fermion number,
keeping the fermion mass $\hat{m}_f$ fixed. For this we would like to
find the self-gravitating solution of $\hat{N}_F$ fermions in AdS, as
a function of $\hat{N}_F$. However when solving the TOV equations,
rather than fixing the total fermion number $\hat{N}_F$, it is
technically more convenient to fix the local chemical potential
$\hat{\mu}(0)$ at the center of the star and then integrate the
equations outwards. Only after the entire solution has been computed,
can the number of fermions be evaluated from \eqref{bulkfn}.

\begin{figure}[h]
\begin{minipage}[t]{0.5\linewidth} 
 \centering
\psfrag{a}[t]{$\hat{\mu}(0)\hspace{.5cm}$}
\psfrag{b}[r]{$\hat{M}$}
\psfrag{c}{$\hat{N_F}$}
\includegraphics[totalheight=0.18\textheight]{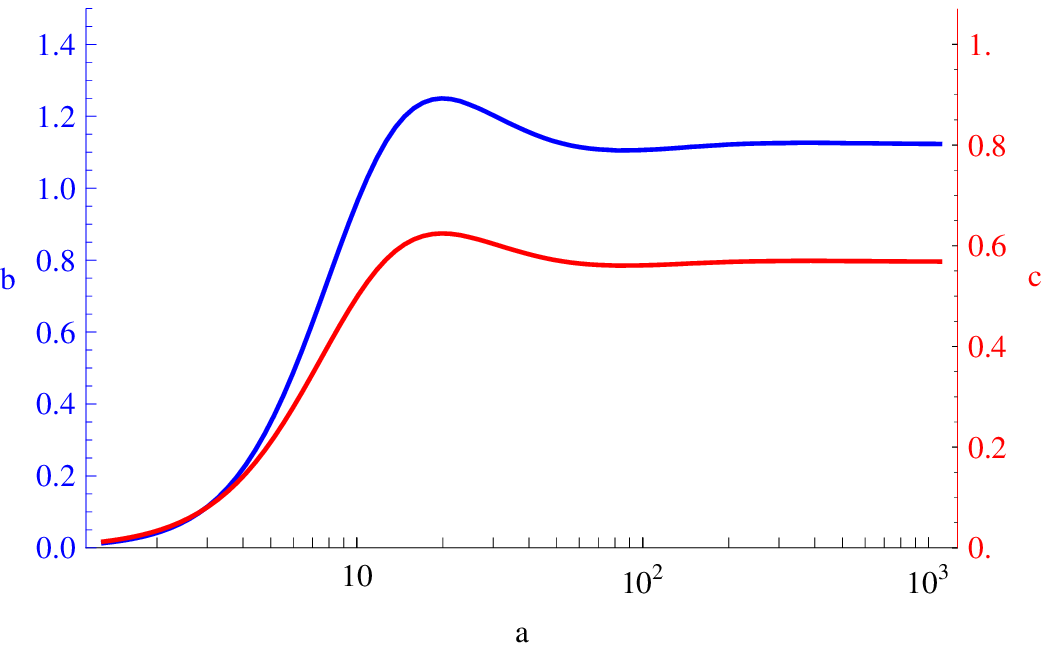}
\caption{Total mass (blue) and fermion number (red) vs chemical potential at the center of the star.}
\label{admmain}
\end{minipage}
\hspace{0.5cm}
\begin{minipage}[t]{0.5\linewidth} 
\centering
\psfrag{a}[t]{$\hat{N}_F$}
\psfrag{b}[r]{$\hat{M} $}
\includegraphics[totalheight=0.18\textheight]{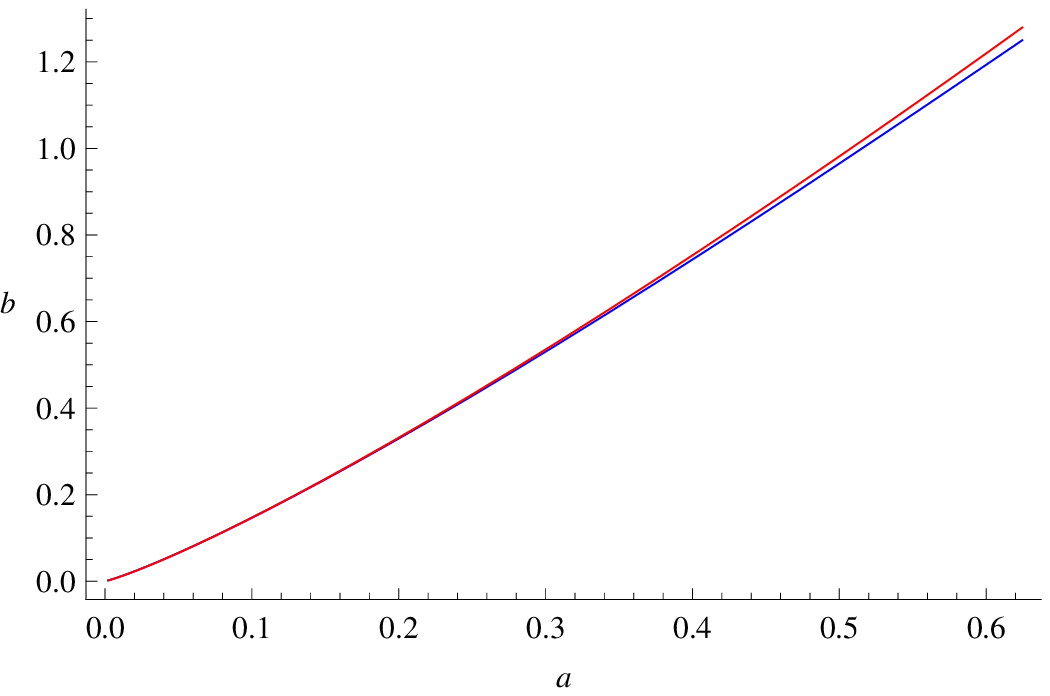}
\caption{ Mass vs Fermion number with (blue) and without (red) backreaction.}
\label{massvsfermion}
\vspace{1.5cm}
\end{minipage}

\begin{minipage}[t]{0.5\linewidth}
\centering
\psfrag{a}[t]{$\hat{N}_F$}
\psfrag{b}[r]{${\hat{M}-\hat{M}_0\over \hat{M}_0}$}
\includegraphics[totalheight=0.18\textheight]{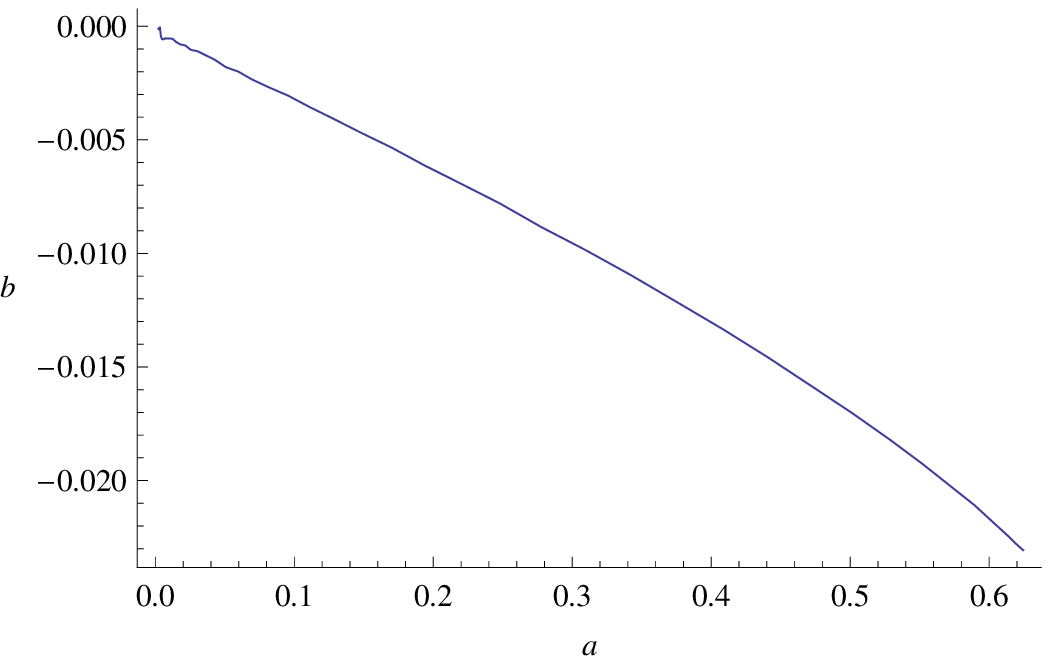}
\caption{Binding energy as a function of fermion number.}
\label{binding}
\end{minipage}
\hspace{.5cm}
\begin{minipage}[t]{0.5\linewidth}
\centering
\psfrag{a}[t]{$\hat{\epsilon}_F$}
\psfrag{b}[r]{$\hat{g}(\hat{\epsilon}_F)$}
\includegraphics[totalheight=0.18\textheight]{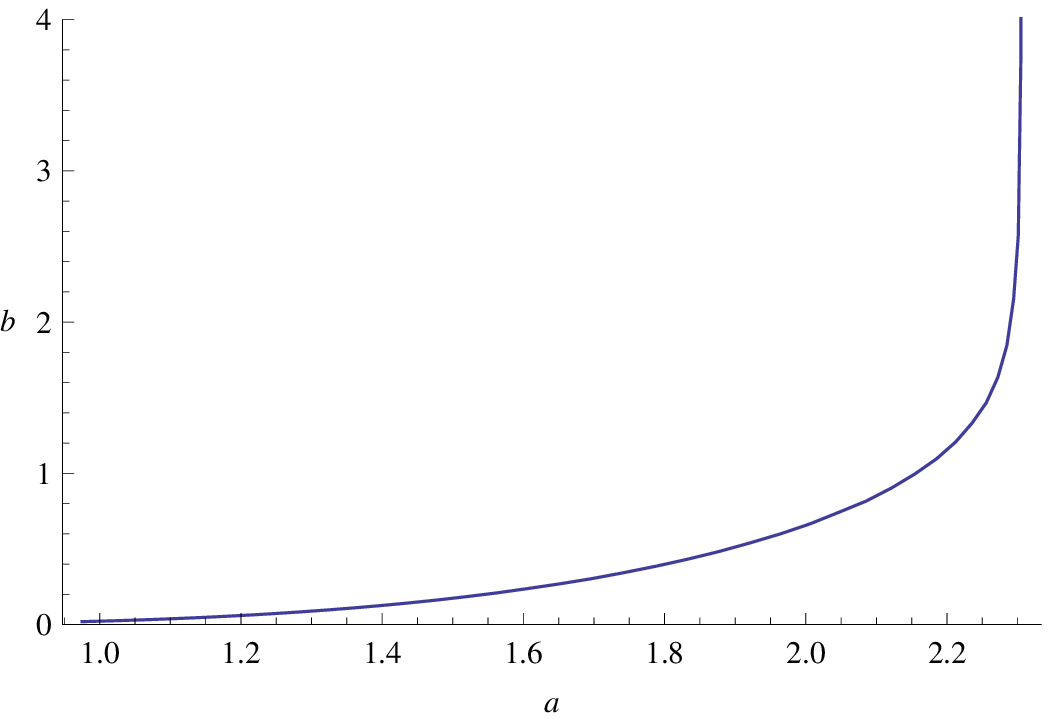}
\caption{Density of states at the Fermi energy $\hat{\epsilon}_F$. The
  value of $\hat{\epsilon}_F$ where the density diverges corresponds
  via relation \eqref{fermireds} to the value of $\hat{\mu}(0)$ where the mass reaches
  its highest value.}
\label{fermiblow}
\end{minipage}
\end{figure}

Following this procedure we plot in figure \ref{admmain} the mass and
fermion number\footnote{Let us mention that while $\hat{N}_F$ is of
  order one in this particular example, it does not mean that we have
  a small number of fermions, since $\hat{N}_F$ is a rescaled
  variable.  The actual number of fermions is $N_F = \,\hat{N}_F\,
  c^{4/5}$ which is large for $c\rightarrow \infty$.} as a function of
the local chemical potential at the center of the star.

The most striking feature of the diagram is the existence of a
critical value for the mass and fermion number which is achieved for a
certain value $\hat{\mu}_c$ of the local chemical potential at the
center of the star. This is the Chandrasekhar, or Oppenheimer-Volkoff
limit. It will be discussed in more detail in the next subsection but
for now there are two important points that we want to emphasize:
first, the right way to read figure \ref{admmain} is that there are no
static, spherically symmetric solutions of the TOV equations with a
larger number of fermions than that achieved by the solution at
$\mu_c$. If we insist on starting with initial conditions on a
spacelike slice, corresponding to a larger number of fermions, the
solution will inevitably become time-dependent and will presumably
collapse. Second, as we will explain below the solutions with
$\hat{\mu}(0)>\hat{\mu}_c$ are unstable under radial perturbations and
hence in a certain sense they should be considered unphysical. We will
return to these points in the next subsection.

In figure \ref{massvsfermion} we show the total mass as a function of
the total number of fermions (keeping $\hat{m}_f$ fixed). In the same
graph we can see what would be the total energy $\hat{M}_0$ of the
same number of fermions if they were placed in AdS without taking the
self-gravitation into account, as computed from \eqref{NEF} and
\eqref{Mtot}. The difference between the two curves is very small. In
figure \ref{binding} we see the binding energy relative to
$\hat{M}_0$. These diagrams reveal a surprising feature of our system:
the self-gravitating solution breaks down and becomes unstable at a
point where the binding energy is only a few percents of the total
energy.

Finally we consider the interpretation of our solution in the
Hartree-Fock picture on the boundary. The Fermi energy on the boundary
$\epsilon_F$ is defined by equation \eqref{redshift} so we have
\be
\label{fermireds}
\hat{\epsilon}_F = \hat{\mu}(0) \sqrt{g_{00}}(0)
\ee In figure \ref{fermiblow} we plot the density of states at the
Fermi energy as a function of the Fermi energy $\hat{\epsilon}_F$,
which can be computed from \eqref{bbbb}. We see that approaching the
critical point the density of states blows up, which
signifies a break-down of the Hartree-Fock approximation.

\subsubsection{Dependence on the fermion mass }

We now consider how the previous results depend on the fermion mass
$\hat{m}_f$. In figure \ref{admmainb} we plot the total mass as a
function of the fermion mass and the chemical potential. We see that
the critical mass and the value $\hat{\mu}_c$ at which criticality is
achieved depend on the fermion mass $\hat{m}_f$. The qualitative
behavior is the following: for small fermion mass criticality is
achieved for higher values of the chemical potential, where the
fermions are in the relativistic regime, and the radius of the star
(which we have not plotted) is large relative to the AdS radius. For
heavier fermions we have criticality in a regime where the fermions
are non-relativistic and the size of the star is smaller than the AdS
radius. In figures \ref{massvsfermionb} and \ref{3dadm} we plot the
mass and density of states respectively, as a function of the Fermi
energy and fermion mass.

\begin{figure}[h]
\begin{minipage}[t]{0.5\linewidth} 
 \centering
\includegraphics[totalheight=0.22\textheight]{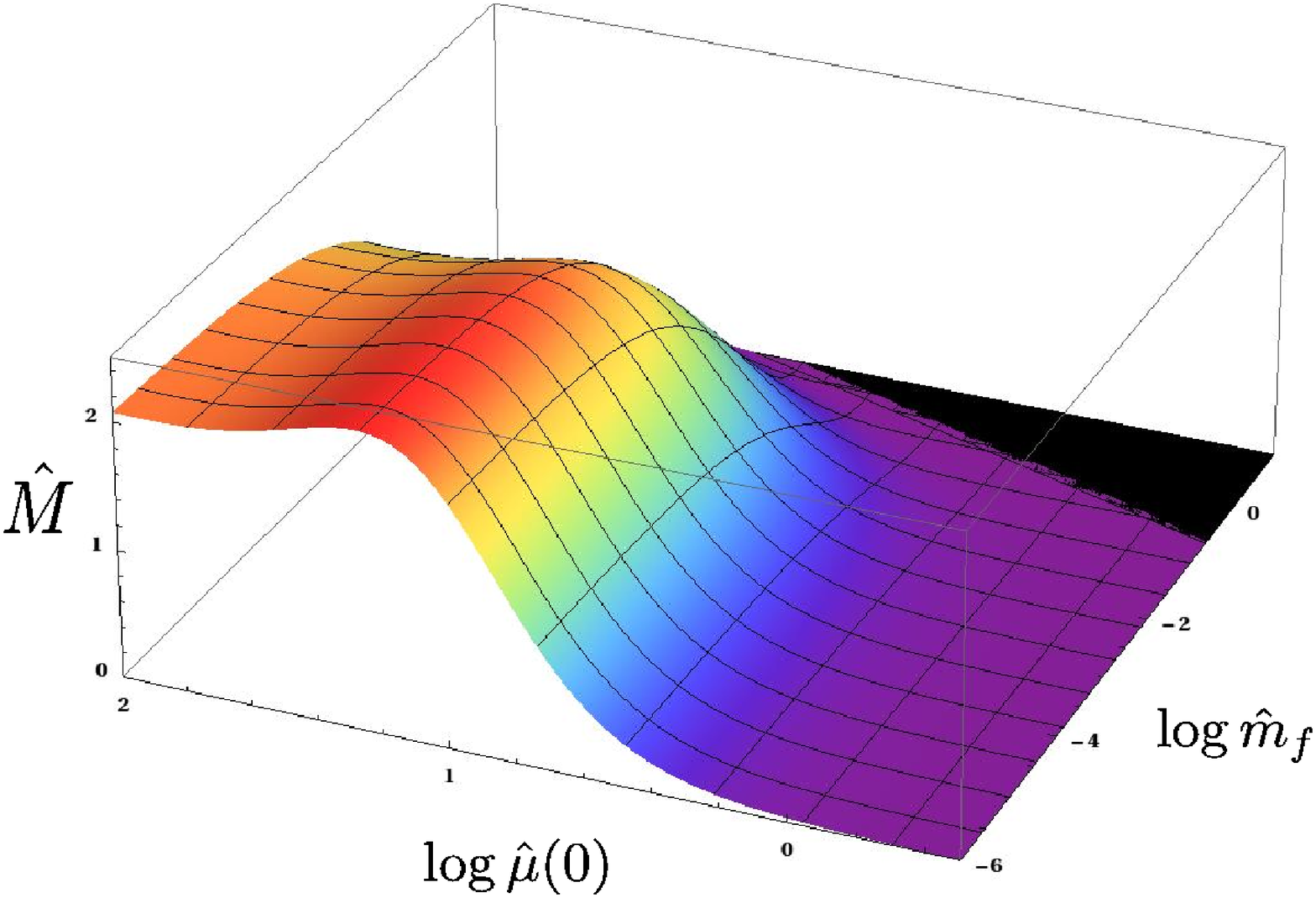}
\caption{Total mass vs chemical potential and fermion mass.}
\label{admmainb}
\end{minipage}
\hspace{0.5cm}
\begin{minipage}[t]{0.5\linewidth} 
\centering
\includegraphics[totalheight=0.22\textheight]{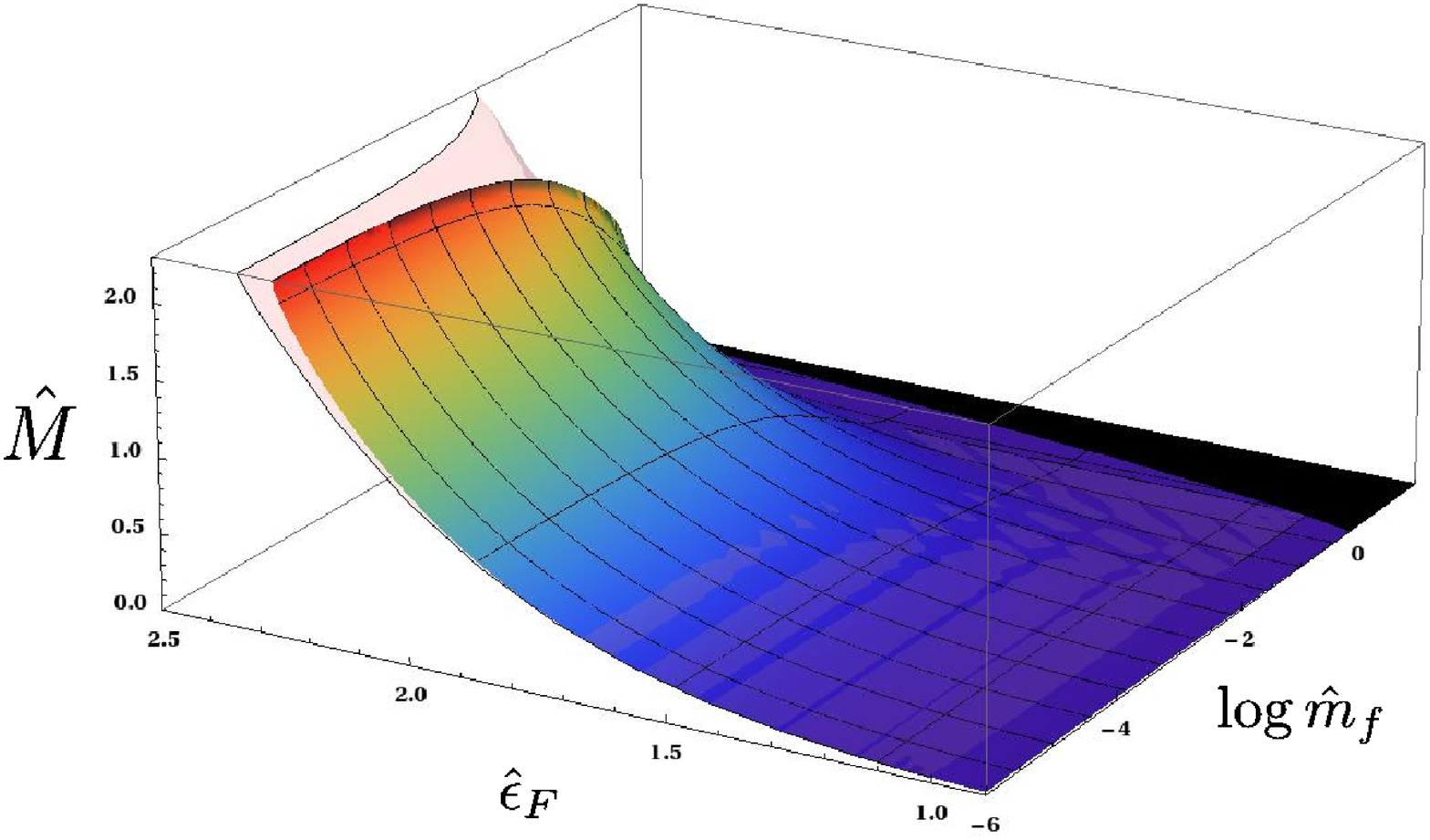}
\caption{Mass vs Fermi energy and fermion number. The underlying pink graph
is the result without self-gravity.}
\label{massvsfermionb}
\vspace{1.5cm}
\end{minipage}
\end{figure}

\begin{figure}[h]
 \centering
\includegraphics[totalheight=0.22\textheight]{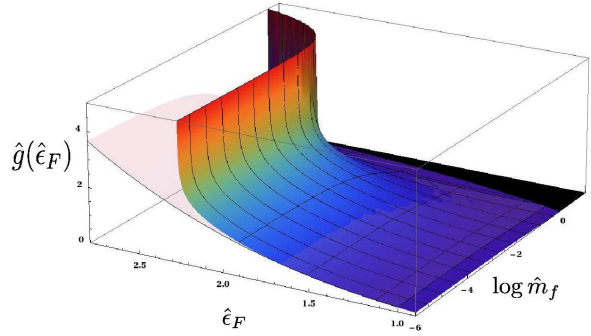}
\caption{Density of states at the Fermi energy vs Fermi energy and fermion mass.
plotted against the graph without backreaction.}
\label{3dadm}
\end{figure}

\subsection{The Chandrasekhar limit}
\label{ssecCL}

The most important qualitative feature of our numerical results is the
existence of a maximum value for the mass of a degenerate fermionic
star. This value is achieved for a specific critical local chemical
potential $\hat{\mu}_c$ at the center of the star. It is possible to
find static spherically symmetric solutions with higher values of
density at the center if we take $\hat{\mu}(0)>\hat{\mu}_c$, however
it is important to notice that increasing $\hat{\mu}(0)$ above
$\hat{\mu}_c$ does not correspond to increasing the total number of
particles. As we can see from figure \ref{admmain} the number of
particles has a maximum at the same value of $\hat{\mu}(0)$ where the
mass reaches its critical value. Solutions with
$\hat{\mu}(0)>\hat{\mu}_c$ correspond to a different radial
distribution of roughly the same number of particles. Moreover as we
will argue below solutions with $\hat{\mu}(0)> \hat{\mu}_c$ are
unstable under radial perturbations.

What happens if we insist on looking for solutions with a larger
number of particles? We could do that by starting with initial data on
a spacelike hypersurface with a total number of fermions higher than
the critical one. Then it is clear that these initial data will evolve
into a time-dependent solution describing a collapsing star. This is
the point where the degeneracy pressure of the fermions can no longer
balance the gravitational attraction, in other words the
``Chandrasekhar limit''.

Let us now turn to the stability of solutions with
$\hat{\mu}(0)>\hat{\mu}_c$. To check the stability we have to linearize
the equations of motion around the solution and compute the
frequencies of the normal modes $\omega_n$, which describe small
oscillations around the solution with time dependence of the form
$e^{-i\omega_n t}$. In general the frequencies are real and the
perturbations are simply oscillating with time, while their amplitude
remains bounded. If however there is a linearized mode with imaginary
frequency, that is with $\omega^2<0$, then we see that the perturbation
will grow exponentially with time implying that the initial solution
was unstable.

While we have not performed the linearized analysis, we summarize a
general argument (see for example \cite{Teuko}) which indicates that
the solutions with $\hat{\mu}(0)>\hat{\mu}_c$ are unstable. Let us say
that we have a one-parameter family of static solutions characterized
by the central density $\hat{\mu}(0)$. If we have a critical point
${d\hat{M} \over d\hat{\mu}}=0$ at $\hat{\mu}_c$, then around that
value, to first order, we have two different solutions one with
$\hat{\mu}_c$ and one with $\hat{\mu}_c + \delta \hat{\mu}$ with the
same total mass. The change of the profile induced by
$\delta\hat{\mu}(0)$ can be thought of as a linearized perturbation of
the solution at $\hat{\mu}_c$. From the fact that the total mass does
not change we conclude that for this perturbation at $\hat{\mu}_c$ we
have $\omega^2 = 0$, i.e. we have a zero mode. Generically we can
assume that $\omega^2$ changes sign as we cross $\hat{\mu}_c$. If we
assume that the solutions for $\hat{\mu}(0)<\hat{\mu}_c$ are stable then
we conclude that $\omega^2<0$ for $\hat {\mu}(0)>\hat{\mu}_c$. Hence
these solutions will develop a tachyonic mode past the critical point
$\hat{\mu}_c$. We expect that at the subsequent critical points more
modes may become unstable.

\subsection{Endpoint of the collapse}
\label{ssecEndPoint}

In a realistic theory the endpoint of the collapse of a degenerate
star depends on the details of the particle spectrum, interactions,
equation of state etc.  For example in the real world a white dwarf
can collapse to a neutron star which in turn may collapse to more
(hypothetical) exotic states such as quark stars and eventually a
black hole. Since we have been mainly concerned with generic features
of degenerate stars in AdS we cannot answer this question in
detail. However we notice that, at least for light fermions, the
critical mass is of the same order as that of a big black hole in
AdS. It is reasonable to expect that such an object is the most
entropic one, for given total mass. This suggests that the final
endpoint of the collapse will be a big black hole in AdS.

Nevertheless we should mention that while the endpoint of the
gravitational collapse is very likely a black hole, the onset of the
collapse itself does not have to do with black hole physics. The onset
of the collapse is marked by the failure of hydrostatic equilibrium
for a large number of fermions and can be understood in terms of the
low-lying (super)gravity modes.

In principle it would be straightforward to compute numerically the
time-dependent solution corresponding to the collapse of a fermionic
fluid in AdS towards a black hole. In practice this problem is significantly
harder than what we have done so far, since it involves solving partial
differential equations instead of ordinary ones (since the fields will now
depend on time and radius). We postpone the analysis of the time-dependent
solutions and the black hole formation to future work.

\subsection{Scaling regime}

In this section we will explore some features of the solutions deep
inside the unstable regime $\hat{\mu}(0)\gg\hat{\mu}_c$.

\subsubsection{Limiting solution}

\begin{figure}[h]
\centering
\psfrag{a}[t]{$\hat{k}_F(r)$}
\psfrag{b}[r]{${r\over \ell}$}
\includegraphics[totalheight=0.18\textheight]{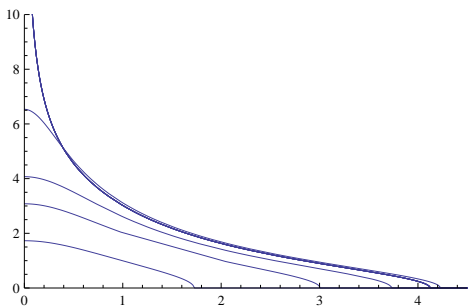}
\caption{Profile of the local Fermi momentum $\hat{k}_F(r)$
  for increasing values of $\hat{\mu}(0)$ and the convergence to a
  limit curve.}
\label{limitingcurve}
\end{figure}

At very large values of the chemical potential at the center one finds
that the solution starts to exhibit an interesting behavior: most of
the radial profile of the chemical potential approaches a limiting
shape, while a sharp ``spike'' develops at the center of the
star. This is shown in figure \ref{limitingcurve}, where we 
plot the local Fermi momentum as a function of the radius. The
Fermi momentum is very large at the center but rapidly drops to
the limiting profile. While the spike becomes sharper and sharper as we
increase $\hat{\mu}(0)$ the total mass contained in it is finite (the
same is true about the number of fermions in the spike), hence the
total mass of the solution remains finite.

One might worry that the spike is an artifact of the numerics, but one
can actually find an analytic solution which describes its limiting
shape. In the limit where $\hat{\mu}(0)\rightarrow \infty$ we can
assume that all quantities have a simple scaling behavior as
$r\rightarrow 0$.  Plugging a power-law ansatz into the TOV equations
and looking for a solution in the regime $r\rightarrow 0$ (where the
cosmological constant can be neglected) one can fix the scaling
solution as
\begin{equation}
\label{scalings}
\begin{split}
& \rho(r) = {45\over 112\,\pi\, G}\,{1\over  r^2} + {\cal O}(r) \cr
& B^2(r) = {7 \over 4} +{\cal O}(r)\cr\
& A^2(r) = r^{1\over 5} b_3 +{\cal O}(r)
\end{split}
\end{equation}
where the constant $b_3$ cannot be fixed by a local
analysis\footnote{More generally, the scaling solution in $d+1$ bulk
  dimensions is
$$ 
\rho(r) = {1 \over r^2}\left(b_1 + {\cal O}(r)\right),\qquad B^2(r)
  = b_2 + {\cal O}(r),\qquad A^2(r) = r^{1\over d+1} \left(b_3+ {\cal
    O}(r)\right)
$$
where
$$
b_1 = {1\over 4\pi G} \,\,{(d-1)^2 (d+1) \over (d-1)^3 + d-3},\qquad
b_2 = {(d-1)^3 +d-3 \over (d-1)^3 -3d +1}
$$
and the constant $b_3$ is arbitrary.}. One can check that the numerical solution
of the spike approaches this analytic form in the limit $\hat{\mu}(0)\rightarrow\infty$.

The fact that in the small $r$ limit the TOV equations admit exact
scaling solutions in which the central density goes to infinity is a
well known fact, see for example \cite{MTW}.

\subsubsection{Oscillations of the mass}
\label{ssecOsc}

The second interesting feature of the regime $\hat{\mu}(0) \gg
\hat{\mu}_c$ is the following\footnote{This section was added to our paper after
  similar calculations were carried out in collaboration with
  S. Bhattacharyya and S. Minwalla in \cite{Bhattacharyya:2009uu}.
  K.P. would like to thank S. Bhattacharyya and S. Minwalla for very
  useful discussions on these issues during the collaboration in
  \cite{Bhattacharyya:2009uu}. We would also like to thank V. Hubeny
  and M. Rangamani for very useful comments and suggestions regarding
  the scaling regime and for bringing relevant literature to our
  attention.} : in figure \ref{admmain} we see that
in the limit $\hat{\mu}(0)\rightarrow \infty$ the mass $\hat{M}$
approaches a limiting value $\hat{M}_c$ \footnote{This limiting value
  depends on the fermion mass $\hat{m}_f$.}. However we notice that
the convergence is not monotonic, but rather the function
$\hat{M}(\hat{\mu}(0))$ undergoes small damped oscillations around the
critical value $\hat{M}_c$. In figure \ref{oscil1} we plot the same
function at a different scale, in which the oscillations become more
visible. This qualitative behavior appears in many problems of similar
kind.
\begin{figure}[h]
\begin{minipage}[t]{0.5\linewidth} 
 \centering
\psfrag{a}[t]{$\log\hat{\mu}(0)$}
\psfrag{b}[r]{$\hat{M}$}
\includegraphics[totalheight=0.18\textheight]{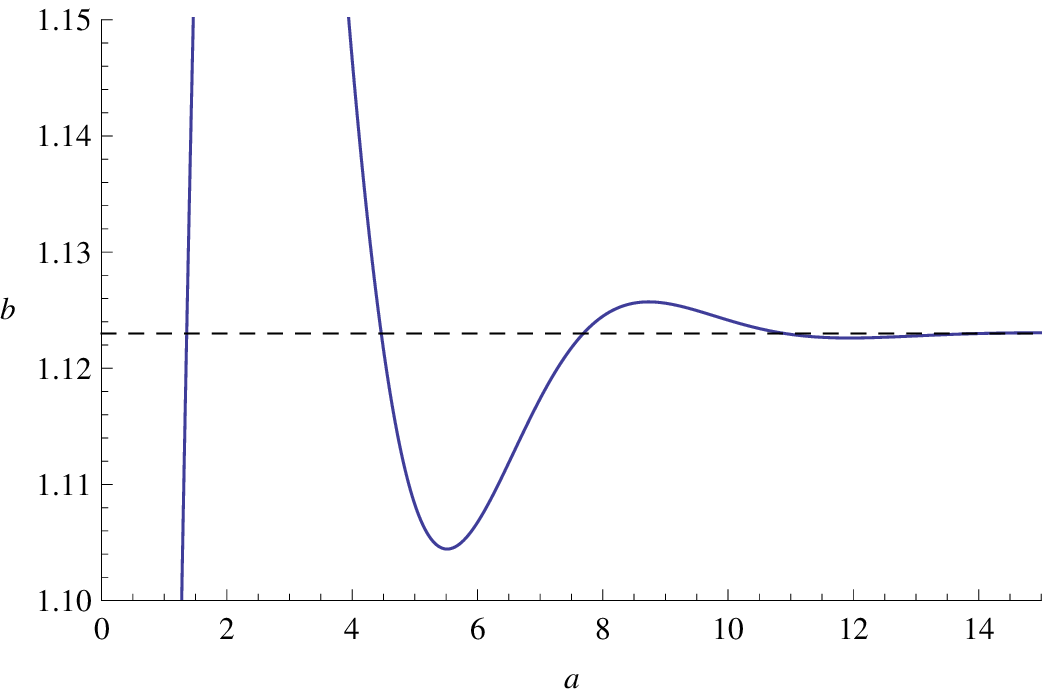}
\caption{Detail of \ref{admmain} where we see the damped oscillations
  around $\hat{M}_c \approx 1.1230$. This plot corresponds to
  $\hat{m}_f \approx 0.95$.}
\label{oscil1}
\end{minipage}
\hspace{0.5cm}
\begin{minipage}[t]{0.5\linewidth} 
\centering
\psfrag{a}[t]{$\log\hat{\mu}(0)$}
\psfrag{b}[r]{${\hat{M}-\hat{M}_c\over e^{-{3\over 5} \log \hat{\mu}(0)}}$}
\includegraphics[totalheight=0.18\textheight]{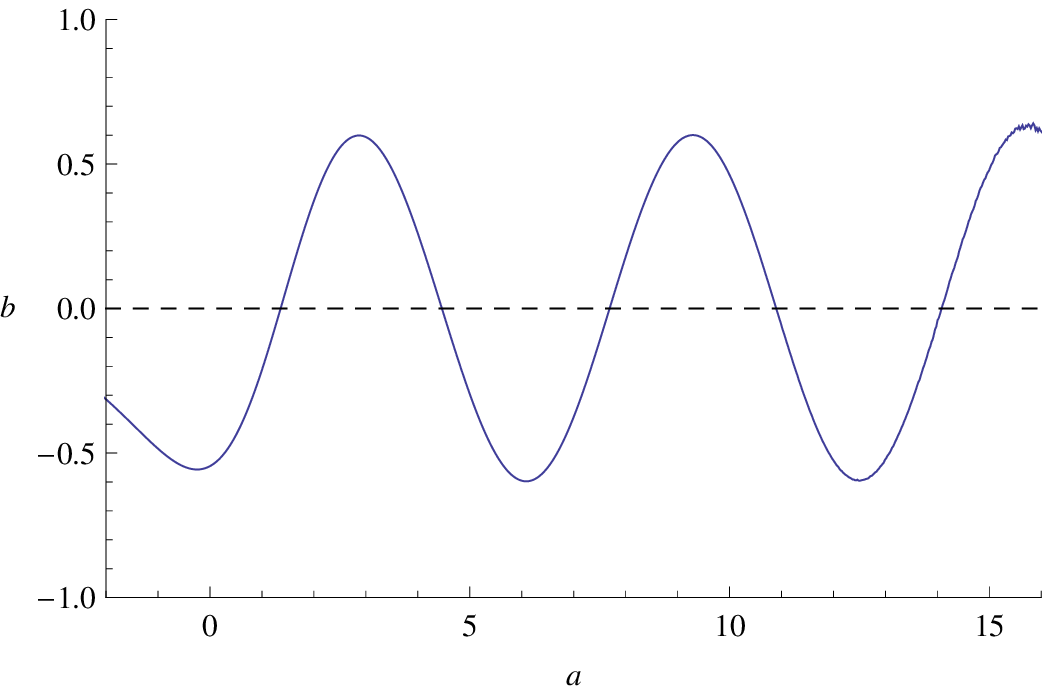}
\caption{In this plot the oscillations become more visible.}
\label{oscil2}
\end{minipage}
\end{figure}

In the limit $\hat{\mu}(0) \rightarrow \infty$ these oscillations can be
described by the following formula
\begin{equation}
\label{oscilm}
\hat{M}(\hat{\mu}) \approx \hat{M}_c  + A e^{-\gamma \log\hat{\mu}(0)} \cos(\omega \log \hat{\mu}(0) + \delta)
\end{equation}
By fitting to the numerical data we find the following values for
these constants\footnote{The precise values of the constants
  $\hat{M}_c,A,\delta$ will in general depend on the value of the
  fermion mass $\hat{m}_f$. The reported values are for $\hat{m}_f
  \approx 0.95$. On the other hand, the constants $\gamma,\omega$ are
  fixed in an $\hat{m}_f$-independent manner and can be determined
  analytically as explained in the rest of this section.}
$$ 
\hat{M}_c \approx 1.230,\qquad A\approx 0.582,\qquad \gamma\approx
0.596,\qquad \omega \approx0.973, \qquad \delta \approx3.508
$$

We will now explain that the formula \eqref{oscilm} can be justified
by analytic methods and the damping constant $\gamma$ and frequency
$\omega$ can be determined exactly. Since this aspect is not central
to the rest of the paper we will be brief and refer the reader to
\cite{Heinzle:2003ud, Vaganov:2007at, Hammersley:2008zz,
  Bhattacharyya:2009uu} for more details. The main
point is to consider the TOV equations as a dynamical system with
respect to the variable $r$. In this sense, the critical solution
mentioned in the previous subsection acts as an attractor fixed point
(in the language of dynamical systems) in the space of solutions of
the TOV equations. The critical solution corresponds to
$\hat{\mu}(0)\rightarrow \infty$. Solutions with very large but finite
$\hat{\mu}(0)$ are in a sense small perturbations of the critical
solution and hence their behavior can be understood by analyzing the
TOV equations in a linearized approximation around the critical
solution. For small $r$ the critical solution is known analytically
\eqref{scalings} and similarly the eigenvalues of small perturbations
around it can be analytically determined by performing a linearized analysis
of the TOV equations around the scaling solution. For our system 
the eigenvalues turn out to be
\begin{equation}
\label{eigenosc}
\lambda = -{6\over 5} \pm i {4 \sqrt{6}\over 5}
\end{equation}
By performing the matching carefully we find that these eigenvalues
translate into the following analytic values for the damping constant
and frequency of the oscillations of the mass
$$
\gamma={3\over 5}\qquad ,\qquad \omega={2\sqrt{6}\over 5}
$$ which are in good agreement with the numerical values mentioned
above.  The small discrepancies are presumably due to various
inaccuracies in our numerics.

Before we close this section let us mention that while this
oscillatory behavior is clearly interesting, one should keep in mind
that it takes place in the regime $\hat{\mu}(0) \gg \hat{\mu}_c$ where
the solutions are unstable. It would be fascinating if there was a
boundary interpretation of the scaling behavior and the oscillations of the mass.

\section{BOUNDARY CFT}
\label{sec:boundary}

In this section we return to the boundary interpretation of the
gravitational interactions between the fermions. As we discussed in
section \ref{sec:freelimit}, in the free limit the star is dual to a
composite multitrace operator \eqref{Phi}. If we first fix the size of
the operator and then send $c$ to infinity, this correspondence is
precise. While the multitrace operator \eqref{Phi} is not protected by
supersymmetry (i.e. it is not a chiral primary), it is protected due
to large $c$ factorization: each of its constituents are (descendants
of) chiral primaries and are thus protected. Moreover at infinite $c$
the conformal dimension of composite operators is simply given by the
sum of individual dimensions.

The interactions between the constituents start to become important in
the regime where we scale the (bare) dimension of the operator as
$\Delta_0 = \varepsilon \,c$ and keep $\varepsilon$ finite.  Now we
want to understand these interactions from the boundary point of
view. We want to work in a regime where the boundary theory has a
semiclassical gravity dual, which implies that the gauge coupling must
be large (for example large $\lambda\gg 1$ in the ${\cal N}=4$
SYM). This means that we cannot analyze the boundary theory
perturbatively in the gauge coupling. However even at strong 't Hooft
coupling the theory has an expansion in ${1\over c}$. As we will see
the ${1\over c}$ expansion together with certain basic assumptions
about the boundary CFT can be used to capture some of the qualitative
features of the system that we found in section \ref{sec:numerical}.

In the free theory the conformal dimension of the operator \eqref{Phi}
is dual to the energy of the fermionic gas in the bulk. The correction
to the energy of the gas due to gravitational interactions, that is
the binding energy, is related to the anomalous dimensions of
operators of the form \eqref{Phi}, once ${1\over c}$ corrections are
taken into account. When $\varepsilon \rightarrow 0$ these corrections
become negligible but they are important when $\varepsilon$ is of
order 1.  In a certain sense these corrections are controllable as we
turn on $\varepsilon$ slowly.

The precise form of the ${1\over c}$ corrections to the composite
operators will depend on the details of the conformal field theory. So
far we have mainly considered the neutron star in AdS, in a model
where the only degrees of freedom consist of a massive fermion coupled
to general relativity. Of course, one of our main interests was to
determine the dual description of the neutron star and in particular
to understand some microscopics of the neutron star instability and
its presumed collapse into a black hole. In order to achieve this, we
need to embed our discussion in a proper AdS/CFT duality, like
e.g. the duality between ${\cal N}=4$ SYM theory and type IIB string
theory on AdS$_5\times $S$^5$. However, one immediately see that this is
potentially difficult. The mass of the fermion $\ell\Delta_0$ scales
like $N^{2/5}\ell$ in the limit we consider, which is larger than the
typical mass of an excited string state $1/\ell_s\sim (\lambda)^{1/4}
\ell$ in the 't Hooft limit. Therefore, one expects that both massive
string degrees of freedom, as well as a large number of Kaluza-Klein
modes of the massless string degrees of freedom, should be included to
properly discuss the physics of the neutron star.

Before discussing these issues in more detail, it is instructive to
imagine a situation where these additional degrees of freedom have
somehow been decoupled. Though perhaps unrealistic it still
interesting to see how much physics such a toy model could capture. In
the rest of this section we will focus on a ``toy model'' CFT, where
the only relevant fields are the graviton and the fermion, up to a
cutoff of the order of the Planck scale in the bulk.

\subsection{Graviton exchange and first correction to anomalous dimensions}

In the toy model setup the bulk physics is completely described by a
massive fermion coupled to general relativity. The corresponding field
theory statement is that the only low-lying ``single trace operators''
in the field theory are the operator $\Psi$ and the stress tensor
$T_{\mu\nu}$, together with their conformal descendants. Of course we
also have to include their multi-trace composites, as required by
crossing symmetry. However these do not correspond to new fields in
the bulk, but rather to multi-particle excitations of the basic
fields. For simplicity we will restrict attention to the case $d=4$
only.

In this simple context, it is possible to estimate the point where the
instability will appear, using known properties of the operator
expansion of conformal field theories in four dimensions.  The idea is
to consider the anomalous dimension $\Delta$ of the composite operator
${\bf \Phi}$ as a function of particle number $N_F$, and to determine
when $d \Delta / d N_F$ changes sign, which is the direct analogue of
the computation we did in the bulk. Now in order to compute the
anomalous dimension of the operator ${\bf \Phi}$, we are going to use
the fact that it is built out of operators $\Psi$ and their
derivatives, and the relevant anomalous dimensions can be extracted
from correlations functions of $\Psi$ with itself. To compute these
correlation functions, we will use the fact that in the OPE of $\Psi$
with itself only the stress-tensor can appear to leading order in
${1\over c}$, apart from multi-particle composites made out of
$\Psi$. This simplification may not be accurate in a complete unitary
CFT but by assumption this will happen in our toy model. Altogether
this computation will therefore be a direct field theory analogue of
graviton exchange, where instead of graviton exchange we have the
exchange of a stress tensor in the intermediate channel of an OPE.

For future purpose we discuss the relevant computations in a slightly
more general setting. Thus, we consider two self-adjoint conformal
primaries $O_1,O_2$. Using the OPE we can define a composite operator
$:O_1 O_2:$ whose conformal dimension is $\Delta_1 +\Delta_2$ to
leading order in $1/c$
\begin{equation}
  O_1(x)O_2(y) = ...+:O_1O_2:(y)+...
\end{equation}
This form of the OPE, and the presence of an operator with the quantum
numbers and dimension of $:O_1O_2:$ can be extracted by performing a
conformal block decomposition of the 4-point function, if we assume
that at infinite $c$ it factorizes to a product of 2-point functions.

At the first non-trivial order in $1/c$ the dimension of the operator
$:O_1O_2:$ will be
\begin{equation}
  \Delta_{12} = \Delta_1 + \Delta_2 + {\delta_{12} \over c}
\end{equation}
where $\delta_{12}$ is the ``anomalous dimension''. To extract
$\delta_{12}$, we consider the 4-point function
\begin{equation}
  \langle O_1(x_1)O_1(x_2) O_2(x_3)O_2(x_4)\rangle =
{1\over |x_{12}|^{2\Delta_1} |x_{34}|^{2\Delta_2}} + {\cal O}({1\over
c})
\end{equation}
Here we assumed that the two-point function of $O_1$ and $O_2$
vanished to leading order in $1/c$. Now let us see what happens at
order ${1\over c}$. As explained above, we will only consider the
effect of a stress-tensor in some intermediate channel, e.g. the
$(12)\rightarrow (34)$ channel. The appearance of $T$ in the OPE of
$O_1O_1$ is controlled by the Ward identities and we have
schematically \cite{Dolan:2000ut}
\begin{equation}
  O_1(x) O_1(y) \sim...+ \Delta_1 {T \over |x-y|^{2\Delta_1-4}} +...
\end{equation}
while
\begin{equation}
  O_2(x) O_2(y) \sim ...+ \Delta_2 {T \over |x-y|^{2\Delta_2-4}}+...
\end{equation}
and also
\begin{equation}
  T (x) T(y) \sim {c \over |x-y|^8}
\end{equation}

\begin{figure}[h]
\begin{minipage}[t]{0.5\linewidth} 
 \centering
\psfrag{a}[r]{${\cal O}_1(x_1)$}
\psfrag{b}[r]{${\cal O}_1(x_2)$}
\psfrag{c}[l]{${\cal O}_2(x_3)$}
\psfrag{d}[l]{${\cal O}_2(x_4)$}
\psfrag{e}[t]{$T_{\mu\nu}$}
\includegraphics[totalheight=0.14\textheight]{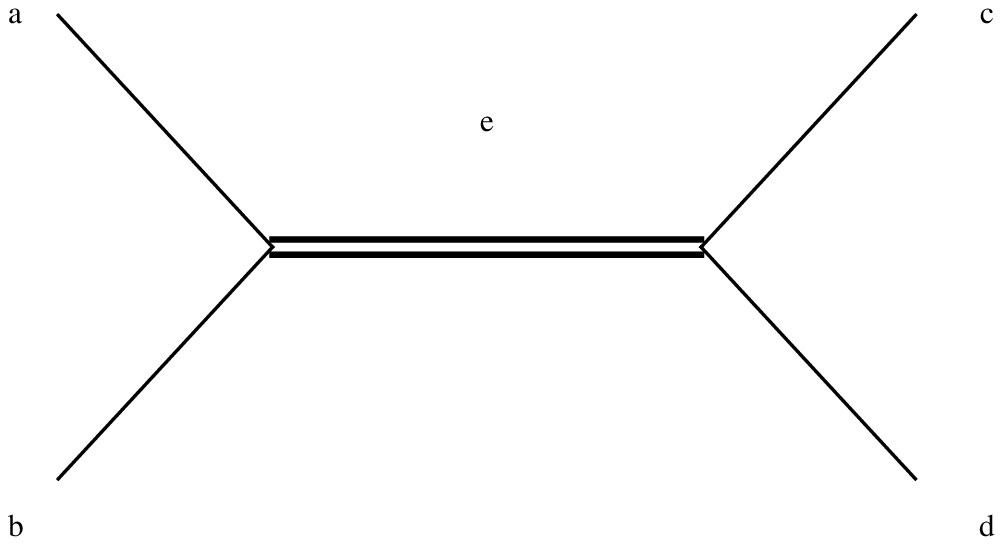}
\caption{Conformal partial wave corresponding to the exchange of $T_{\mu\nu}$
in the double OPE.}
\label{cpw1g}
\end{minipage}
\hspace{0.5cm}
\begin{minipage}[t]{0.5\linewidth} 
\centering
\psfrag{a}[r]{$\phi_1(x_1)$}
\psfrag{b}[r]{$\phi_1(x_2)$}
\psfrag{c}[l]{$\phi_2(x_3)$}
\psfrag{d}[l]{$\phi_2(x_4)$}
\psfrag{e}[t]{$g_{\mu\nu}$}
\includegraphics[totalheight=0.18\textheight]{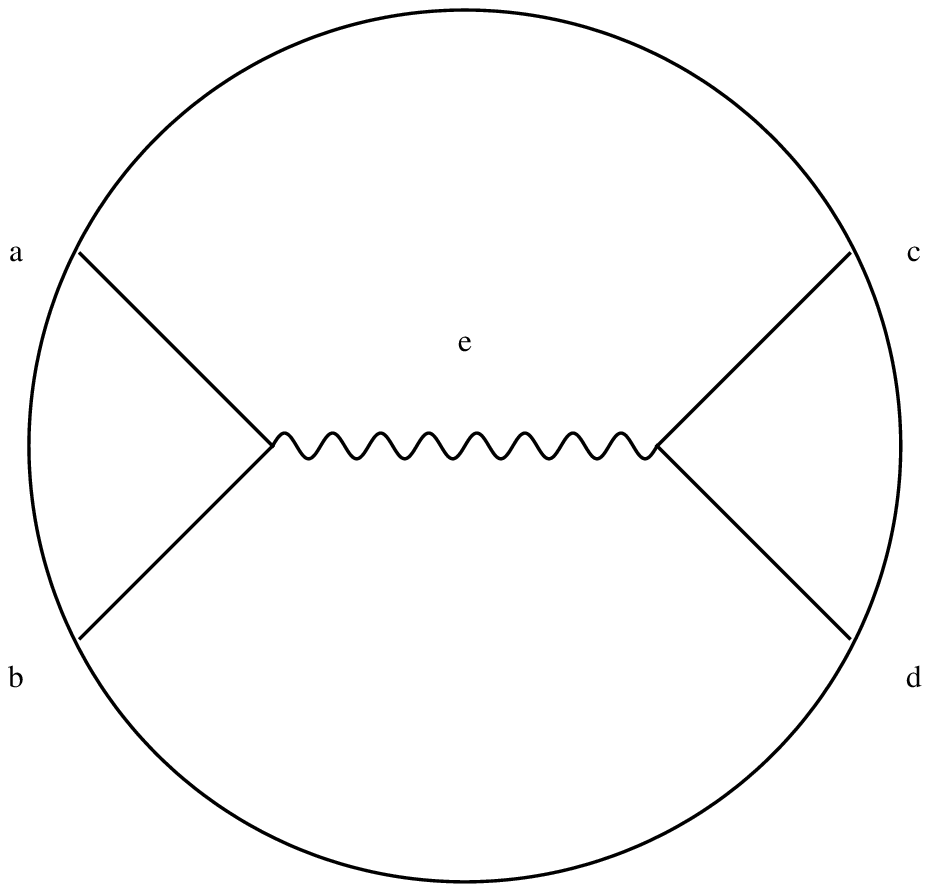}
\caption{Graviton exchange Witten diagram.}
\label{cpw2g}
\end{minipage}
\end{figure}

From this we can compute the contribution of $T$ and all of its
descendants in the double OPE $(12)\rightarrow (34)$ and we can write
the result in terms of a ``conformal partial wave.'' The final result
is given in eqn (4.3) in \cite{Dolan:2000ut}. Taking this result,
expanding it in the $(13)\rightarrow (24)$ channel, and isolating the
logarithmic singularity in that channel leads to
\begin{equation}
  \delta_{12} =- {\Delta_1 \Delta_2 \over 4c } 
\end{equation}
For e.g. $N=4$ SYM we have in the conventions of
\cite{Dolan:2000ut} $c = {N^2-1 \over 4}$ so the anomalous dimension is
\begin{equation}
  \delta_{12} \approx - {\Delta_1 \Delta_2  \over N^2} .
\label{ann}
\end{equation}

This has exactly the same dependence on $\Delta_1$, $\Delta_2$ and $N$
as one would expect from a bulk graviton exchange between two
particles of mass $\ell \Delta_1$ and $\ell \Delta_2$ separated a
distance $r\sim \ell$ i.e.
$$
G {m_1 m_2 \over \ell^2}
$$ This strongly suggests that one should also be able to determine
that to leading order in $1/c$ the conformal dimension of ${\bf \Phi}$
behaves as \be \label{deldel} \Delta(\Phi) \sim \Delta -a
\frac{\Delta^2}{N^2} \ee where $a$ is some number of order unity. The
result (\ref{deldel}) would support the bulk result that the neutron
star becomes unstable at $\Delta \sim N^2$. It is however not
straightforward to establish (\ref{deldel}). One first needs to
generalize the above computations to correlations functions involving
$\Psi$'s and their derivatives, and the latter are no longer conformal
primaries. In addition, in these four-point functions not only $T$ but
also bilinears in $\Psi$ and its derivatives will appear in the
intermediate channel, so that additional conformal partial waves need
to be included in the computation. These additional bilinear
contributions to the double OPE are necessary in order to have a
crossing-symmetric 4-point function at order ${1\over c}$: the
conformal partial wave corresponding to $T_{\mu\nu}$ exchange is not
crossing symmetric by itself. It has to be ``dressed up'' with
contributions from the exchange of fermion bilinears in order to
combine into a crossing-symmetric contribution which basically becomes
the graviton exchange Witten diagram. Finally it would be important to
clarify possible complications from the mixing of the composite
operators with other operators of the same quantum numbers, once
interactions are taken into account.

Apart from these complications it seems likely that the single
graviton exchange in the bulk can be reproduced by the $T$ exchange in
the double OPE. As we mentioned, the latter is fixed by conformal
invariance and the Ward identities, so it can be reliably computed
even at strong 't Hooft coupling if we know the conformal dimensions
$\Delta_1,\Delta_2$ and the central charge $c$. Thus the gravitational
force between two particles in the bulk can be understood on the
boundary via the $T$ exchange, which is a universal property of
CFTs. To compute the full backreaction in our scaling limit we need to
sum over all pairs of particles, but also to include the effects of
3-body, 4-body etc. interactions in the bulk as we discussed in
section \ref{subsec:qsads}. It is not clear to what extent these
higher order forces can be explained on the boundary by some universal
argument, as was the case for 2-body forces.

On the other hand, if we have established that the 2-particle force in
the bulk is reproduced by $T$ exchange, then to compute the total
backreaction from 2-particle forces we simply have to sum over all
pairs (i.e. all pairs of single trace constituents of the multi-trace
operator ${\bf \Phi}$).  This summation is an involved combinatoric
problem but it has nothing to do with strong coupling effects in the
gauge theory, so in principle it should be tractable. This part of the
backreaction is the ``Newtonian'' limit of general relativity, in
the sense that we ignore self-interactions of the gravitons. While
this is not the full answer, we would like to emphasize that the
existence of a maximum mass and the Chandrasekhar limit can also be
estimated by treating gravity as Newtonian.

\subsection{Boundary description of the instability and the collapse}

The degenerate star represents a state in the conformal field theory
with a large number of fermionic ``glueballs'', placed in a
configuration with lowest possible energy. It is a high-density state
at zero temperature.  If we keep increasing the density by adding more
glueballs, then at some point the state becomes unstable and
presumably undergoes a phase transition towards a deconfined
``quark-gluon plasma'' thermal state, the dual of a black hole in the
bulk.

It would be interesting to further explore the physics of this
instability and the meaning of the tachyonic mode in the CFT. We
expect that when the number of single trace operators multiplied
together becomes very large the resultant operator mixes very strongly
with other operators in the CFT and eventually with the operators dual
to the black hole microstates, which have a very large entropy.
Presumably the time scale set by the condensation of the tachyonic
mode contains information about the thermalization time of the dual
field theory.

\section{VALIDITY AND EMBEDDING IN STRING THEORY}
\label{sec:validity}

Our discussion so far has been general, without reference to a
specific AdS/CFT setup. In this section we would like to discuss a
couple of issues related to the validity of our approximations and the
possibility of embedding our story in a precise holographic duality.

There are various possible sources of corrections to our analysis. We
will address them in order of complexity as follows: first we will
assume that the theory in the bulk consists only of a massive fermion
and the graviton and check the validity of our approximations. Then we
will include the effect of other massless fields, Kaluza-Klein modes
on the sphere as well as stringy degrees of freedom.

\subsection{Toy model with a single fermionic field and the graviton}

A possible complication that we have not dealt with so far is that
without a fermion conservation law, the number of fermions does not
have to be constant in time as there may be processes under which
fermions convert to gravitons and vice versa.  Without a conservation
law it is natural to expect that if we wait long enough then part of
the fermions will be converted to gravitons and the final state will
be a finite temperature thermal gas of gravitons and fermions in
equilibrium.

Nevertheless we will now argue that the timescale for this
thermalization can be made parametrically large, in such a way that
the degenerate fermionic star is a good description of the system for
quite a long timescale. This is based on the ${1\over c}$ suppression
of the interactions which are responsible for changing the number of
fermions and gravitons. Let us now explain this point in more detail.
We want to estimate the decay rate of the fermion gas to a mixed gas
of fermions and gravitons. Notice that provided that the typical
wavelength of the fermions is small (i.e. $\hat{\mu}(0)$ is large),
which is true in our limit, this question can be studied locally by
looking at a small box in AdS containing the fermionic gas and asking
how quickly it converts to a mixture of fermions and gravitons.  In
other words it is a question about the validity of our equation of
state, which can be analyzed even in flat space. So we consider a box
of the gas at chemical potential $\mu$ and we try to estimate its
``lifetime'' $\tau$. The relevant scales in the problem are the
fermion mass $m_f$, the chemical potential $\mu$ and the planck mass
$m_p$ which controls the interactions.  For simplicity we will only
check the validity in the regime where $\mu>>m_f$, in which the
particles are extremely relativistic, and thus interactions are more
energetic. If the approximation turns out to be reliable in this limit
then we expect the same to be true for less energetic particles i.e. when
$\mu \approx m_f$.

In this limit we have only two dimensionful parameters $\mu$ and
$m_p$. We write the lifetime of the fluid as
$$
\tau = f\left({\mu \over m_p}\right) {1\over \mu}
$$ for some function $f(x)$ of the dimensionless ratio involved in the
problem.  Our scaling limit corresponds to $\mu \sim c^{1\over d+1}$
while in general $m_p \sim c^{1 \over d-1}$. In other words
$x\rightarrow 0$. Let us assume that $f$ has a smooth limit in that
regime and expand
$$
f(x) = x^a  + {\rm subleading}
$$ How can we determine the exponent $a$? This limit can also be
understood as one where we keep $m_p$ fixed and send $\mu$ to
zero. This is a limit of an infinitely dilute gas, where we can use
Feynman diagrams to estimate the lifetime. As a typical example
consider the leading Feynman diagram of two fermions going into two
gravitons. From such diagrams it is easy to determine the dependence
of the lifetime on Newton's constant and we find that to leading order
$\tau$ goes like ${1\over G}$, which implies $a=-d+1$, or the lifetime
$$
\tau \sim {m_p^{d-1} \over \mu^d}+ {\rm subleading}, \qquad {\rm for}\qquad 
{\mu \over m_p } \rightarrow 0
$$
or
$$
\tau \ell \sim c^{1 \over d+1} + {\rm subleading}, \qquad {\rm for} \qquad c \rightarrow \infty
$$ Hence the star becomes long-lived (in AdS radius units) at large
$c$. Similar scaling holds for other possible diagrams\footnote{Notice
  that in the regime where $\mu\gg m_f$ the particles are effectively
  massless, so from energy momentum conservation there are no diagrams
  with 2 incoming and 1 outgoing particles.}, if we assume that the
correlators scale with $c$ in the way expected from a theory with a
standard large $c$ expansion\footnote{This means that the connected
  correlator of $n$ ``single trace'' operators scales like
  $c^{{2-n\over 2}}$, in a normalization where the 2-point functions
  of single trace operators are of order one.}.

\subsection{Including other modes}

The effects of these new fields can be roughly divided in two
categories: first they may lead to a modification of the equation of
state for the fermions (or even to a more drastic breakdown of the
approximation of a fermionic fluid in the bulk) and second they may
introduce new long range forces (for example electromagnetic ones if
the fermions carry R-charge). The effect of new long range forces is
easier to deal with: one has to solve the TOV equations coupled to
additional massless modes. For example we will discuss in more detail
the inclusion of an electric field in the next section. As we will see
the qualitative behavior of our system remains the same. In general
the number of massless modes which can be sourced by the fermions
will be small and we expect that generically the gravitational
coupling will be the dominant one\footnote{One way to explain this
  expectation is that other long range forces will only couple to the
  ``charges'' of the basic fermion $\Psi$, while gravity couples to
  the mass of $\Psi$ as well as the kinetic energy of the fermions
  i.e. the ``derivatives'' acting on $\Psi$.}. So we do not expect
that the inclusion of new long range forces will modify our results
qualitatively.

On the other hand the corrections to the equation of state may be more
drastic.  To be concrete, we will first discuss whether these
corrections are under control in the usual IIB AdS$_5\times$ S$^5$
background dual to ${\cal N}=4$ SYM. In this case we run into a
difficulty.  The mass of the fermion $\ell\Delta_0$ and the chemical
potential $\mu$ scale like $N^{2/5}\ell$ in the limit we consider,
which is parametrically larger than the typical mass of an excited
string state $1/\ell_s\sim (g_s N)^{1/4} \ell$ at weak string
coupling. Therefore, we expect that massive string states will be
produced by the interactions between the fermions. Notice that our
scaling arguments of the previous section do not apply since we can
also have diagrams of two fermions producing a single massive string
state, for which the ${1\over N}$ suppression is not sufficiently
large. Moreover the Hagedorn growth in the density of string states
(and thus in the density of possible final states) suggests that the
longevity of the star cannot be guaranteed. Instead the star may
quickly convert into a very complicated mess of stringy
modes\footnote{We would like to thank S. Minwalla for very helpful
  comments about the validity of our approximations.}.

An alternative AdS/CFT realization of our star would be in M-theory
backgrounds, i.e in AdS$_4\times$ S$^7$ or AdS$_7\times$ S$^4$. The
main advantage of these theories is that all new physics beyond
supergravity is related to the single UV scale $m_{11}$, the
eleven-dimensional Planck mass. Since we are working in a limit where
all energy scales are parametrically smaller than $m_{11}$ we can
safely ignore all M-theoretic objects beyond those of
eleven-dimensional supergravity.  To see this more precisely, let us
from the start consider the full eleven-dimensional picture. In this
case the massive fermion corresponds to a certain Kaluza-Klein mode of
the supergraviton multiplet on the sphere. Now let us try some
specific expressions. For simplicity we ignore factors of order 1 in
the rest of this section. In the case of M-theory on AdS$_4\times$
S$^7$ the AdS radius $\ell$ is related to the eleven-dimensional
Planck mass $m_{11}$ by
\begin{equation}
\label{ads4c}
\ell = {N^{1/6} \over m_{11}}
\end{equation}
where $N$ is the number of $M2$ branes, or the units of 4-form
flux. In other words the central charge is $c=N^{3/2}$. It is not hard
to see that in order to have nontrivial gravitational backreaction
then we need an energy density (from the 11 dimensional point of view)
which scales like
$$
\rho = N^{3/2} {1\over \ell^{11}}
$$
So the 11d chemical potential of our fluid is of the order
$$
\mu = N^{3/ 22} {1\over \ell}
$$
Plugging into the formula for the lifetime for $d=10$ we find
$$
\tau \ell= \ell {(m_{11})^9 \over \mu^{10}} +... =  N^{3/22}+...
$$ so it increases with $N$ (though very slowly). This shows that if
we have no new degrees of freedom at scales parametrically smaller
than $m_{11}$ then in the large $N$ limit our fluid becomes stable.

Let us now explain why we do not expect any such degrees of freedom in
M-theory on AdS$_4\times$ S$^7$. Consider the sequence of
compactifications of M theory on AdS$_4\times$ S$^7$ labeled by the
integer $N$. In the large $N$ limit the two scales that we are
familiar with are the AdS scale $\ell$ and the 11d planck mass
$m_{11}$ which are related by \eqref{ads4c}. Let us now assume that
there exists another mass scale, which lies somewhere between $1/
\ell$ and $m_{11}$ where new massive degrees of freedom appear, in
analogy with the string scale in IIB. Let us call this scale
$m_{new}$. We now consider how this scale behaves, relative to the AdS
scale, in the large $N$ limit. In other words, what is the limit of
the product $\ell\, m_{new}$ at large $N$? According to the AdS/CFT
duality for this system we have
$$ \ell\,m_{new} \rightarrow 0
$$ otherwise the statement that in the large $N$ limit we can
approximate M-theory on AdS$_4\times$ S$^7$ by 11d supergravity would
not make sense.

So assuming this condition, we conclude that $m_{new}$ will
necessarily have to be a new ``UV scale'' (i.e. a small length scale
relative to the size of AdS). So we learn that in the large $N$ limit
we have two UV scales $m_{new}$ and $m_{11}$. Notice that as we make
$N$ larger and larger both of these scales move even more towards the
UV, and become more and more tiny as length scales compared to the AdS
scale.  If in the large $N$ limit we had two such UV scales in
M-theory on AdS it is natural to assume that the same would be true in
flat space, since it is hard to imagine how the finite radius of AdS
would introduce a new UV scale.

On the other hand we know that 11d flat space M-theory has only one
scale $m_{11}$, so $m_{new}$ has to be proportional to $m_{11}$ in the
large $N$ limit and not parametrically separated from it. This is
because there cannot be a free parameter to tune the ratio between the
two as there are no moduli in 11d flat space M-theory. So essentially
there is only one UV scale $m_{11}$ in the large $N$ limit. Then we
can apply the arguments of the previous sections as before.

This argument suggests that our neutron star can probably be reliably
embedded in M-theory on AdS$_4\times$S$^7$.

\section{CHARGED STARS}
\label{sec:charge}

In this section we analyze the degenerate star in the case that the
fermions are charged under a $U(1)$ gauge field. One motivation to do
so is the following: in most known examples of AdS/CFT dualities the
light fields in the bulk correspond to chiral primaries on the
boundary i.e. they have nontrivial R-charge. Hence the degenerate star
will source the corresponding electromagnetic field, which in turn
will induce an additional electromagnetic force on the fermions. We
would like to understand to what extent the inclusion of these forces
will modify our results. Another motivation is the recent analysis of
bulk systems at finite charge density, usually in the presence of a
black hole, in the context of ``holographic
superconductivity''\footnote{While this draft was being prepared the
  work \cite{Hartnoll:2010gu} appeared, which has some overlap with
  the results of this section. That work focuses on bulk
  configurations of a charged fermionic fluid in the Poincare patch
  with planar sections (in contrast our stars are in global AdS and
  have spherical symmetry). Presumably the solutions of
  \cite{Hartnoll:2010gu} can be recovered by taking the planar limit
  of the solutions constructed here, while scaling the fermion mass
  and charge appropriately at the same time. We would like to thank
  S. Hartnoll for discussions about their work.}.

How can we determine the equilibrium configuration of a large
number of charged fermions which are interacting gravitationally and
electromagnetically? We would like to work directly in the
hydrodynamic description and derive the analogue of the TOV equations
for charged matter. First let us understand what is the equation of
state that we have to use. Does the addition of charge modify the
equation of state \eqref{textbook}, \eqref{identity}, \eqref{rhon}? An
equation of state is defined as the relation between various intensive
quantities in the thermodynamic (infinite volume) limit. However when
we have long-range forces sourced by the fluid it is impossible to
take the infinite volume limit (the system suffers from the analogue
of the Jeans instability).

To see what is the right way to include the effect of the charge let
us remember how we incorporated the gravitational interactions of the
fermions: we used the standard equation of state, but instead of
simply solving the hydrodynamic equations $\nabla_\mu T^{\mu\nu}=0$ on
a fixed background, we solved the Einstein equations using the fluid
as a source, which led to the TOV equations. Similarly in the case of
a charged fluid what we have to do is to solve the Maxwell-Einstein
equations, where the fluid sources both the Einstein equations by its
stress energy tensor and the Gauss law for the $U(1)$ gauge field by its
charge density.

\subsection{Charged TOV equations}

Here we derive the TOV equations for charged degenerate stars. We have
the Einstein equations
$$
R_{\mu\nu} - {1\over 2} R g_{\mu\nu} +\Lambda g_{\mu\nu} = 8 \pi GT_{\mu\nu} 
$$ where the total energy-momentum tensor is now that of a perfect
Fermi fluid plus that of the electromagnetic field
\begin{equation}
T_{\mu \nu} = (\rho + p) u_\mu u_\nu + p g_{\mu\nu} + \frac{1}{4 \pi}
\left( F_{\mu \rho} F_{\nu}^{\rho} - \frac{1}{4} g_{\mu \nu} F_{\rho
  \sigma} F^{\rho \sigma} \right)
\end{equation}
Gauss's law reads
\begin{equation}
  \partial_\mu(\sqrt{g}F^{\mu\nu}) = 4\pi\sqrt{g} J^\nu
\end{equation}
Turning on only the radial electric field $E(r)$, we have $F_{01} = -
F_{10} = \sqrt{g_{tt} g_{rr} } E(r) $ as the only non-vanishing
components of the electromagnetic field strength. Inserting this
$T_{\mu \nu}$ into the Einstein equations merely has the effect of
shifting the energy density and pressure by the following
$$ 
\rho \rightarrow \rho + \frac{ E(r)^2 }{8 \pi} \qquad \mbox{and}
\qquad p \rightarrow p - \frac{ E(r)^2 }{8 \pi}
$$ This implies that the  form of the first of the equations
in (\ref{toveq}) gets modified to
\begin{equation}
{dM\over dr} =   V_{d-1}r^{d-1}\left(\rho  + \frac{E^2 }{8 \pi}  \right)  
\label{mass-int}
\end{equation}
whereas the other two equations in (\ref{toveq}) are only implicitly
affected through $M(r)$, which enters the functions $A(r)$ and $B(r)$
in (\ref{definitions}). Now, conservation of $T_{\mu \nu}$ yields
\begin{equation}
\frac{dp}{dr} = - (p + \rho) \frac{A^{\prime} (r)}{A (r)} + \frac{E
  (r) }{4 \pi r} \left[ \left( d - 2 \right) E (r) + r E^{\prime} (r)
  \right]
\label{ch-con}
\end{equation}
The electric field enters the system of equations as a new variable,
hence we need another equation in order to uniquely specify the
solution. This comes in the form of Gauss' law, written as
\begin{equation}
E^{\prime} (r) = - \frac{\left( d - 2 \right) E (r) }{r} + V_{d-1} \,
\rho_{ch} \, B (r)
\label{gauss}
\end{equation}
where the charge density $\rho_{ch} (r) = q_f \, {\bf n} (r) $ for
fermion charge $q_f$ and the number density $ {\bf n} (r) $ can be
obtained from the thermodynamic identity (\ref{identity}).  All the
primes denote derivatives with respect to $r$. Finally substituting
eq.(\ref{gauss}) into eq.(\ref{ch-con}) gives us the charged analog of
the last equation in (\ref{toveq}).
\begin{equation}
\frac{dp}{dr} = - (p + \rho) \frac{A^{\prime} (r) }{A (r) } +
\frac{V_{d-2}}{4 \pi } \, \rho_{ch} \, E (r) \, B (r)
\end{equation}
The boundary conditions remain the same as in section \ref{ssec:TOV}
for the uncharged case.  Also eqs. (\ref{sradius}) and (\ref{bulkfn})
for the radius $R$ and total fermion number $N_F$ remain unchanged in
form, except that the functions $M(r)$ and $B(r)$ therein are now
replaced by their charged counterparts. We now proceed to solve this
system of coupled differential equations numerically.

\subsection{Numerics of charged solutions}

Extending our results of the last section, we now look for numerical
solutions of spherically symmetric and static, but charged TOV
equations for degenerate stars in AdS$_5$.  As discussed in section
\ref{subsec:numericalresults} we are working with parameters scaled as
$G_5/\ell^3 \rightarrow 0$ and $M \ell \rightarrow \infty$.  Therefore
other physical quantities get the following rescaling
\begin{equation}
\label{chargescalingc}
\hat{E}\equiv \left({8 G_5 \over \pi \ell^3}\right)^{1/2} E,\qquad
\hat{q}_f \equiv  \left({8 G_5 \over \pi \ell^3}\right)^{-3/10} q_f
\end{equation}
where the hatted quantities are kept fixed as $c\rightarrow \infty$.
Notice that in this normalization the charge includes a factor of the
$U(1)$ coupling constant which goes like $G_5^{1/2}$. This means that
in the normalization where the charge is integral it scales like
$c^{1/5}$ which is consistent with the scaling $c^{1/5}$ of the
fermion mass $m_f$ for the case where the fermions are (descendants
of) chiral primaries.

\subsubsection{A charged degenerate star}
\label{ssecCStar}

In this section we present solutions of typical degenerate stars
comprised of charged fermions. In order to study the effects due to
varying charge, we keep both the total fermion number $\hat{N}_F$ as
well as the fermion mass $\hat{m}_f$ fixed. Then naturally, upon
varying the charge of the star, we expect to see a variation in the
total mass as well as the chemical potential profile. Figure
\ref{nschmass} shows the radial mass distribution for four stars with
different charges. The mass profiles approach the total mass as $r/
\ell \rightarrow \infty$. Notice that turning on fermion charge
gradually increases the mass even though the total number of fermions
is held fixed. This is what we expect from eq.(\ref{mass-int}), where
the square of the electric field enters the mass integral. What
happens here is that electrostatic repulsion acts against
gravitational attraction and that reduces the binding energy compared
to the uncharged case.

\begin{figure}[ht]
\begin{minipage}[t]{0.5\linewidth} 
\centering
\psfrag{cx1}[t]{ $r/\ell$}
\psfrag{cy1}[r]{$ \hat{M}(r) $}
\includegraphics[totalheight=0.18\textheight]{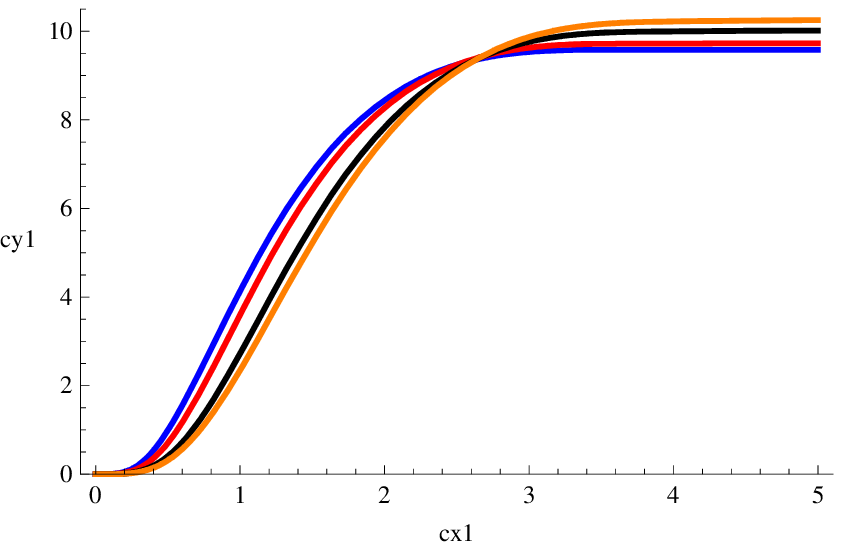}
\caption{Radial mass distributions of charged degenerate stars. Blue
  profile denotes the uncharged case; red, black, orange denote
  profiles with succesively increasing charges.  }
\label{nschmass}
\end{minipage}
\hspace{0.5cm} 
\begin{minipage}[t]{0.5\linewidth}
\centering
\psfrag{cx3}[t]{$r / \ell$}
\psfrag{cy3}[r]{$ \hat{k}_F(r)/\hat{m}_f $}
\includegraphics[totalheight=0.18\textheight]{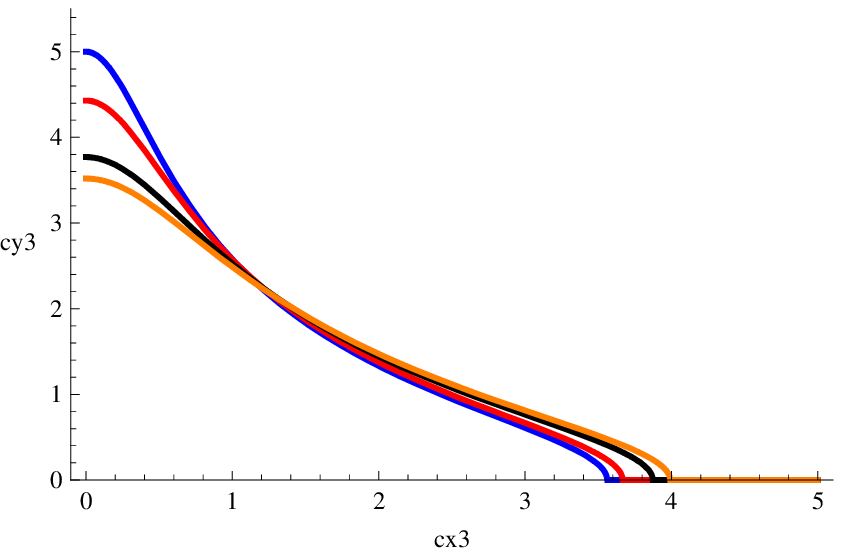}
\caption{ Radial functions of Fermi momentum in ascending order of
  charge, with uncharged case shown in blue. }
\label{nschy0}
\end{minipage}
\begin{minipage}[t]{0.5\linewidth} 
\centering
\psfrag{cx4}[t]{$r/ \ell$}
\psfrag{cy4}[r]{$r^3 \cdot \hat{E}(r) / \ell^3 $}
\includegraphics[totalheight=0.18\textheight]{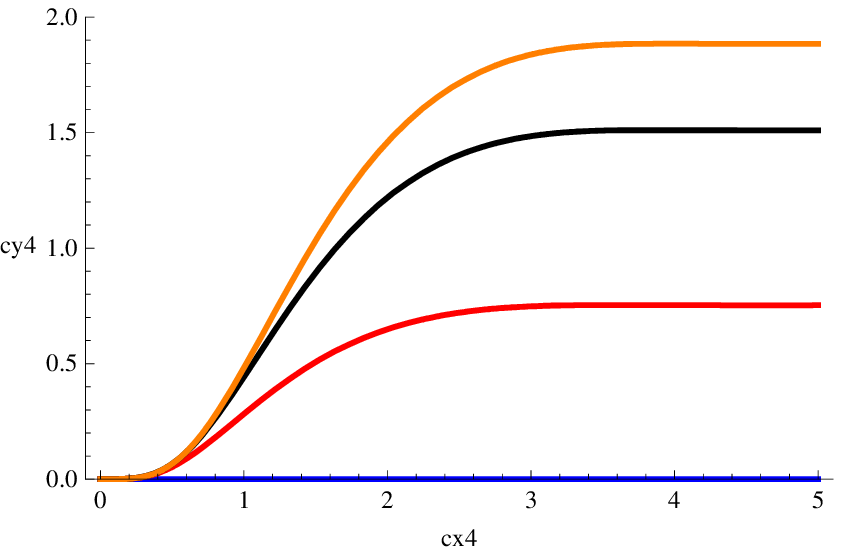}
\caption{ Electric fields inside and outside the stellar mass, scaled
  by $r^3$. Flat line (blue) denotes uncharged case, orange curve
  represents maximally charged case.  }
\label{nschE}
\end{minipage}
\hspace{0.5cm}
\begin{minipage}[t]{0.5\linewidth}
\centering
\psfrag{cx5}[t]{$r/ \ell$}
\psfrag{cy5}[r]{$g_{00}(r) $}
\includegraphics[totalheight=0.18\textheight]{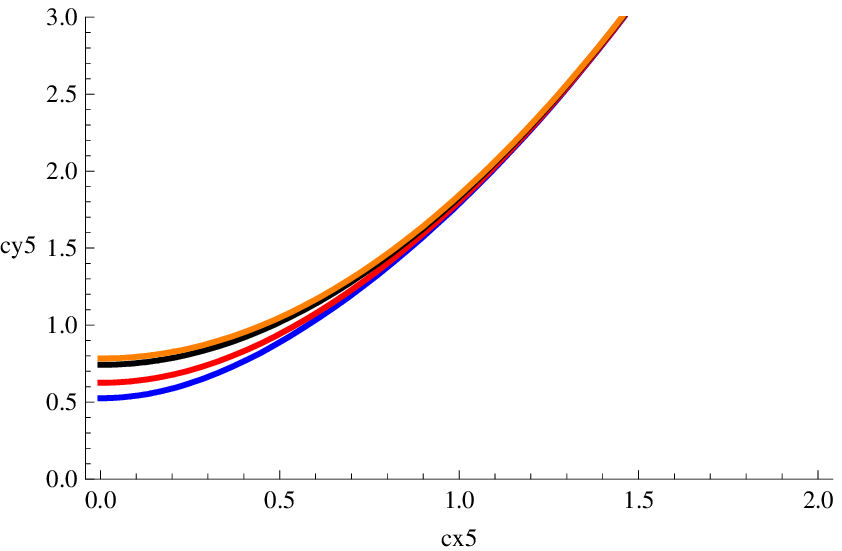}
\caption{Gravitational potential of the charged degenerate stars. }
\label{nschA_2}
\end{minipage}
\end{figure}

In figure \ref{nschy0} we plot radial profiles of the local Fermi
momentum in the bulk for the same set of charges as in figure
\ref{nschmass}. The radius of the star can be read-off from the
$x$-intercept, where the pressure, energy and number densities all
vanish.  Compared to the uncharged star (denoted by the blue curve),
we see that increasing fermion charge has the effect of increasing the
overall size of the star. Also the fermionic momentum at the core of
charged stars is less than that of the uncharged one. Both these
effects can be understood as the electrostatic repulsion acting
against the gravitational attraction. Remarkably, as we see from these
plots, this cancellation effect is only small, given that we have
determined solutions for a wide range of fermionic charges.

Figure \ref{nschE} gives the solution of the electric field of the
same stars, scaled by $r^3$. The blue profile is zero everywhere, as
it should. For the other profiles, we note that they exactly flatten
outside the radius of the respective stars. For a spherically
symmetric metric in five dimensions, this means that the electric
field outside the star behaves like one resulting from a point charge
that falls off as $1/r^3$. Finally in figure \ref{nschA_2} we see the
$g_{00}$ component of the metric. As expected, the minima of the
charged solutions lie slightly above that of the uncharged.

What we notice from the results presented in this section is that
deviations from the uncharged star upon tuning on fermion charge are
relatively only a small fraction and comparable to the difference
between the self-gravitating and unbackreacted star in the previous
section. 

\subsubsection{Family of charged TOV solutions and critical mass}

We now study the characteristics of a family of charged
solutions. This will help us identify stable solutions for static,
spherically symmetric charged degenerate stars. Furthermore this
analysis also reveals some interesting features of the solution into
the unstable phase (after the critical point). The results are
presented in figure \ref{nschTOV}. Here we have four profiles in order
of increasing fermion charges; violet being the uncharged case and
green being the maximum charge. The fermion mass $\hat{m}_f$ has been
held fixed throughout. Each point on these profiles represents a star
with a given central chemical potential $\hat{\mu}(0)$. Knowing this,
one can easily compute the energy density at the core of a star
$\hat{\rho}(0)$, which is plotted on the $x$-axis with a logarithmic
scaling.

On the vertical axis of figure \ref{nschTOV} we have the mass. Just as
in figure \ref{admmain} in section \ref{ssecFCM} once again we clearly
observe the existence of a critical mass, seen at the overall maximum
of each profile. This is the Chandrasekhar or Oppenheimer-Volkoff
limit for charged stars.  After this limit, the solutions are unstable
against radial perturbations for the same reasons as discussed in
section \ref{ssecCL}.  Compared to the uncharged case, we see that the
critical mass rapidly increases upon turning on a charge. The figure
also shows that the critical chemical potential $\hat{\mu}_c$
gradually increases and so does the amplitude of oscillations in the
unstable region like those discussed in section \ref{ssecOsc}. In
conclusion, we observe that many qualitative features regarding the
onset of gravitational collapse discussed in section \ref{ssecCL}
remain valid also for charged degenerate stars.

These observations also suggest that the holographically dual CFT picture for
charged stars is a relatively straightforward extension of the
uncharged case. In addition to the $T_{\mu\nu}$ exchange between
single-trace constituents we also have to include the exchange of the
conserved current $J_\mu$ dual to the $U(1)$ gauge field in the bulk.

Finally let us also note that the charge of the star cannot be
increased indefinitely, keeping other parameters fixed. We have
checked numerically, that in that event the solution itself breaks
down. This can be understood in the following way. From the discussion
in section \ref{ssecCStar} above we have seen that the electrostatic
force of the star acts against its self-gravity. Once this repulsion
becomes large enough to overtake self-gravity, it inhibits the
formation of a star. On the other hand, there might be a double
scaling limit which would allow us to construct arbitrarily large
charged stars and to take the ``Poincare limit'' of global AdS, which
would most likely connect to the solutions considered in
\cite{Hartnoll:2010gu}. It would be interesting to explore this further.

\begin{figure}[ht]
\centering
\psfrag{cx6}[t]{ $\log \, \hat{\rho}(0)$}
\psfrag{cy6}[r]{$ \hat{M} $}
\includegraphics[totalheight=0.35\textheight]{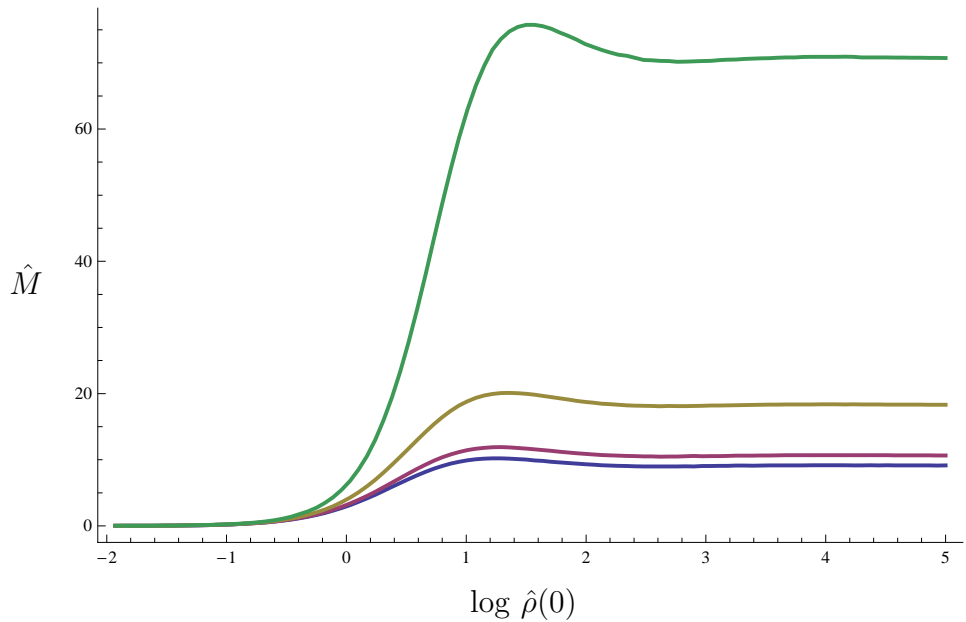}
\caption{Total mass vs central energy density (logarithmically scaled)
  for a family of stars with increasing values of fermion charge, from
  uncharged case (violet) to maximally charged (green).}
\label{nschTOV}
\end{figure}

\subsubsection{Attractor fixed points}

Another feature of the limiting solutions deep into the unstable
regime is the existence of attractor fixed points in the solution
space of TOV equations is shown in figure \ref{nschrad1}, where the
mass has been plotted versus the radius of the star for a family of
stars each with a specific central chemical potential
$\hat{\mu}(0)$. If we interpret this in a dynamical systems language,
the radial direction plays the role of the time parameter. In the
scaling limit $\hat{\mu} (0) \rightarrow \infty$, the solutions
remarkably spiral towards the attractor fixed points of the TOV
equations. In figure \ref{nschrad1}, the different profiles
denote solutions with different fermion charges, with the violet curve
being the uncharged case and the green profile being maximally
charged. These spirals are precisely the mass (and similarly radius)
oscillations that we discuss in section \ref{ssecOsc} below.  Here we
see note that these spiraling attractors/mass oscillations are a
generic feature of the TOV system of equations, both with and without
$U(1)$ charges. Figure \ref{nschrad2} shows a blow-up near one of the
fixed points. In these plots the fermion mass $\hat{m}_f$ was held
fixed. Upon reversing the set-up to vary $\hat{m}_f$ and keep the
fermion charge fixed, we have also observed similar spiraling
attractors. In this sense, the fermion mass and charge fully specify
the parameter space of all fixed points.

\begin{figure}[ht]
\begin{minipage}[t]{0.5\linewidth} 
\centering
\psfrag{cx7}[t]{ $R/\ell$}
\psfrag{cy7}[r]{$\hat{M}$}
\includegraphics[totalheight=0.18\textheight]{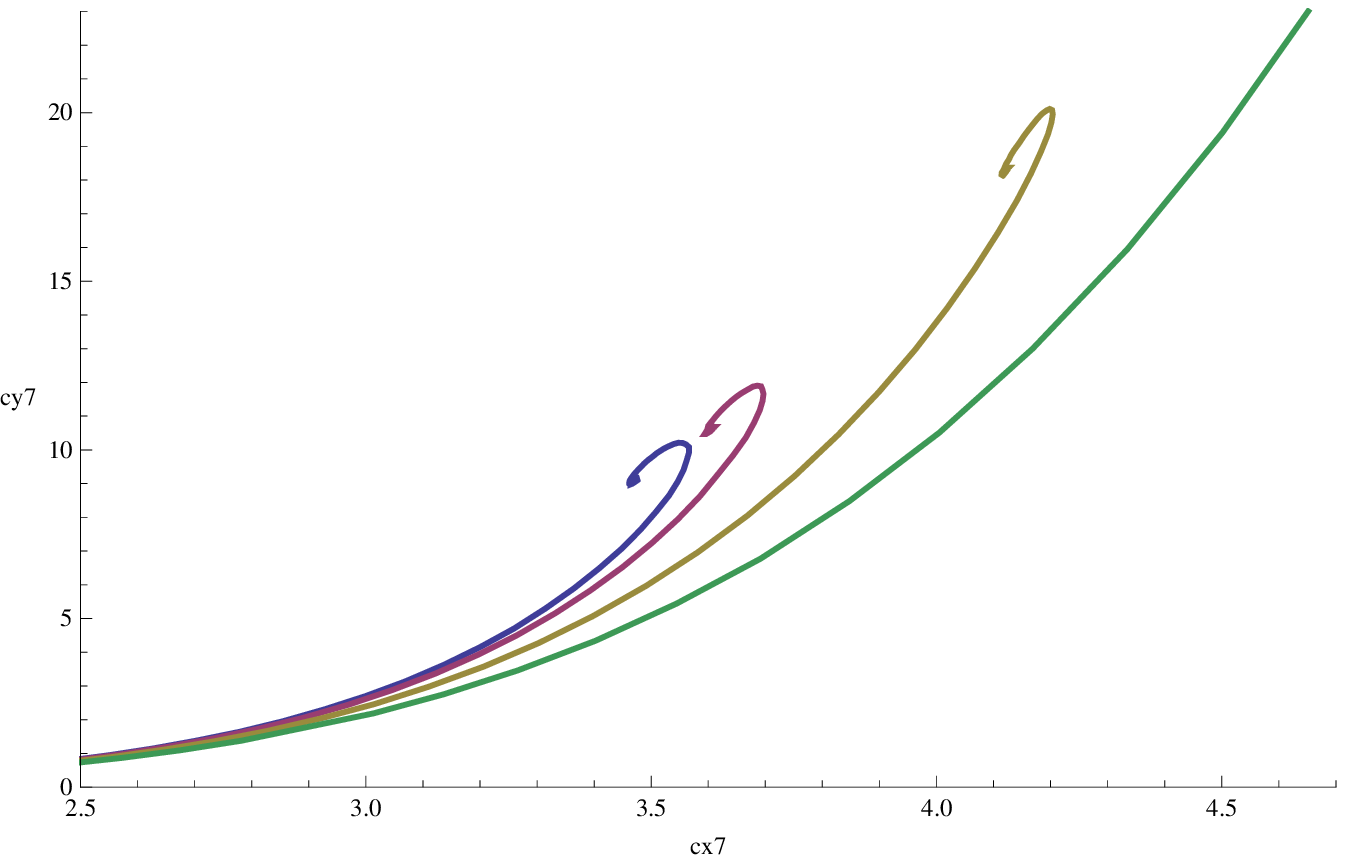}
\caption{Mass vs radius of a family of stars for increasing
  fermion charges from blue curve (uncharged) to green (maximum
  charge).  }
\label{nschrad1}
\end{minipage}
\hspace{0.5cm} 
\begin{minipage}[t]{0.5\linewidth}
\centering
\psfrag{cx8}[t]{$R / \ell$}
\psfrag{cy8}[r]{$ \hat{M}$}
\includegraphics[totalheight=0.18\textheight]{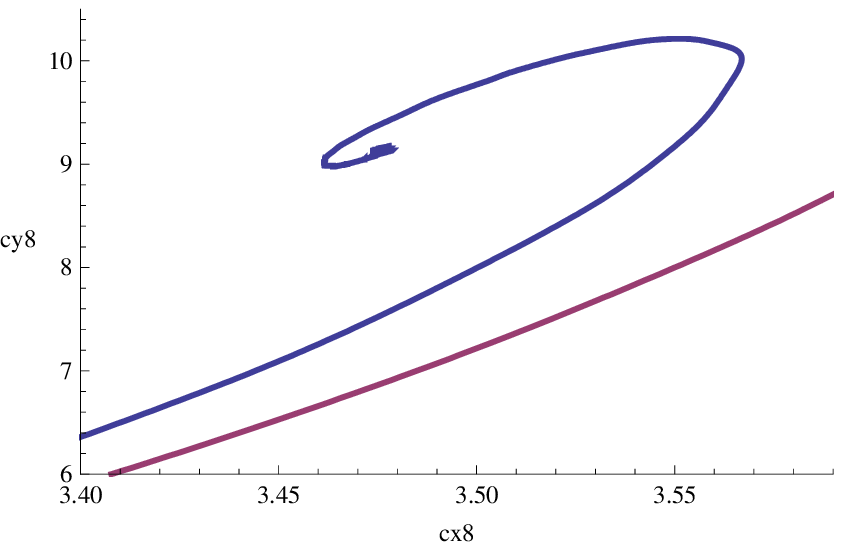}
\caption{Zooming-in the mass vs stellar radius profile near the
  fixed point. The blue curve denotes a family of uncharged stars. }
\label{nschrad2}
\end{minipage}
\end{figure}

\section{THERMAL GRAVITON STAR}
\label{sec:superheated}

In this section we will study the thermal version of a star in an
AdS$_{d+1}\times$S$^k$ background. We will consider a thermal gas of
all\footnote{In this section when we refer to a ``graviton star'' we
  refer to a system where all massless supergravity fields have been
  turned on thermally.} supergravity modes at sufficiently high
temperature so that they backreact to the geometry. Our solution will
be the analogue of the radiation star \cite{Page:1985em}, including
certain modifications due to the internal sphere S$^k$.

The main motivation for considering the thermal star is the following:
if we start with a generic, sufficiently complicated, initial state in
AdS we expect that at late times\footnote{Depending on the choice of
  initial state the thermalization may be slow in units of the AdS
  scale, in the large $N$ limit, as we argued in the previous
  section. However if we wait long enough the system will eventually
  thermalize.}  the system will be approximately described by a
thermal density matrix of an ensemble with the same values of
conserved charges (such as energy or R-charge) as the initial
state. This final thermal state may be a black hole in AdS, or a
thermal gas of all supergravity fields depending on the initial
state. We want to study the thermal gas endpoint in the case where the
temperature is high enough for the gas to backreact.

Moreover, the thermal star may describe a ``superheated'' phase of
certain gauge theories as we now explain. Large $N$ gauge theories
with classical gravity duals undergo deconfinement phase transitions
\cite{Witten:1998qj}, which are the analogue of the Hawking-Page phase
transition between a gas of gravitons at low temperature and a black
hole at high temperature. This phase transition takes place at a
temperature $T_{HP}$ which is of order one (i.e does not scale with
$N$). The low temperature phase has energy of order $N^0$ while the
high temperature has energy of order $N^2$. The jump of energy between
the two phases indicates that the phase transition is of first
order. What would happen if we considered the same system in the
microcanonical ensemble?  Notice that the thermal gas phase, while
thermodynamically subdominant, is locally stable even at
$T>T_{HP}$. This means that if we work in the microcanonical ensemble
we can push the thermal graviton gas into a superheated phase by
pumping in energy so that the effective temperature $T$ of the gas
will go above the Hawking-Page temperature $T_{HP}$\footnote{Of course this
would not make sense in the canonical ensemble.}.

Let us consider what happens to this superheated phase as we increase
the temperature $T$. From general intuition we expect that the
temperature cannot be increased indefinitely since the graviton gas
will undergo gravitational collapse, if sufficiently heavy. The energy
of a thermal gas in AdS$_{d+1}$ grows like $T^{d+1}$ so it will backreact
to the geometry when $G T^{d+1}\ell^2 \sim 1$ where $G$ is Newton's
constant. In terms of the central charge we find that gravitational
backreaction of the thermal gas will start to take place when $T\sim
c^{1/(d+1)} \ell^{-1}$ in the large $c$ limit\footnote{Notice that
  this is parametrically smaller than the Planck scale in the bulk
  which grows like $m_P \sim c^{1\over d-1} \ell^{-1}$.}. So in
theories where there is no other scale up to the Planck mass (for
example in theories without a ``string scale''\footnote{As we
  discussed in the previous section, this phase is not relevant for
  the ${\cal N}=4$ SYM, since the required temperature is much hotter
  than the Hagedorn temperature but it might be realizable in M-theory
  backgrounds.}) we expect that the superheated thermal gas phase will
indeed persist up to temperatures of the order where gravitational
backreaction will become important.  We want to compute the effect of
the backreaction and understand thermodynamic properties of the
superheated phase, for example its equation of state $E(T)$.

To compute the backreaction of a thermal gas we have to solve the
Einstein equations coupled to the stress tensor of a fluid at
temperature $T$. For massless fields in the bulk the relevant equation
of state is that of radiation. If the massless fields have spin we
simply have to multiply the energy and pressure by the appropriate
number of polarizations. This means that a thermal graviton star in
AdS$_{d+1}$ is the same as the radiation star of \cite{Page:1985em},
apart from a difference in a numerical coefficient in the equation of
state, relating the energy density to the temperature, due to the
number of graviton polarizations.  Then we do indeed see that there is
a maximum temperature beyond which the superheated phase ceases to
exist.

In this section we would like to repeat the computation of
\cite{Page:1985em} in backgrounds of the form AdS$_{d+1}\times$S$^k$
i.e with an internal sphere, since they are more relevant for
AdS/CFT. For this purpose we start directly with the $d+k+1$
dimensional supergravity equations, sourced by a $d+k+1$ dimensional
radiation fluid. After describing the structure of the resulting
equations we give a concrete example by solving these equations
numerically for the specific case of M-theory on AdS$_4\times$S$^7$.

Another motivation to consider a backreacted thermal gas on
AdS$_{d+1}\times$S$^k$ background, thus generalizing the work of
\cite{Page:1985em}, is in order to clarify the effect of the KK modes
on the S$^k$ sphere. In the previous sections of the paper we were working in
the $d+1$ dimensional reduced theory on AdS$_{d+1}$, so a natural
question was how much do the other KK modes on the sphere modify our
results. To answer this question qualitatively it is perhaps easier to
go directly to the full $d+k+1$-dimensional theory where there is no
issue about the KK reduction. As we will see the qualitative features
of the solutions are similar to the ones we have been discussing.
\subsection{The equations of motion}

We start by considering a general AdS$_{d+1}\times$S$^k$ compactification. 
We take the action in $d+k+1$ dimensions to be
$$
S = {1\over 16\pi \widetilde{G}} \int dx \sqrt{\widetilde{g}} \left(
\widetilde{R} -{1\over 2}F^2\right) 
$$ where for the $(d+1)$-form we have defined $ F^2={1\over (d+1)!}
F_{\mu_1...\mu_{d+1}}F^{\mu_1...\mu_{d+1}}$. The possible presence of
Chern-Simons terms will not be relevant for our static spherically
symmetric solutions. The equations of motion are
\begin{equation}
\label{EOMSeleven}
\widetilde{R}_{\mu\nu} -{1\over 2}\widetilde{g}_{\mu\nu} \widetilde{R} = {1\over 2\,d!} F_{\mu \mu_2...\mu_{d+1}}F_{\nu}^ {\mu_2...\mu_{d+1}} -{1\over 4}F^2\widetilde{g}_{\mu\nu}
\end{equation}
To fix the asymptotic form of the metric we first look for an
AdS$_{d+1}\times$S$^k$ solution, without the radiation fluid, of the
form
$$
ds^2 = \ell^2 ds_{{\rm AdS}_{d+1}}^2 + R^2 d\Omega_k^2
$$ where $ds^2_{{\rm AdS}_{d+1}}$ and $d\Omega_k^2$ are the metrics of the
unit radius AdS$_{d+1}$ and S$^k$ respectively. We take the form $F$ to be
proportional to the volume form of AdS$_{d+1}$. We introduce the effective $d+1$ dimensional cosmological constant
$\Lambda$ as
\begin{equation}
\label{flamb}
{F^2\over 4} = {(d+k-1)\over (d-1)(k-1)} \Lambda
\end{equation}
and then from the equations of motion we find that
\begin{equation}
\ell^2 = - {d (d-1) \over 2 \Lambda}\qquad,\qquad R^2 = - {(d-1)(k-1)^2 \over2d \Lambda}
\label{radsph}
\end{equation}

In the presence of the radiation fluid the $d+k+1$ dimensional equations of
motion \eqref{EOMSeleven} are modified as
$$
\widetilde{R}_{\mu\nu} -{1\over 2}\widetilde{g}_{\mu\nu} \widetilde{R} = {1\over 2\,d!} F_{\mu\mu_2...\mu_{d+1}}F_{\nu}^{\mu_2...\mu_{d+1}} -{1\over 4}F^2\widetilde{g}_{\mu\nu} +
8\pi \widetilde{G} T_{\mu\nu}^{fluid}
$$
where the stress energy tensor of the fluid in $d+k+1$ dimensions is
$$
T_{\mu\nu}^{fluid} = (\widetilde{\rho} + \widetilde{p}) \widetilde{u}_\mu
\widetilde{u}_\nu + \widetilde{p} \widetilde{g}_{\mu\nu}
$$ 
with the equation of state for radiation
$$
\widetilde{p} = {1\over d+k} \widetilde{\rho}
$$
The energy density is related to the temperature in $d+k+1$ dimensions by
$ \rho = \sigma T^{d+k+1}$ where the constant $\sigma$ is proportional to
total number of bosonic degrees of freedom of the theory. The velocity  satisfies $\widetilde{u}_\mu
\widetilde{u}^\mu = -1$ with respect to the metric
$\widetilde{g}_{\mu\nu}$.

We now focus on static configurations
which also respect the $SO(k+1)$ symmetry of the S$^k$. The size of S$^k$
will not be constant in the radial direction, therefore we consider the
following ansatz for the metric in $d+k+1$ dimensions
$$
ds^2 =  e^{-{2k\over d-1} \phi(x^\mu)} g_{\mu\nu} (x^\mu) + R^2 e^{2 \phi(x^\mu)} d\Omega_k^2
$$ where $g_{\mu\nu}$ is the $d+1$ dimensional metric and the form $F$
is still proportional to the volume form of the $d+1$ dimensional
space.
To proceed we will write the equations in terms of the
$d+1$ dimensional metric $g$ and the scalar $\phi$. The $d+1$ dimensional
Newton's constant $G$ is defined by the relation
$$
 {1\over 16\pi G}={R^k V_k \over 16 \pi \widetilde{G}} 
$$ where $V_k$ is the volume of the unit S$^k$. The velocity
$\widetilde{u}_\mu$ is normalized to square to $-1$ with respect to
the metric $\widetilde{g}_{\mu\nu}$, so we introduce
$$
u_\mu= e^{{k\over d-1}\phi}\widetilde{u}_\mu
$$ which satisfies $g^{\mu\nu}u_\mu u_\nu=-1$. We also define
effective $d+1$ dimensional pressure and density
$$
p = (R^k V_k)^{-1} e^{-{2k\over d-1}\phi}\widetilde{p},\qquad \rho = (R^k V_k)^{-1}
e^{-{2k\over d-1}\phi} \widetilde{\rho}
$$
After some algebra we find that \eqref{EOMSeleven} can be written as $d+1$ dimensional
equations in the form
$$
R_{\mu\nu} -{1\over 2}g_{\mu\nu}R +\Lambda g_{\mu\nu}={k(d+k-1) \over (d-1)}\left(\partial_\mu \phi \partial_\nu \phi - {1\over 2}
g_{\mu\nu} (\partial\phi)^2 - V(\phi)g_{\mu\nu}\right) 
+ 8 \pi G\left[(\rho+p)u_\mu u_\nu + p g_{\mu\nu}\right] 
$$
$$
 \Box\phi = V'(\phi) - 8\pi G p
$$
with
$$
V(\phi) = -\Lambda {d \over k-1}\left({e^{-2{kd\over d-1}\phi}\over k d }-  {e^{-2{k+d-1\over d-1}\phi}\over d+k-1} \right)  -\Lambda {d-1 \over
k(d+k-1)}
$$
and
$$
p = {1\over d+k}\rho
$$

\subsection{The graviton star solution}

Now we consider a spherically symmetric ansatz for the $d+1$ dimensional metric
$$
ds^2 =-e^{2\chi(r)-2\beta(r)} dt^2 + e^{2\beta(r)} dr^2 + r^2 d\Omega_{d-1}^2
$$ with $u_t = e^{\chi(r)-\beta(r)}$ and $\phi(r),p(r),\rho(r)$
functions of $r$ only. The scalar equation becomes
$$
\partial^2_r \phi + 
\left(\chi'-2\beta' + {d-1 \over r} \right) \partial_r \phi =  e^{2\beta}V'(\phi) - e^{2\beta} 8 \pi G  p
$$
The $tt$ equation gives
$$
{d-1 \over 2 r^2} \left( (d-2) (e^{2\beta}-1) + 2 r \beta'\right)e^{-2\beta}
+  {d(d-1)\over2 \ell^2} ={k (d+k-1) \over (d-1)}\left({1\over 2}(\partial_r\phi)^2
e^{-2\beta} + V(\phi)  \right)+ 8 \pi G \rho
$$
Defining
$$
e^{-2\beta} = 1 + {r^2 \over \ell^2} - C_{d+1} {M(r) \over r^{d-2}}
$$
with $C_{d+1} = {16 \pi G \over (d-1) V_{d-1}}$ we find
$$
M' = V_{d-1}r^{d-1} \left[\rho
+ {1\over 8 \pi G}  {k (k+d-1)\over (d-1)}\left({1\over 2}
(\partial_r \phi)^2  e^{-2\beta} + V(\phi)\right)\right]
$$
Adding the $tt$ and $rr$ equations we find
$$
\chi' = {r\over d-1}  \left( {k(d+k-1)\over  (d-1)}
{1\over 2}(\partial_r\phi)^2  + 8\pi G (\rho+p) e^{2\beta}\right)
$$
Finally from the conservation equation of the total $T_{\mu\nu}$ one can
show that the effective $d+1$ dimensional density and pressure are
\begin{equation}
\label{eoseleven}
 \rho(r) = {\widetilde{\rho_0} \over R^k V_k}e^{-{k(d+k+1)\over
    d-1}\phi_0 +(d+k+1)\chi_0}\, e^{{k(d+k-1)\over d-1}\phi(r) -
  (d+k+1)(\chi(r)-\beta(r))},\qquad p(r)= {1\over d+k}\rho(r)
\end{equation}
where $\widetilde{\rho}_0$ is the $d+k+1$ dimensional density at the center.

These equations can be used to find solutions numerically. We have to
specify the $d+k+1$ dimensional density $\widetilde{\rho}_0$ at the
center and the value $\phi_0$ of the scalar field and then integrate
outwards. The value $\chi_0\equiv\chi(0)$ at the center can be
determined as follows: we notice that the differential equations are
invariant under a constant shift of $\chi$ which corresponds to an
overall rescaling of the time coordinate. By demanding that at
infinity the metric is written in the standard parametrization of AdS
space we have to impose $\chi(\infty)=0$. Using the invariance under
constant shifts of $\chi$ this can always be achieved by an
appropriate choice of $\chi_0$. So $\chi_0$ is not really an
independent parameter that we can tune.  At this stage it seems that
the independent initial data are given by the two numbers
$(\widetilde{\rho}_0,\phi_0)$.  However we also have another boundary
condition that we have to satisfy. At large $r$ the scalar field has
to go to the minimum of the potential $\phi=0$, otherwise we would not
have an asymptotically AdS space. This shows that the pairs of initial
data $(\widetilde{\rho}_0,\phi_0)$ are actually not independent.  For
every $\widetilde{\rho}_0$ there is a  value $\phi_0$ such
that after evolving these initial data we get an asymptotically AdS
space. Solving this ``shooting problem'' numerically we can determine
a one-parameter family of graviton-star solutions parametrized by the
density at the center $\widetilde{\rho}_0$.

\begin{figure}[h]
\centering
\psfrag{y3}[t]{$M$}
\psfrag{x3}[r]{$\log \widetilde{\rho}_0$}
\includegraphics[totalheight=0.18\textheight]{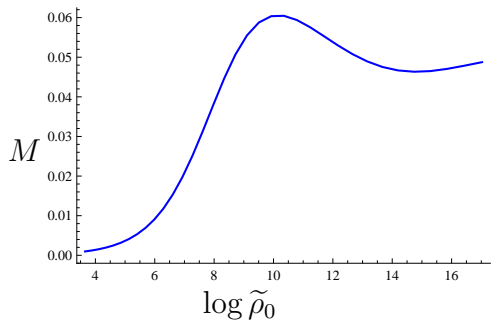}
\caption{Mass of a thermal star in AdS$_4\times$S$^7$ as a function of
  the 11-dimensional energy density $\widetilde{\rho}_0$ at the
  center.}
\label{thermalone}
\end{figure}

In figure \ref{thermalone} we show the mass of the star as a function
of $\widetilde{\rho}_0$. We notice that the qualitative features are
similar to those of a radiation star \cite{Page:1985em} in
AdS$_4$. There is a critical central density $\widetilde{\rho}_c$
beyond which the mass of the star starts to decrease signaling a
Chandrasekhar-type instability. Solutions with $\widetilde{\rho}_0>
\widetilde{\rho}_c$ are unstable under radial perturbations and
presumably collapse. The natural endpoint of this collapse is an
eleven dimensional Schwarzschild black hole in thermal equilibrium
with its Hawking radiation\footnote{We are always working in the
  microcanonical ensemble so the total energy is conserved.}. It would
be interesting to study these thermal stars in more detail and their
possible relevance for finite temperature gauge theories.

\section{DISCUSSIONS}
\label{sec:discussions}

In this paper we studied the holographic interpretation of degenerate
fermionic stars in anti de Sitter space. Clearly our work leaves many
important questions unanswered and is only a first step in a
potentially interesting direction. Let us now mention some possible
further directions.

It may be interesting to study the dynamical process of collapse
towards a black hole by computing numerically the time-dependent
solution in the case that the mass of the star exceeds the
Chandrasekhar bound. In principle this bulk computation is
straightforward, though perhaps tedious, and it might be useful for
answering certain questions about the dual deconfinement transition in
the boundary CFT.  Along these lines it might also be interesting to
explore possible connections with critical scaling during the
collapse. The Choptuik scaling has been discussed in the context of
AdS/CFT in recent works\cite{AlvarezGaume:2006dw,
  AlvarezGaume:2008qs,AlvarezGaume:2008fx, Avgoustidis:2009fn}. The
analogue of Choptuik scaling has been observed in the (driven)
collapse of neutron stars \cite{Novak:2001ck,
  Noble:2007vf}\footnote{We would like to thank L. Alvarez-Gaume for
  bringing this to our attention and for useful comments.}, so it would
presumably be relevant for the collapse of our star and might have an interesting
interpretation in the dual gauge theory.

Another interesting direction would be the computation of boundary
correlation functions in the presence of the star. It would be useful
to understand how the bulk Fermi surface is visible in terms of the
boundary correlators and compare with works \cite{Lee:2008xf,
  Liu:2009dm, Cubrovic:2009ye, Hartnoll:2009ns} where the fermionic
correlators are evaluated in the background of a black hole.  Let us
notice that our star (at least the one without charge) can only exist
in global AdS coordinates and not in the Poincare patch.  In terms of
the boundary theory the star would describe a low-temperature sector
of the theory, at temperatures below the Hawking-Page transition and
the black hole formation. So in a sense it would be the analogue of
the low-temperature phase of the systems studied in the aforementioned
works, at low temperature/chemical potential.

It would be useful to further clarify issues related
to the validity of our approximations. While we tried to address some
of these issues, a more carefully analysis would be desirable. In
particular it it would be interesting to settle conclusively whether a
degenerate fermionic star can be reliably realized in a known AdS/CFT
duality.

Finally, it is clear that the boundary CFT treatment in our paper was
on a rather basic level. We hope that more can be done in the
direction of analyzing the interacting gas of fermions directly on
the CFT side, perhaps under suitable simplifying assumptions about the
spectrum of the theory and the strength of interactions, extending the
basic estimates that we sketched in this paper. Furthermore it would be
interesting to understand how to translate the dynamical evolution of
the collapsing star (after the relevant solution has been calculated
in the bulk) into useful information about the thermalization and
deconfinement of the gas of glueballs in the gauge theory side. We leave
these questions for future work.

\section*{Acknowledgments}

We would like to thank O. Aharony, L. Alvarez-Gaume, I. Bena,
S. Bhattacharyya, E. Floratos, S. Hartnoll, S. Hellerman, G. Horowitz,
V. Hubeny, N. Iizuka, E. Kiritsis, J. Maldacena, P. McFadden,
S. Minwalla, B.  Nienhuis, H. Ooguri, M. Rangamani, L. Rastelli,
K. Schalm, K. Schoutens, M.  Shigemori, A. Strominger, M. Taylor,
S. Trivedi, and H. Verlinde for discussions.  This work was partly
supported by the Foundation of Fundamental Research on Matter
(FOM). K.P.  would like to thank the organizers of the Kolymbari
meeting, the university of Crete, the theory group at CERN and the
Saclay string theory group for hospitality during various
stages of this work.

\end{document}